\documentclass[final,3p,times,twocolumn]{myelsarticle}
\usepackage{amssymb}
\usepackage{verbatim}
\usepackage{slashed}
\usepackage{epsfig}
\usepackage{graphicx}
\usepackage{amsmath}
\usepackage{mathrsfs}
\usepackage{bm}
\usepackage{ulem}
\usepackage{epstopdf}
\usepackage{float}

\newcommand{\eqs}[1]{\begin{equation} \begin{split} #1\end{split} \end{equation} }
\newcommand{\RCP}{\mbox{$R_{CP}$}}
\newcommand{\dAu}{\mbox{d-Au}}
\newcommand{\jpsi}{$J/\psi$}
\newcommand{\sigabs}{\mbox{$\sigma^{\mathrm{abs}}_{J/\psi}$}}
\newcommand{\cf}[1]{{Fig.~\ref{#1}}}
\newcommand{\ce}[1]{Eq.~(\ref{#1})}

\newcommand{\ie}{{\it i.e.~}}
\newcommand{\eg}{{\it e.g.~}}
\newcommand{\etal}{{\it et al.}}
\def\pp{pp}
\def\pA{pA}
\def\dA{dA}
\def\AA{AA}
\def\AB{AB}
\def\dAu{d{\rm Au}}
\def\pPb{p{\rm Pb}}
\def\AuAu{{\rm AuAu}}
\def\PbPb{{\rm PbPb}}
\def\InIn{{\rm InIn}}
\def\SU{{\rm SU}}
\def\J{$J/\psi$}

\def\X{$\chi_c$}
\def\x{\chi}
\def\P{$\psi'$}
\def\p{\psi'}
\def\U{$\Upsilon$}

\def\C{c{\bar c}}

\def\e{\epsilon}

\newcommand{\Br}{{\rm Br}}

\def\NP{{ Nucl.\ Phys.\ }}
\def\PL{{ Phys.\ Lett.\ }}

\def\PR{{ Phys.\ Rev.\ }}
\def\PRep{{ Phys.\ Rep.\ }}

\def\ZP{{ Z.\ Phys.\ }}

\def\be{\begin{equation}}
\def\ee{\end{equation}}
\def\lsim{\raise0.3ex\hbox{$<$\kern-0.75em\raise-1.1ex\hbox{$\sim$}}}
\def\gsim{\raise0.3ex\hbox{$>$\kern-0.75em\raise-1.1ex\hbox{$\sim$}}}

\usepackage{lineno}

\journal{elsevier: }

\usepackage{ifpdf}
\ifpdf
\usepackage[pdftex]{hyperref}
\else
\usepackage[hypertex]{hyperref}
\fi

\hypersetup{
  pdftitle={},%
  pdfauthor={},%
  pdfsubject={},%
  pdfkeywords={},%
  pdfstartview={},%
  bookmarksopen=true, breaklinks=true, debug=true, %
  colorlinks=true, linkcolor=red, citecolor=blue, urlcolor=blue
}

\begin{document}
\begin{frontmatter}
\title{Quarkonium production in high energy \\ proton-proton and proton-nucleus collisions}
\author[address1,address9]{Z.~Conesa del Valle}
\author[address2]{G.~Corcella}
\author[address3]{F.~Fleuret}
\author[address4]{E.G.~Ferreiro}
\author[address5]{V.~Kartvelishvili}
\author[address6]{B.Z.~Kopeliovich}
\author[address7]{\\J.P.~Lansberg}
\author[address9]{C.~Louren\c co}
\author[address10]{G.~Martinez}
\author[address11]{V.~Papadimitriou}
\author[address12]{H.~Satz}
\author[address13]{E.~Scomparin}
\author[address15]{T.~Ullrich}
\author[address16]{\\O.~Teryaev}
\author[address17,address18]{R.~Vogt}
\author[address19]{J.X.~Wang}
\address[address1]{Institut Pluridisciplinaire Hubert Curien (IPHC), Universit\'e de Strasbourg, CNRS-IN2P3, Strasbourg, France}
\address[address9]{European Organization for Nuclear Research (CERN), Geneva, Switzerland}
\address[address2]{INFN, Laboratori Nazionali di Frascati, Via E.Fermi 40, I-00044, Frascati,
Italy}
\address[address3]{LLR, \'Ecole polytechnique,  CNRS/IN2P3, Palaiseau, France}
\address[address4]{Departamento de F\'{\i}sica de Part\'{\i}culas and IGFAE, Universidad de Santiago de Compostela, Santiago de Compostela, Spain}
\address[address5]{Lancaster University, Lancaster LA1 4YB,United Kingdom}
\address[address6]{Departamento de F\'{\i}sica Universidad T\'ecnica Federico Santa Mar\'{\i}a, Instituto de Estudios Avanzados en Ciencias e Ingenier\'{\i}a and Centro Cient\'ifico-Tecnol\'ogico de Valpara\'iso, Casilla 110-V, Valpara\'iso, Chile}
\address[address7]{IPNO, Universit\'e Paris-Sud 11, CNRS/IN2P3, F-91406 Orsay, France}
\address[address10]{SUBATECH, Ecole des Mines de Nantes, Universit\'e de Nantes, CNRS-IN2P3, Nantes, France}
\address[address11]{Fermi National Accelerator Laboratory, P.O. Box 500, Batavia, Illinois, 60510, U.S.A}
\address[address12]{Fakult\"at f\"ur Physik, Universit\"at Bielefeld, Germany}
\address[address13]{INFN Torino, Via P. Giuria 1, Torino, I-10125, Italy}
\address[address15]{Brookhaven National Laboratory, Upton, New York 11973, USA}
\address[address16]{Bogoliubov Laboratory of Theoretical Physics, JINR,
Dubna 141980, Russia}
\address[address17]{Physics Divsion, Lawrence Livermore National Laboratory, 
Livermore, CA 94551}
\address[address18]{Physics Department, University of California at Davis, 
Davis, CA 95616}
\address[address19]{Institute of High Energy Physics, Chinese Academy of Sciences,  P.O. Box 918(4), Beijing, 100049, China}
\begin{abstract} 
 We present  a brief overview of the most relevant current issues related to quarkonium
production in high energy proton-proton and proton-nucleus collisions along with some perspectives. After reviewing
recent experimental and theoretical results on quarkonium production in $\pp$ and $\pA$ collisions, 
we discuss the emerging field of polarisation studies. Thereafter, we report on issues
related to heavy-quark production, both in $\pp$ and $\pA$ collisions, complemented by 
$\AA$ collisions. To put the work in a broader perspective, we emphasize the need for new observables to investigate quarkonium 
production mechanisms and reiterate the qualities that make quarkonia a unique tool for many investigations in particle and nuclear physics.

\end{abstract}
\end{frontmatter}
\tableofcontents
\section{Introduction}
\label{Introduction}
The attention devoted to heavy quarkonium states started with the discovery 
of the $J/\psi$ charmonium meson in 1974 followed by the $\Upsilon$ bottomonium
meson in 1977. From the theoretical point of view, quarkonium bound states
offer a solid ground to probe Quantum Chromodynamics (QCD), due to the high scale
provided by the large mass
of the heavy quarks. 

The application of
perturbative techniques together with the factorization principle 
gave birth to the Color-Singlet Model. 
Since then, the appearance of puzzling measurements has never stopped and has led
to new challenges for theorists.
These puzzles required the introduction of new ideas providing
new probes for the understanding of QCD.
Fifteen years ago, the observation of an excess in
charmonium production reported by the CDF Collaboration,
by orders of magnitude over the theoretical predictions available at that time, 
 gave rise to the theory of Non-Relativistic QCD.

Data collected at the
Tevatron, at HERA, and at low energy $e^+e^-$ colliders 
has never ceased to challenge the existing theoretical models: 
the apparent violation of universality arising
when comparing data from the hadron-hadron and the lepton-hadron colliders, the disagreement
between the predictions for the polarization of the $J/\psi$ 
produced in hadronic
collisions and the
current data, as well as the excess of double charmonium production first observed by
Belle. The solution to these
puzzles 
requires new theoretical developments on the underlying production
mechanism(s), as the computation of higher-order corrections for characteristic
processes and the study of new production processes, in addition to investigations of new observables  
not analyzed so far.

The interest in this field and its developments are not limited to the 
issue of
the production mechanisms.
The progress in lattice calculations and effective field theories have converted
quarkonium physics
into a powerful tool to measure the mass of the heavy quarks and the strength
of the QCD coupling.
The properties
of production and absorption of quarkonium in a nuclear medium 
provide quantitative
inputs for the study of QCD at high density and temperature.
%
In fact, charmonium production off nuclei is one of the most
promising probes for studying properties of matter created
in ultrarelativistic heavy-ion collisions. Since quarkonium is heavy,
it can be used as a probe of the properties of the medium
created in these collisions, such as the intensity of interactions
and possible thermalization.

Lattice QCD calculations predict that, at sufficiently large energy
densities, hadronic matter undergoes a phase transition to a plasma of
deconfined quarks and gluons. 
Since 1986, substantial
efforts have been dedicated to
the research of high-energy heavy-ion collisions in order to reveal the
existence of this phase transition and to analyze the properties of strongly
interacting matter in the new phase.
The study of quarkonium production and
suppression is among the most interesting investigations in this field since
calculations indicate that the QCD binding potential is screened in the QGP
phase. The level of screening increases with the energy density/temperature 
of the system.
Given the existence of several quarkonium states, each of them 
with different binding
energies, it is expected that they will sequentially melt into open
charm or bottom mesons above succesive energy density thresholds. 
%
Moreover, the SPS and RHIC data on charmonium physics have brought to
light further interesting results,
among them puzzling features in proton-nucleus data, 
which highlighted new aspects of charmonium physics in nuclear reactions,
namely the role of cold nuclear matter effects.

The startup of the LHC and the opening of
the new energy frontier will offer new and challenging possibilities for the study of quarkonia.

All the above reasons have motivated the organization of 
the Quarkonium
2010 workshop held
at the Ecole Polytechnique (Palaiseau, France) 
from July 29 to July 31, 2010.
This workshop, gathering both experimentalists and theorists, was devoted 
to finding
answers to the numerous quarkonium-hadroproduction puzzles at the dawn of the
LHC era and the concurrent apogee of the Tevatron and RHIC. 
Introductory and review talks
focusing upon recent theoretical and experimental results were presented, in addition to 
six topical Working Group (WG) discussion sessions, each one 
devoted to a specific issue:

\begin{itemize}

\item WG1: Quarkonium production in $\pp$ 

\item WG2: Quarkonium production in $\pA$

\item WG3: Polarisation and Feeddown 

\item WG4: Open Heavy Flavour (vs hidden)

\item WG5: Quarkonium (production) as a tool

\item WG6: New Observables in Quarkonium production 

\end{itemize}

We present here a comprehensive review addressing
all these matters.
The following sections summarize the working group discussions, together with the achieved conclusions. 
In section 8 we prioritize directions for
ongoing and future developments.

\section{Quarkonium production in $pp$ collisions}\label{WG1}

Among the wealth of quarkonium production measurements, the 
first CDF analyses of {\it direct} $J/\psi$ and $\psi(2S)$
production\footnote{``Prompt production'' excludes quarkonium production from weak decays of more
massive states, such as the $B$ meson. ``Direct production'' further
excludes quarkonium production from feeddown, via the electromagnetic
and strong
interactions, from more massive states, such as higher-mass quarkonium states.}
at $\sqrt{s}=1.8$~TeV \cite{Abe:1997jz,Abe:1997yz} are likely the most
important to date. They revealed that the measured
rates were more than an order of magnitude larger than leading order (LO),
 color-singlet model (CSM) calculations~\cite{CSM_hadron}, believed  at that
time -- the mid 1990's-- to be the most straightforward application of 
perturbative QCD to quarkonium production.
Both these discrepancies
motivated a number of theoretical investigations on quarkonium hadroproduction,
including within the NRQCD factorization framework~\cite{Bodwin:1994jh}.  In
the latter approach, quarkonium production can also proceed via creation of
color-octet $Q \overline Q$ pairs, present in higher-Fock states, whose
effects are believed to be suppressed by powers of the relative $Q \overline
Q$ velocity, $v$.

Despite advances, there is still no clear picture of the quarkonium
hadroproduction mechanisms. Such mechanisms would have to explain both the 
cross section and polarization measurements at the Tevatron
\cite{Abe:1997jz,Abe:1997yz,Abachi:1996jq,Affolder:2000nn,Acosta:2004yw,Abulencia:2007us,Aaltonen:2009dm}
and at RHIC
\cite{Adare:2006kf,Adler:2003qs,Atomssa:2008dn,Abelev:2009qaa,daSilva:2009yy,Adare:2009js}. Obviously, the 
mechanisms involved in hadroproduction should also comply with the constraints from photo/electro-production 
(see \cite{Butenschoen:QQ10} and references therein) and production in $e^+e^-$ annihiliation 
(see \cite{Wang:QQ10} and references therein). 
Here we discuss the hadroproduction cross section only and leave the discussion
of polarization to Working Group 3~(Sec.~\ref{WG3}). We only note that the NRQCD approach, 
based on the dominance of the color-octet transition, so 
far fails to provide a consistent description of production and polarization as well as inclusive production 
in $e^+e^-$ annihiliation. 
This failure may be because the charmonium system is too light for 
relativistic effects to be small and thus the velocity expansion 
\cite{Bodwin:1994jh} may have been truncated at too low an order.  If
the convergence of the velocity expansion of NRQCD is the issue,
then one would expect better agreement between theory and data for 
the $\Upsilon$.  This could explain why only color-singlet contributions
\cite{Artoisenet:2008fc} (LO in $v$) appear to be in better agreement with 
$\Upsilon$ \cite{Affolder:1999wm,Acosta:2001gv,Abazov:2005yc,Abazov:2008za} than
$\psi$ production \cite{Aaltonen:2009dm}. 

At collider energies, quarkonium production occurs
predominantly through $gg$ channels.  Higher order $\alpha_S$ corrections
to the $S$ states have only recently been
calculated~\cite{Artoisenet:2007xi,Campbell:2007ws,Gong:2008sn,Gong:2008hk,Gong:2008ft}.
The results show that the total cross
sections do not increase much when these corrections 
\cite{Campbell:2007ws,Brodsky:2009cf} are included so that the
perturbative series seems to converge, except perhaps at large $\sqrt{s}$ where initial-state
radiation effect may need to be resummed~\cite{Lansberg:2010cn}. However, the color-singlet corrections
at high $p_T$ are very large because the QCD corrections to the CSM open
new production channels important at high $p_T$. Similar behavior has 
also been seen in photoproduction
\cite{Kramer:1995nb}.

At LO ($\alpha_s^2$), the cross section differential in $p_T^2$ scales as $p_T^{-8}$
while several different diagrams which contribute at NLO decrease as $p_T^{-6}$
or $p_T^{-4}$.  At NLO in color-octet production, the higher-order corrections
to $S$ state production do not substantially harden the $p_T$ distributions.
The LO color octet channel, gluon fragmentation, already scales with the
smallest possible power of $p_T$, $p_T^{-4}$.  However, there are substantial 
fragmentation contributions to $^1S_0$ production which enhance high $p_T$
production of these states.  At NNLO ($\alpha_s^5$), important new
channels also appear.  Color singlet gluon fragmentation is relevant in the
limit $p_T \gg m$.  Conversely, the exchange of two gluons in the $t$-channel
in the high energy limit $s \gg \hat{s}$ where $s/\hat{s}$ is the square of
the four-momentum of the colliding hadrons/partons has been studied in
the $k_T$-factorization approach.  In either limit, the
expansion can be reorganized to simplify the evaluation of the dominant
contribution.  Away from this regime, corrections to each of these approaches
may be important.  Instead of calculating the full NNLO contributions to the
CSM, NNLO$^\star$ calculations consider only real gluon emission at NNLO and
control the divergences using a cutoff.  This method has large uncertainties
arising from the sensitivity to the cutoff and the renormalization scale
\cite{Artoisenet:2008fc}.  We briefly summarize recent developments here.  For a more
complete discussion, see Refs.~\cite{Brambilla:2010cs,Lansberg:2008gk}.

\subsection{Current progress:  Tevatron and RHIC}

The contributions of the NLO color-singlet corrections reduce the discrepancy
between the inclusive CSM cross sections and the CDF $\Upsilon$ data 
\cite{Aaltonen:2009dm}.  However, the NLO rate still falls too steeply at
high $p_T$ to describe this region successfully.  The NNLO$^\star$ contribution
may be able to fill the gap between calculations and data at high $p_T$.
While the large theoretical uncertainties do not place strong constraints
on the color-octet contribution, the data no longer require them 
\cite{Lansberg:2008gk}.

Higher-order color singlet contributions to $J/\psi$ and $\psi(2S)$
production have also been calculated.  The calculation is simpler for the
$\psi(2S)$ because there is no feeddown from higher charmonium states.
CDF extracted direct $\psi(2S)$ production \cite{Aaltonen:2009dm}. The rates were 
compared to the CSM calculations.  While the higher order CSM corrections
significantly reduce the discrepancy between the calculation and the
Tevatron data, they are insufficient to remove it entirely.  At intermediate
$p_T$, the upper limit of the NNLO$^\star$ calculation is in agreement with the
data while, for $p_T > 10$ GeV, there is a gap between the calculations and
the data.  The same is true for the $J/\psi$ \cite{Lansberg:2008gk}.

NLO corrections to $^1S_0$ and $^3S_1$ color octet $J/\psi$ 
\cite{Gong:2008ft} and $\Upsilon$ \cite{Gong:2010bk} production have
been calculated. In both cases, these
corrections are small for $p_T < 20$~GeV. 
Values of the NRQCD $\langle O^{J/\psi} \big( ^3S_1^{[8]}\big)
\rangle$ and $\langle O^{J/\psi} \big( ^1S_0^{[8]}\big) \rangle$ matrix 
elements were obtained by fitting the prompt production
rate measured by CDF \cite{Acosta:2004yw} and were found to be compatible
with those extracted at LO \cite{Gong:2008ft}. Feeddown was ignored and
the $P$ state color octet contribution set to zero.
Since a satisfactory fit could not be
obtained for $p_T<6$~GeV,
these points were not included in the fit.  Note that NRQCD factorization
may break down at low $p_T$ so that resummation effects may be necessary to
describe the data in this region.

Two papers \cite{Ma:2010yw,Butenschoen:2010rq} recently appeared
with complete calculations of NLO $\alpha_s$  color-octet
production through order $v^4$ for the $^3S_1$, $^1S_0$, and $^3P_J$ channels. 
Although the two calculations agree at the level of partonic cross section,
the values of the NRQCD matrix elements extracted from the two different
fitting procedures are inconsistent. In any case, all these computations 
fail to comply with the constraint obtained in~\cite{Zhang:2009ym}
by assuming that the $e^+e^-\to J/\psi+X_{{\rm non-}c\bar{c}}$ rate measured by Belle~\cite{Belle:2009nj} 
comes {\it only} from the color-octet processes: 
\eqs{\langle 0| {\cal
O}^{J/\psi}[{}^1S_0^{(8)}]|0\rangle + 4.0\,\langle0| {\cal
O}^{J/\psi} [{}^3P_0^{(8)}]|0\rangle/m_c^2 \\ \leq (2.0 \pm 0.6)\times
10^{-2}~{\rm GeV}^3\label{eq:constraint-CO-ee}} 
at NLO in $\alpha_s$. If one keeps in mind that the color-singlet
contribution --assumed to be zero in the equation above-- saturates the experimental
measurement by Belle, it is clear that this upper bound is in fact
quite conservative, even though this measurement may be affected by
the 4 charged track requirement. The violation of this bound by these recent NLO analyses, where
color-octets dominate,
should be taken rather seriously, especially because $e^+e^-$ annihilation may be  a cleaner probe
of the inclusive charmonium production mechanisms.

On top of its earlier $J/\psi$ analyses, PHENIX has shown preliminary measurements of the $p_T$ 
dependence of the $\psi(2S)$ cross section at 200~GeV 
\cite{daSilva:2009yy}. This is the first measurement of the $p_T$ 
dependence of an excited charmonium state at RHIC. PHENIX measured the 
feeddown contribution of the $\psi(2S)$
to the $J/\psi$ to be $8.6 \pm 2.3 \%$, in good agreement with the 
world average. 

STAR has recently published~\cite{Abelev:2009qaa} measurements
 of the $J/\psi$ 
cross section in 200~GeV $pp$ collisions for $5<p_T<13$~GeV/$c$. 
This greatly extends the $p_T$ range over which $J/\psi$ data are available at 
RHIC. Although PHENIX can trigger at all $p_T$, it has so far
been limited to $p_T$ below about 9~GeV/$c$ \cite{daSilva:2009yy} 
because of its much smaller acceptance.

STAR has also recently presented a measurement of the fraction of $J/\psi$
produced from  $B$ decays~\cite{Abelev:2009qaa}.
The results appear to be consistent with the CDF measurement at an order of
magnitude higher energy. 

Unfortunately, little is known about $J/\psi$ feed down from higher charmonium resonances
such as the $\chi_c$ at RHIC.
These may significantly contribute to the
yield  of  $J/\psi$ mesons observed by PHENIX and STAR. For instance, it is likely that the
fraction of $J/\psi$ from $\chi_c$ should be independent of $p_T$, $y$  and $\sqrt{s}$. 
This may be relevant for the interpretation of the high $p_T$ STAR data \cite{Abelev:2009qaa} which 
favor NRQCD over CSM 
production in calculations with NRQCD at LO \cite{Nayak:2003jp} 
and CSM contributions up to NNLO$^\star$ (see \cite{Artoisenet:2008fc}). Both these
calculations do not include feeddown. At lower $p_T$,  PHENIX $J/\psi$ data
\cite{daSilva:2009yy,Adare:2009js} agrees with LO NRQCD including 
feeddown~\cite{Chung:2009xr}. However, in the latter case, the low $p_T$
values may call the validity of the perturbative calculation into question, especially
since the  LO NRQCD spectrum diverges as $p_T\to 0$.
Recently, NLO corrections to the CSM were evaluated as well and were shown to close the gap between the LO CSM
and the PHENIX data  at $p_T=1-2$ GeV. When the NNLO$^\star$ contribution is included at $p_T > 5$
GeV, the upper limit agrees with data \cite{Lansberg:2010vq}. We note here that the  NLO CSM
do not show any divergences at low $p_T$. However, resummation of initial state radiation may 
further improve the agreement with PHENIX data.

\subsection{First LHC results}

\begin{figure}[hbt!]
\begin{center}
\includegraphics[width=\columnwidth]{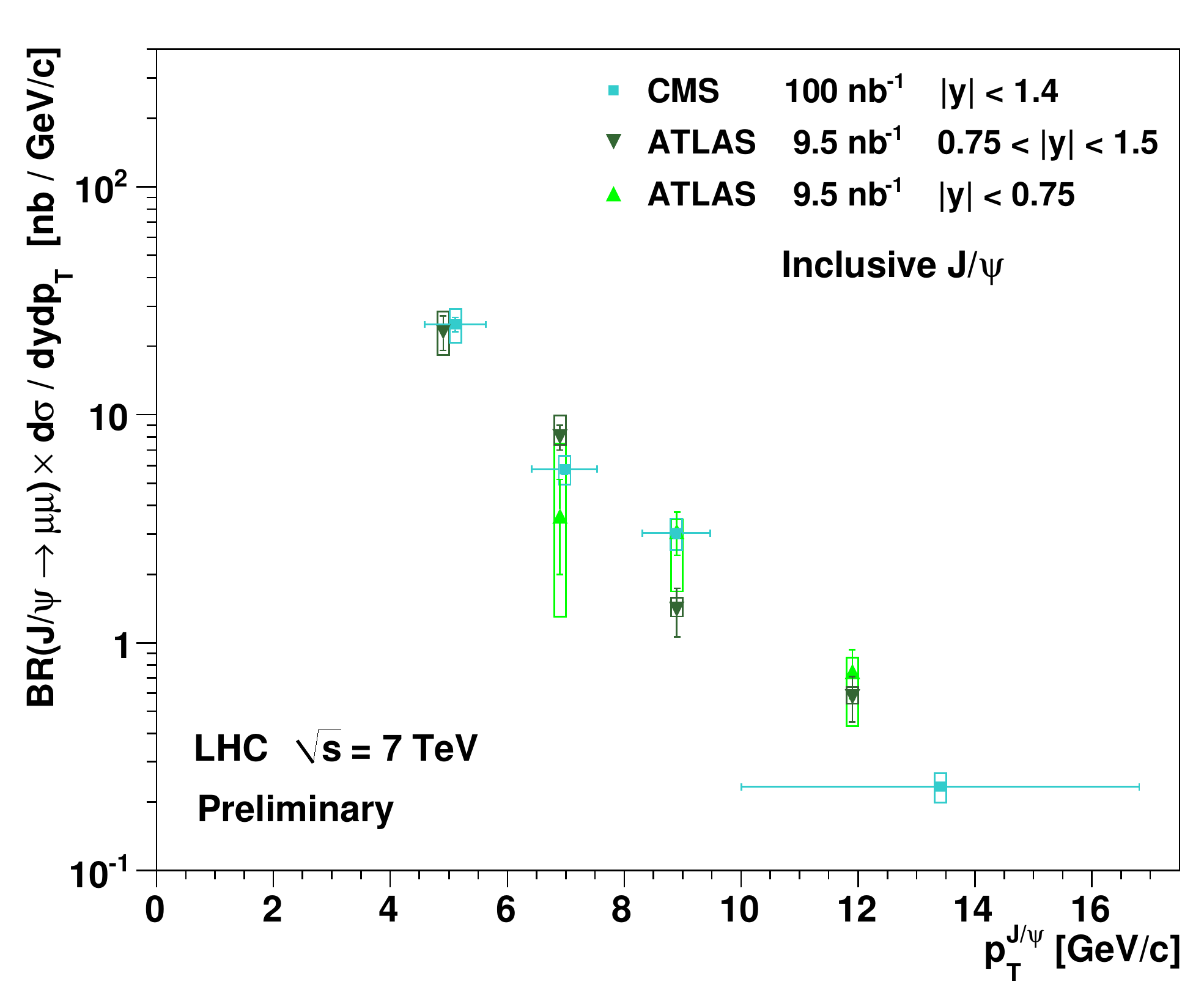} 
\end{center}
\caption{The preliminary mid rapidity inclusive $J/\psi$ $p_T$ distributions
measured by CMS ($|y|<1.4$) and ATLAS ($|y|<0.75$ and $0.75<|y|<1.5$).
Data compilation courtesy of H. W\"{o}hri.}
\label{LHCmid}
\end{figure}

The first LHC $pp$ run, at 7 TeV, has already produced superior
quality data after a short time.  The CMS~\cite{Dahms}, ATLAS~\cite{Price:QQ10},
ALICE~\cite{Arnaldi:HP10} and LHCb~\cite{Robbe:QQ10} experiments reported inclusive $J/\psi$ $p_T$ 
distributions from integrated luminosities of 100 nb$^{-1}$, 9.5 nb$^{-1}$, 11.6 nb$^{-1}$ and 
14.2 nb$^{-1}$, respectively. 
Before ALICE differential cross sections at mid-rapidity are available\footnote{
ALICE has released the inclusive differential $p_T$ cross sections at large-rapidity, but only the invariant yields are available at mid-rapidity at the time of this writeup. 
 }, no low $p_T$ data will be available at mid-rapidity, see Fig.~\ref{LHCmid}.
ATLAS and CMS experiments are equipped with large magnetic fields, so their single muon $p_T$ thresholds are too high for 
reconstruction of low $p_T$ $J/\psi$'s. However, the data taken
at $p_T > 4$ GeV/$c$ are consistent with each other even though the rapidity
ranges differ slightly. The CMS rapidity bin, $|y|<1.4$, overlaps the two
midrapidity ranges reported by ATLAS, $|y|<0.75$ and $0.75<|y|<1.5$.

\begin{figure}[hbt!]
\begin{center}
\includegraphics[width=\columnwidth]{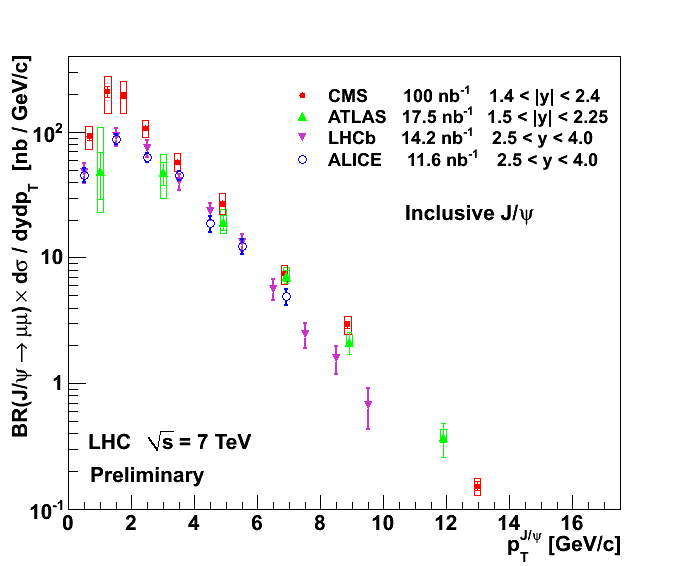} 
\end{center}
\caption{The preliminary forward rapidity inclusive $J/\psi$ $p_T$ distributions
measured by CMS ($1.4<|y|<2.4$), ATLAS ($1.5<|y|<2.25$), LHCb ($2.5<y<4.0$)
and ALICE ($2.5<y<4.0$).
Data compilation courtesy of E. Scomparin and H.~W\"{o}hri.}
\label{LHCfor}
\end{figure}

At more forward rapidities, the single muon $p_T$ from $J/\psi$ decay is high 
enough for $J/\psi$ reconstruction at $p_T \rightarrow 0$. A compilation of
the forward $J/\psi$ data is shown in Fig.~\ref{LHCfor}. The CMS and ATLAS
data, taken in similar rapidity intervals of $1.4 < |y|<2.4$ and $1.5 <|y|<2.25$
respectively, agree rather well for $p_T > 3$ GeV/$c$. They seem to differ
somewhat at lower $p_T$, but the difference is not statistically significant.
The more forward LHCb and ALICE data agree quite well with the shape of the
somewhat more central rapidity CMS and ATLAS data.  

\begin{figure}[hbt!]
\begin{center}
\includegraphics[width=\columnwidth]{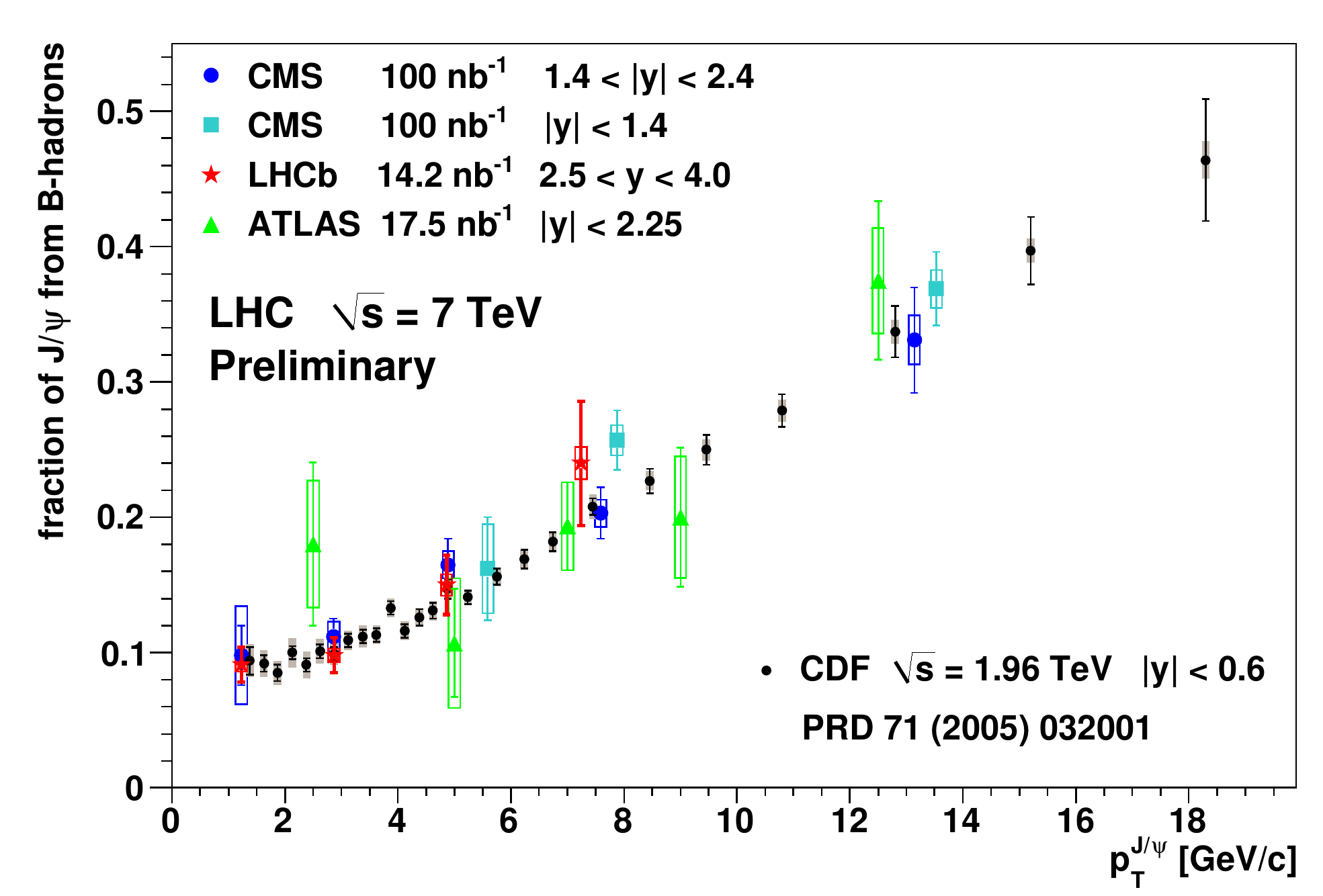} 
\end{center}
\caption{The preliminary fraction of non-prompt $J/\psi$ as a function of $p_T$ 
measured by CMS ($|y|<1.4$ \& $ 1.4<|y|<2.4$), ATLAS ($|y|<2.25$) and LHCb ($2.5<y<4$).
Data compilation courtesy of H. W\"{o}hri.}
\label{Bfrac}
\end{figure}

The data shown in Figs.~\ref{LHCmid} and \ref{LHCfor} are for inclusive
$J/\psi$ production which includes feeddown from the $\chi_c$ and $\psi(2S)$
states and semileptonic $B$ meson decays.  The fraction of $J/\psi$
resulting from $B$ meson decays,
\eqs{
B \, {\rm fraction} \equiv \frac{B \rightarrow J/\psi \, X}{{\rm any \, inclusive  \,} J/\psi}}
is compiled in Fig.~\ref{Bfrac}.  The preliminary LHC results at $\sqrt{s} = 
7$~TeV are compared with the high statistics CDF midrapidity data 
\cite{Acosta:2004yw} at $\sqrt{s} = 1.96$~TeV.  
The result seems to be remarkably
independent of both center-of-mass energy and rapidity.  The lower statistics
and lower $p_T$ STAR data also agree with the magnitude of the
higher energy data shown in Fig.~\ref{Bfrac}.

\begin{figure}[hbt!]
\begin{center}
  \includegraphics[width=\columnwidth]{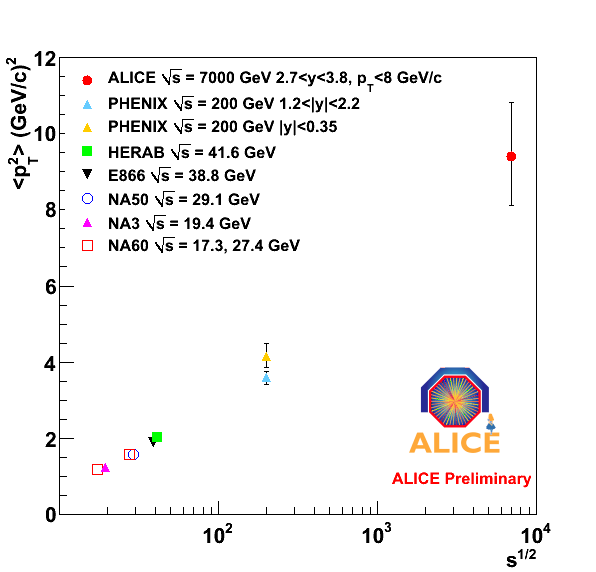} 
\end{center}
\caption{The preliminary measurement of $\langle p_T^2 \rangle$ from ALICE
compared with lower energy data \protect\cite{Scomparin:QQ10}.}
\label{ALICEpt2}
\end{figure}

Although the level of the $B$ meson contribution to $J/\psi$ production seems
to be remarkably independent of $\sqrt{s}$, the average $p_T^2$ of the measured
inclusive $J/\psi$'s has been seen to increase with $\sqrt{s}$. The value of
both $\langle p_T \rangle$ and $\langle p_T^2 \rangle$ increases approximately
linearly with $\ln\sqrt{s}$ from $\sqrt{s} \sim 17$~GeV to 200~GeV.  The 
preliminary ALICE measurement reported here \cite{Scomparin:QQ10}, seems to follow
this trend.  The dependence of $\langle p_T^2 \rangle$ with $\sqrt{s}$ is
shown in Fig.~\ref{ALICEpt2}.  Note that the fixed-target results are centered
around midrapidity while two rapidity ranges, $|y|<0.35$ and $1.2<|y|<2.2$,
measured at RHIC, suggest that the average $p_T^2$ is higher at midrapidity
than forward rapidity, not a surprising result. The ALICE measurement, using
the muon spectrometer, is
also at forward rapidity.

We note that comparisons
of these data with model calculations using standard DGLAP evolution for the gluon distribution with
perturbative QCD agree rather well (see \cf{fig:CSMvsLHC} for a comparison with the CSM) 
suggesting that, even at rather forward
rapidity, in a region of $x$ and $Q^2$ where the parton densities are not 
well measured, no alternative factorization schemes involving {\it e.g.}
saturation effects, are required.

\begin{figure}[hbt!]
\begin{center}
\includegraphics[width=\columnwidth]{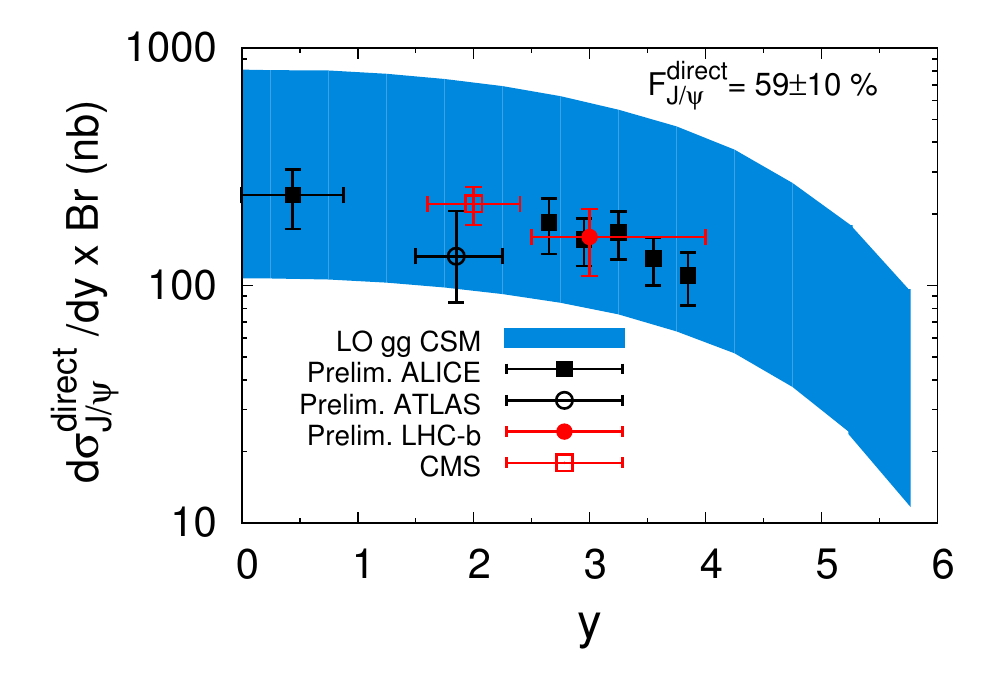} 
\end{center}
\caption{Comparison between the LO CSM prediction for $d\sigma^{direct}_{J/\psi}/dy\times \Br$ 
from $gg$ fusion LO contributions in $pp$ collisions at $\sqrt{s}=7$ TeV compared
to ALICE~\protect\cite{boyer}, ATLAS~\protect\cite{ATLAS-JPsi}, CMS~\protect\cite{CMS:2010yr}
and LHCb~\protect\cite{LHCb-JPsi}  results multiplied by the direct fraction (and correcting for the non-prompt component, 
assumed to be 10\%,  if applicable). Adapted from~\cite{Lansberg:2010cn}.}
\label{fig:CSMvsLHC}
\end{figure}

Finally, CMS also reported a preliminary $\Upsilon(1S)$ measurement based on 
$678 \pm 38$ events.  The
$\Upsilon(1S)$, $\Upsilon(2S)$ and $\Upsilon(3S)$ peaks are clearly separated
in the data.

\subsection{What's next ?}

The LHC quarkonium data shown here is just the tip of the iceberg.  More data
will be forthcoming in the next runs.  Presumably direct $J/\psi$ and
$\Upsilon(1S)$ production will also be measurable, {\it i.e.} the feeddown
contributions will be distinguished and separated.

On the other hand, after September 2011, the Tevatron will unfortunately 
be shut down.  There are 8.5~fb$^{-1}$ of data from Run II on tape with $\sim
10$~fb$^{-1}$ expected by the end of the run.  This should be enough data to
measure the $J/\psi$ and $\psi(2S)$ up to $p_T > 30$~GeV/$c$.  The three $\Upsilon$
$S$ states will also be measured with relatively high precision.  Not only
should the final Tevatron measurements include the feeddown of $\chi_c$ to
$J/\psi$ and $\chi_b$ to $\Upsilon$, it is expected that the $\chi_c$ and
$\chi_b$ $P$ states with large branching ratios to the $S$ states can be 
separated from each other using photon conversions.  Even though Run II will
come to an end, the data will continue to be analyzed for some time with more
exciting results forthcoming.

HERA has already ended its physics run.  However, analyses are still in 
progress.  The H1 and ZEUS data have been combined in these final analyses,
increasing the available statistics.  These fits will better improve our
understanding of quarkonium photoproduction.

Indeed, utilizing the entire Tevatron and HERA data sets to study the balance
between production of color singlet and color octet states in hadro- and
photoproduction will provide more insight into the production mechanisms than
fitting to either data set alone. Last but not least, the $B$-factory  analyses on charmonium
production, not only from $B$ decays, are detailed enough nowadays to separate out the
different inclusive contributions, imposing constraints on the production 
mechanisms at $B$ factories, $ep$ colliders and $pp$ colliders. For instance, it starts 
to be rather clear that $C=+1$ color-octet transitions are much less
probable than initially thought, which suppresses the leading color-octet yield 
at $p_T<5$ GeV in hadroproduction.

In 200 GeV $pp$ collisions, PHENIX has measured cross sections for production 
of the (unresolved) $\Upsilon$ states and the $\psi(2S)$.  A $\chi_c$ 
measurement will be published soon. 
All of these measurements involve low yields and 
will benefit greatly from improved luminosity in the next few years. PHENIX 
has also measured the polarization of the $J/\psi$ in 200 GeV $pp$ collisions 
and will also do so with existing 500 GeV data.  PHENIX and STAR will also
produce unpolarized $J/\psi$ and $\Upsilon$ results at 500 GeV, a new energy
midway between the 200 GeV data and the Tevatron energy.

RHIC is vigorously pursing a long term plan of detector and machine upgrades.
The ongoing RHIC luminosity upgrades will be completed by the 2013 run. 
The introduction in 2011 and 2012 of silicon vertex detectors into PHENIX at 
mid and forward rapidity and of the STAR Heavy Flavor Tracker 
in 2014, will enable open charm and open bottom to be measured independently 
with greatly improved precision.  The detector upgrades will also allow 
improved measurements of quarkonium states due to improved mass resolution 
and background rejection. 

The RHIC run plan for the next 5 years or so will be centered on exploiting 
the capabilities of the new silicon vertex detectors and other upgrades, 
combined with the increased RHIC luminosity.  There will likely be long $pp$ 
runs at 200 GeV, plus shorter runs at 62 GeV to explore 
the energy dependence of open and hidden heavy flavor production. 
The luminosity increase and the
PHENIX detector upgrades will enhance the PHENIX heavy quarkonium program
with increased $p_{T}$ reach for the $J/\psi$ and low statistics measurements 
of the combined $\Upsilon$ states. The increased luminosity will
enable the large acceptance STAR detector to extend its $p_T$ reach to 
considerably higher values for $\Upsilon$ and $J/\psi$ measurements than 
previously possible. 

The RHIC program for the period beyond about 5 years is still under development.
The RHIC experiments are presently engaged in preparing a decadal plan that 
will lay out their proposed science goals and detector upgrades for 2011 to 
2020. 

\section{Quarkonium production in $\pA$ collisions}\label{WG2}

The study of charmonium production is an interesting 
test of our understanding of strong interaction physics.
For $pp$ collisions, various theoretical approaches have been proposed, but a
satisfactory description of \jpsi\ hadroproduction is still 
missing~\cite{Brambilla:2010cs}. 
In spite of this situation, the study of quarkonium production in the apparently more difficult proton-nucleus environment has 
also attracted considerable interest. In fact, in $\pA$, the heavy-quark pair is created in the nuclear medium, and
the study of its evolution towards a bound state can add significant constraints to the models.
For example, the strength of the interaction of the evolving $c\overline c$ pair 
with the target nucleons, that can lead to a break-up of the pair and consequently to a 
suppression of the \jpsi\ yield, may depend on its quantum states at the production level
(color-octet or color-singlet), and on the kinematic variables of the pair~\cite{Kopeliovich:1991pu,Vogt:2001ky}.
In addition to final-state effects, also initial-state effects may influence the observed \jpsi\ yield in $\pA$. 
In particular, parton shadowing in the target nucleus~\cite{deFlorian:2003qf,Eskola:1998iy,Eskola:1998df,Eskola:2008ca,Eskola:2009uj} may suppress (or enhance, in case 
of antishadowing) the probability of producing a \jpsi, and its effect depends on the kinematics of the $c\overline c$ pair production~\cite{Ferreiro:2008wc}.
Moreover, the energy loss in the
nuclear medium of the incident parton~\cite{Gavin:1991qk}, prior to $c\overline c$ production, may
significantly alter the \jpsi\ cross section and its kinematic distribution.  
Finally, a suppression of the \jpsi\ has been proposed a long time ago as a signature
of the formation, in ultrarelativistic nucleus-nucleus collisions, of a state where 
quarks and gluons are deconfined (Quark-Gluon Plasma)~\cite{Matsui:1986dk}. 
Results from $\pA$ collisions, taken in the same kinematical conditions of 
$\AA$, and properly extrapolated to nucleus-nucleus collisions, 
can be helpful to calibrate the contribution of the various cold nuclear 
matter effects to the overall observed suppression~\cite{Alessandro:2004ap,Arnaldi:2007zz}.

\subsection{Basic phenomena and observables} 
\subsubsection{Hierarchy of time scales}
Nuclear effects are controlled by the characteristic time scales of charmonium production.
A colorless $\bar cc$ pair created in a hard reaction is not yet a physical charmonium state since it does not have a fixed mass.
According to the uncertainty relation it takes time to disentangle the ground state charmonium from its excitations,
\begin{equation}
t_f=\frac{2E_{\bar cc}}{m_{\psi'}^2-m_{J/\psi}^2}.
\label{2.20}
\end{equation}
This time scale can be also interpreted as formation time of the charmonium wave function.
At charmonium energies $E_{\bar cc}<25$ GeV in the nuclear rest frame this time is shorter than the mean nucleon spacing in a nucleus, $t_f<2 $fm, so one can treat the formation process as instantaneous, and the attenuation  in nuclear matter is controlled by the inelastic charmonium-nucleon cross section, 
\begin{equation}
S(L)=\exp\left[-\sigma^{J/\psi N}_{in}L\,\rho_A\right],
\label{2.40}
\end{equation}
where $L$ is the path length in the medium of density $\rho_A$. 

At much higher energies, $E_{\bar cc}\gg 100$ GeV, the formation time is long, $l_f\gg R_A$, even for heavy nuclei, the initial size of the $\bar cc$ dipole is "frozen" by Lorentz time dilation during propagation through the nucleus, so the attenuation factor gets the simple form,
\begin{equation}
S(L)=\left\langle\exp\left[-\sigma(r_T)L\,\rho_A\right]\right\rangle_{r_T},
\label{2.60}
\end{equation}
where the exponential is averaged over transverse
separation $r_T$ of the dipole weighted with the initial and final distribution amplitudes \cite{Kopeliovich:1991pu}.
This regime is relevant for $J/\psi$ production at RHIC at the mid rapidity, where $E_{\bar cc}=300$ GeV.
The $J/\psi$  energy rises with $y$ as $e^y$ relative to the target nucleus, but decreases as $e^{-y}$ relative to the beam nucleus. Therefore, the asymptotic regime is relevant to any positive rapidity in $\dA$ collisions,
but in $\AA$ collisions at forward rapidities nuclear effects turn out to be in different regimes for the two colliding nuclei.

Another important time scale characterizes the hard process of $\bar cc$ pair creation.
Although the proper time is short, $t^*\sim 1/m_c$, it is subject to Lorentz time dilation in a frame, where the charm quarks have high energy.
\begin{equation}
t_c=\frac{2E_{\bar cc}}{M_{\bar cc}^2}
\label{2.80}
\end{equation}
If this time is short, $t_c<2$ fm, one can consider production as instantaneous. 
This regime corresponds to charmonium energy $E_{\bar cc}<50\,$ GeV. This estimate is relevant for production of $\chi$, but the
characteristic energy may be about twice higher for $J/\psi$, because in the color singlet model the invariant mass of $\bar cc$ is $\langle M_{\bar cc}^2\rangle=2m_{J/\psi}^2$.
This time scale is usually short for $J/\psi$ produced at mid rapidity in fixed-target experiments. For instance at $\sqrt{s}=40\,$GeV $t_c=1.2$ fm if $x_F=0$.

\begin{figure}[htb!]
\begin{center}
\includegraphics[width=6cm]{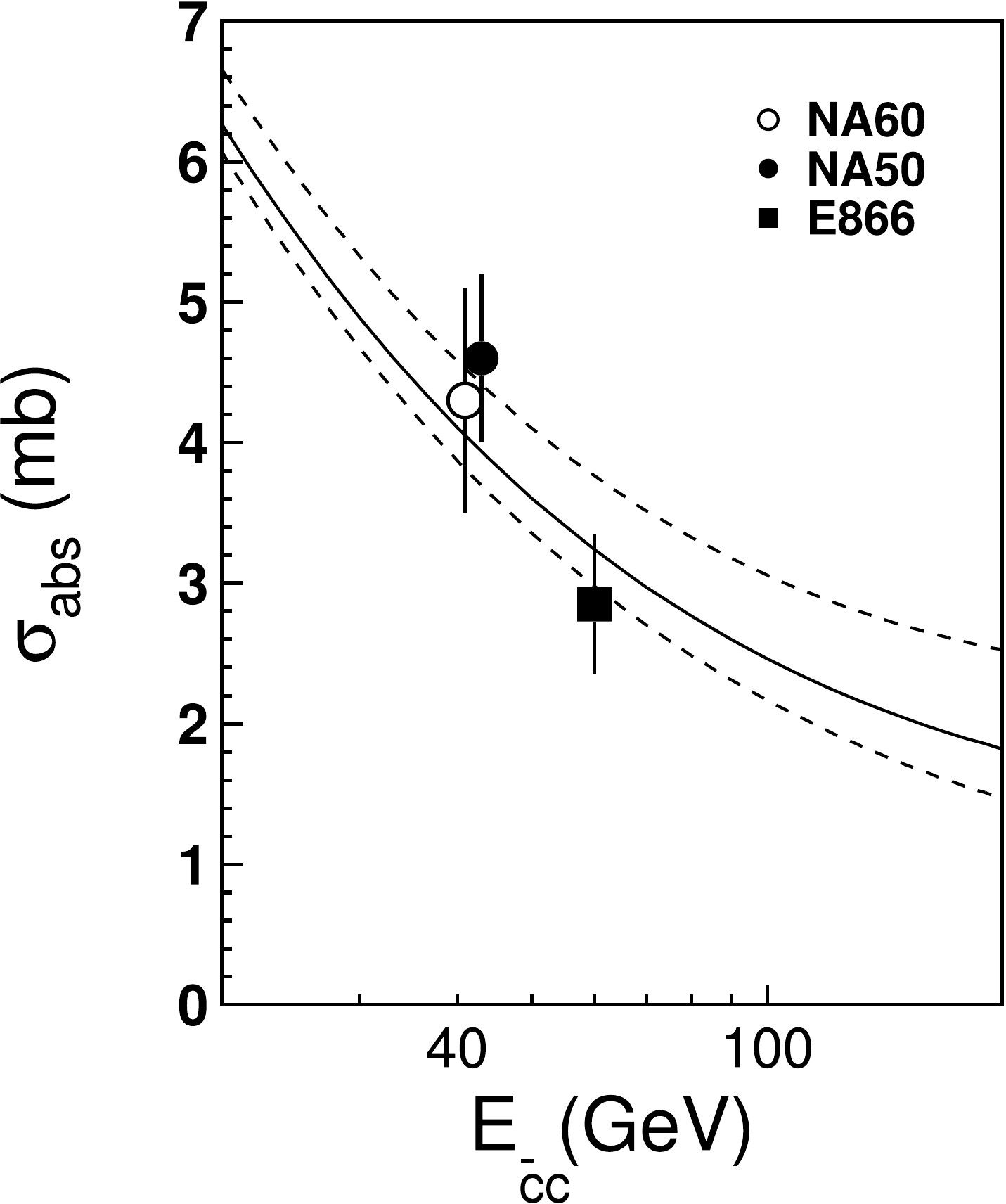}
\end{center}
\caption{\label{sig_abs} The effective break-up cross section as function of $J/\psi$ energy. 
The solid curve shows theoretical expectation \cite{Kopeliovich:2010jf}.
The 
dashed curves show the estimated theoretical uncertainty.
}
 \end{figure}

The most difficult for calculations is the intermediate energy regime, where either 
$t_c$ or/and $t_f$ are of the order of the nuclear size. Theoretical tools for this 
case, the path-integral technique, was developed in \cite{Kopeliovich:1991pu,Kopeliovich:2010jf}.
The $\bar cc$ dipole is produced with a separation $\sim1/m_c$ and ends up with the large size of $J/\psi$.
The time scale for such an expansion, $t_f$, rises with energy, therefore the effective absorption, or break-up, cross 
section is expected to drop with energy \cite{Kopeliovich:1991pu}. Indeed, this effect was observed in fixed target experiments, as is plotted in Fig.~\ref{sig_abs}. Data are well explained by color transparency for expanding $\bar cc$ dipoles.

\subsubsection{Higher twist charm quark shadowing}

In the regime of long time scale for $\bar cc$ production, of $t_c\gg R_A$, the process, $g\to\bar cc$, is subject to shadowing. This is apparently higher twist, since the mean $\bar cc$ transverse separation is $\langle r_T\rangle\sim 1/m_c$. Same is true for the break-up cross section of the produced colorless $\bar cc$ propagating through the nucleus.  

The amplitude of $\bar cc$ production at a point with impact parameter $b$ and longitudinal coordinate $z$, averaged over the dipole size, reads \cite{Kopeliovich:2001ee,Kopeliovich:2010nw},
\begin{eqnarray}
&&S_{pA}(b,z)=\int d^2r_T\,W_{\bar cc}(r_T)
\label{140}
\\ &\times&
\exp\left[-{1\over2}\sigma_{\bar ccg}(r_T)T_-(b,z)
-{1\over2}\sigma_{\bar cc}(r_T)T_+(b,z)\right].
\nonumber
\end{eqnarray}
Here $T_-(b,z)=\int_{-\infty}^z dz'\rho_A(b,z')$;~ $T_+(b,z)=T_A(b)-T_-(b,z)$, and $T_A(b)=T_-(b,\infty)$. 
Shadowing for $\bar cc$ production over the nuclear thickness $T_-(b,z)$ occurs with the shadowing cross section
corresponding to a 3-body dipole, gluon and $\bar cc$, which for equal momenta of $c$ and $\bar c$ equals to $\sigma_{\bar ccg}(r_T)={9\over4}\sigma_{\bar cc}(r_T/2)-{1\over8}\sigma_{\bar cc}(r_T)$.

For $t_c\gg R_A$ the weight factor in (\ref{140}) has the form \cite{Kopeliovich:2001ee},
$W_{\bar cc}(r_T)\propto K_0(m_c r_T)\,r_T^2\,\psi_{J/\psi}(r_T)$,
where one factor $r_T$ comes from the amplitude of $\bar cc$ production, and another one either from the amplitude of gluon radiation in the case $J/\psi$ production, or from the radial wave function of $\chi_2$.

With the survival probability amplitude Eq.~(\ref{140}) the nuclear ratio reads,
\begin{equation}
R_{\pA}={1\over A}\int d^2b\int\limits_{-\infty}^{\infty}dz\,
\left|S_{pA}(b,z)\right|^2.
\label{145}
\end{equation}

\begin{figure}[htb!]
\begin{center}
\includegraphics[width=6cm]{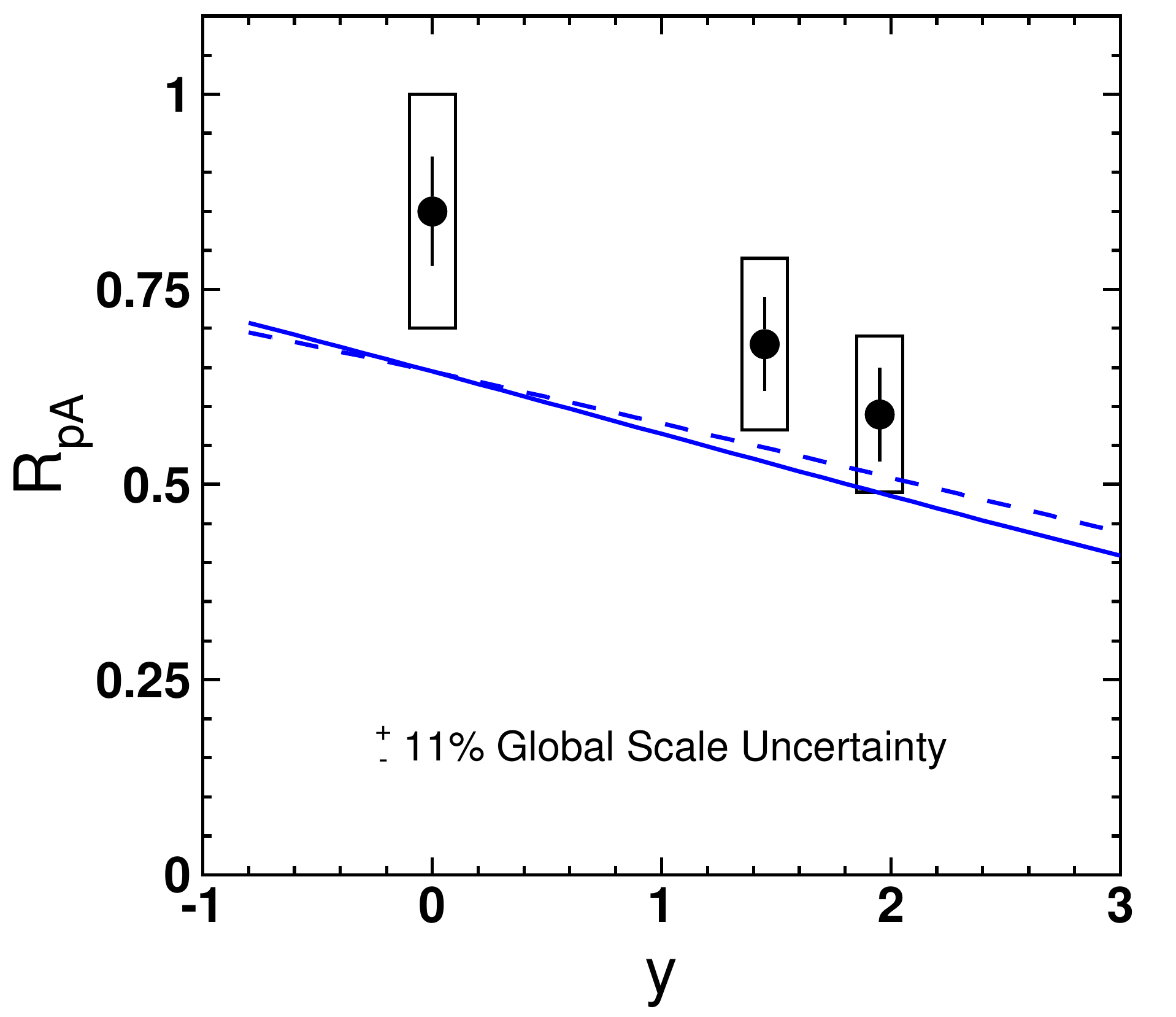}
\end{center}
\caption{\label{d-A-y} Dashed curve presents nuclear suppression 
of $J/\psi$ as function of rapidity in $\pA$ collisions. 
Solid curve is corrected for gluon shadowing. Data 
are for $\dAu$ collisions at $\sqrt{s}=200$ GeV~\cite{Adare:2007gn}.}
 \end{figure}

The results at $\sqrt{s}=200$ GeV are plotted by dashed curve in Fig.~\ref{d-A-y} as function of $y$. 
\cite{Kopeliovich:2010nw}. 

\subsubsection{Leading twist gluon shadowing}

In terms of the Fock state decomposition gluon shadowing corresponds to higher Fock components in the projectile hadron, containing at least one gluon. Multiple interactions of this gluon in the nuclear target are the source of shadowing. 

Theoretical expectations for gluon shadowing are quite diverse. 
A weak shadowing was predicted within the dipole approach \cite{Kopeliovich:1999am} 
and in an analysis \cite{deFlorian:2003qf} of DIS data developed at leading and
next-to-leading order -- nDS LO and NLO parameterizations\footnote{We note that the same DIS analysis provided 
a constrained NLO fit with a stronger gluon shadowing -- nDSg --. The $\chi^2$ of this fit is however not as good as that of the 
unconstrained fit --nDS.} --. However, a rather strong shadowing and correspondingly antishadowing for gluons was evaluated in \cite{Strikman:2010ew} based on the hadronic representation \cite{Gribov:69zzz,Karmanov:1973va} and 
in the global analyses EKS98~\cite{Eskola:1998iy,Eskola:1998df}, EPS08~\cite{Eskola:2008ca} and EPS09~\cite{Eskola:2009uj}. 
We emphasize that the two last analyses included data on proton-nucleus collisions, which makes them model dependent. We
finally note that nDSg and EKS98 parameterizations are compatible for $x<10^{-3}$. 

Also, the choice of the adequate partonic production mechanism 
-- either via a \mbox{$2\to 1$} or a \mbox{$2\to 2$} process -- may affect both the way to 
compute the nuclear shadowing and its
expected impact on the production \cite{Ferreiro:2008wc}.

This controversy should be settled after nuclear effects for charmonium production at LHC will be studied experimentally.

\subsubsection{Non trivial transition from $\pA$ to $\AA$}

The usual strategy in search for a signal of final state attenuation of charmonia in the created dense medium, is extrapolation to $\AA$ collisions of the cold nuclear matter effects observed in $\pA$. This is how the "anomalous" suppression of $J/\psi$ was observed in the NA50/60 data \cite{Lourenco:1996wn,Gonin:1996wn,Abreu:1997jh,Scomparin:2009tg,Arnaldi:2009ph}: one fits the effective absorption cross section $\sigma_{abs}$ to describe the nuclear suppression observed in $\pA$ collisions. Some results of such a procedure are shown in Fig.~\ref{sig_abs}. Then with this cross section one predicts the normal cold nuclear matter effects in nuclear collisions, and whatever deviates from such an expectation is called anomalous and is related to the
final state suppression.

\begin{figure}[htb!]
\begin{center}
\includegraphics[width=5cm]{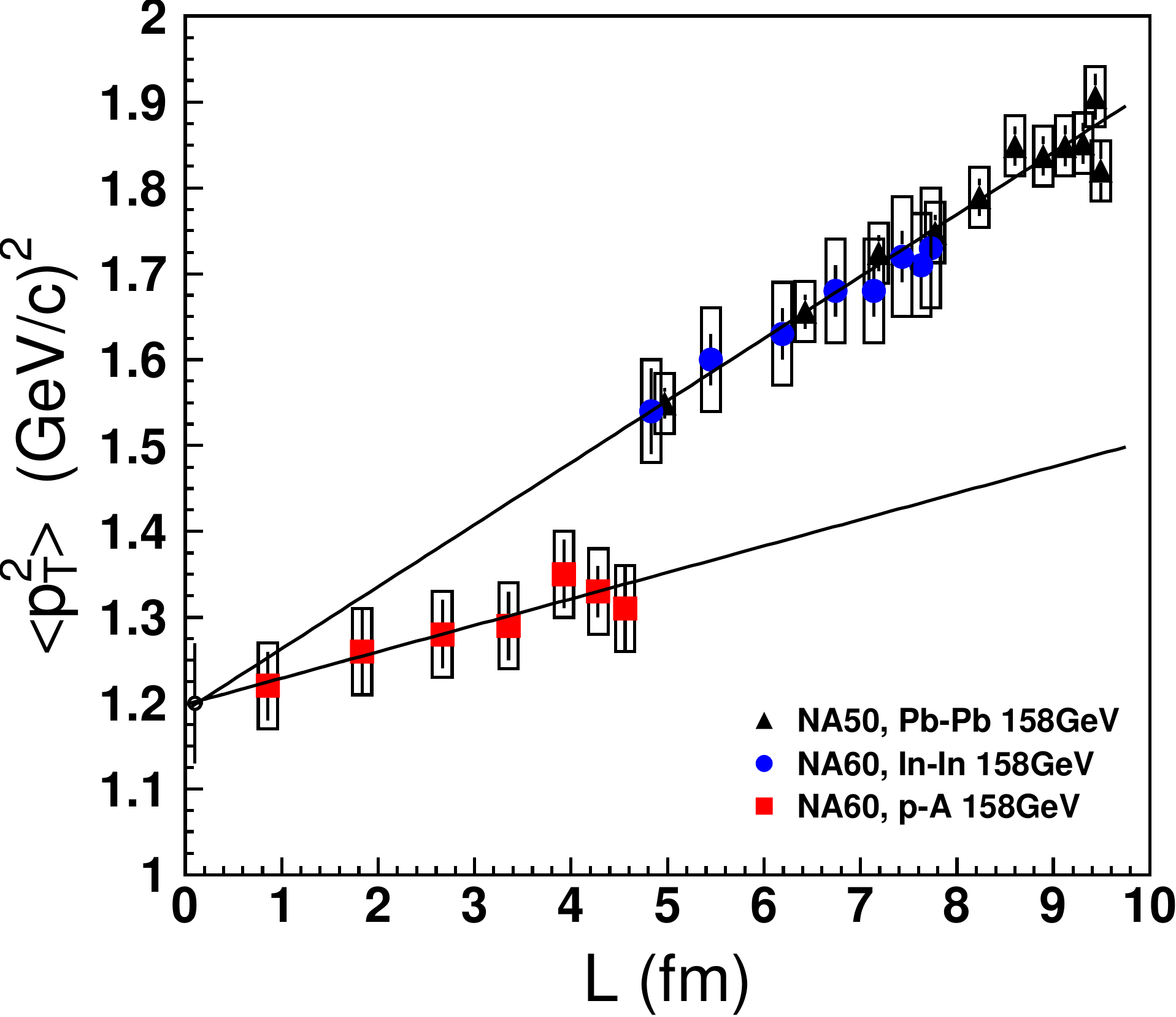}
\end{center}
\caption{\label{pa-aa}$J/\psi$ $p_T$ broadening  measured at SPS as a function of the path length $L$ in nuclear matter.}
 \end{figure}

However, recent data \cite{Cortese:HP08} from the NA60 experiment for $p_T$ broadening of $J/\psi$  depicted in Fig.~\ref{pa-aa} demonstrate that such an extrapolation may not be straightforward, since the "cold" nuclear matter in $\AA$ collisions turns out not to be cold at all.

\begin{figure*}[htb!]
\begin{center}
\includegraphics[width=5cm]{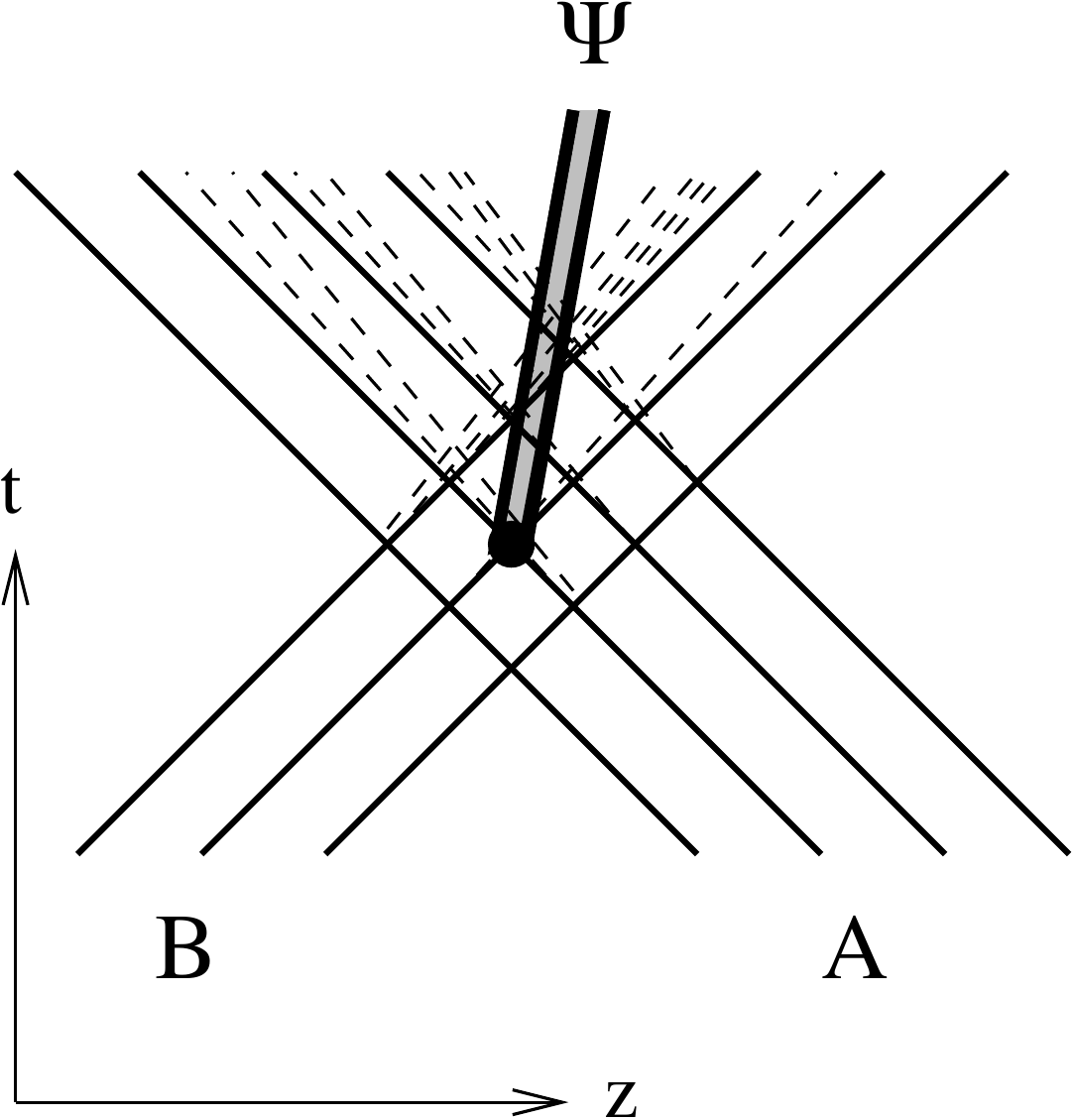}
\includegraphics[width=5cm]{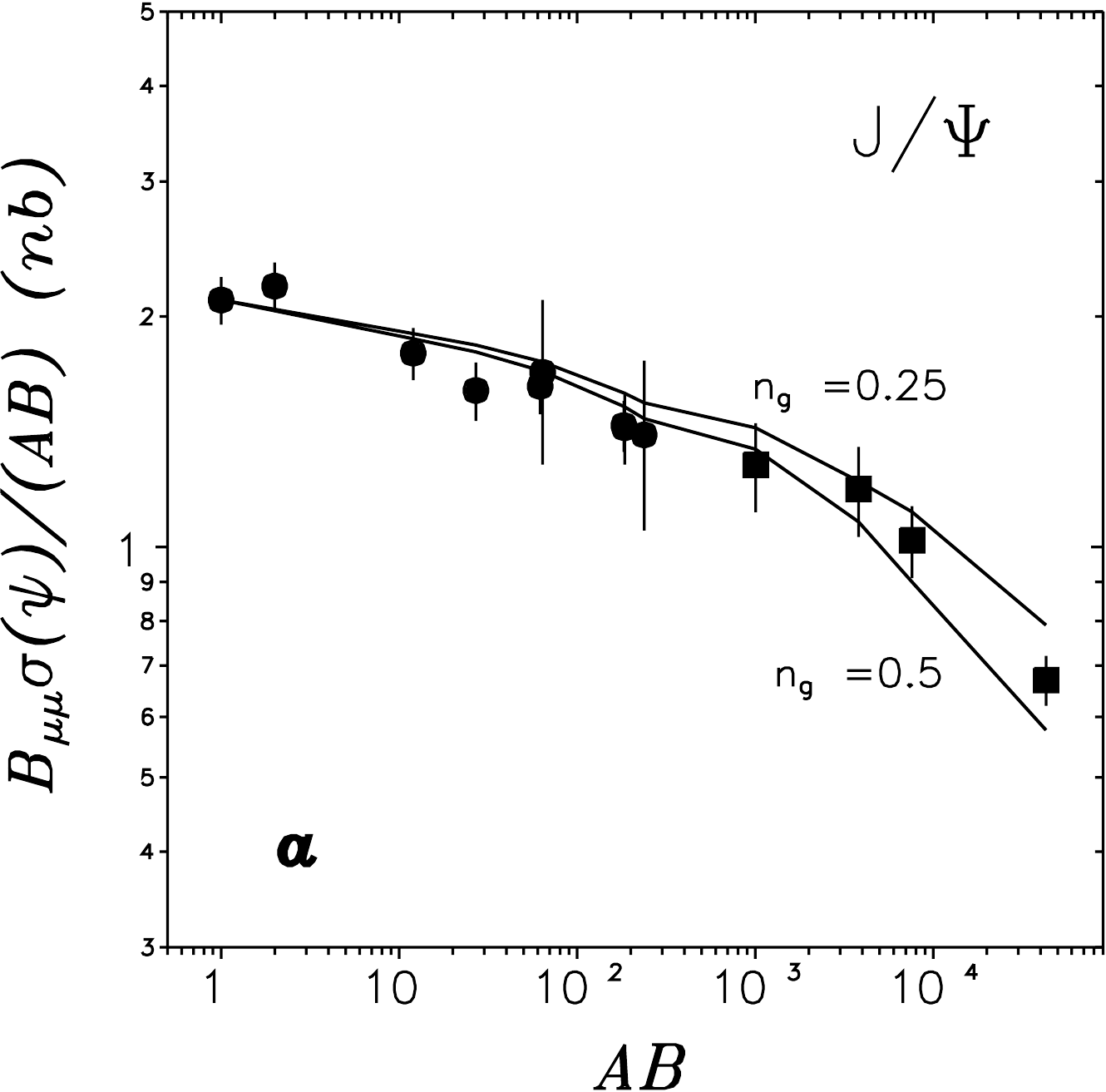}
\end{center}
\caption{\label{prompt} {\it Left:} A two-dimensional (time - longitudinal coordinate) plot for charmonium
production in a collision of nuclei $A$ and $B$ in the c.m.  of the colliding
nucleons.  The solid and dashed lines show the nucleon and gluon trajectories,
respectively. {\it Right:}
Nuclear
suppression of $J/\psi$ production in $\pA$ and $\AB$ collisions as function of
the product $A\times B$.  The two curves demonstrate theoretical uncertainty of calculations~\cite{Hufner:1998hf}.
The circles and squares correspond to $\pA$ and $\AB$ data, respectively.  The
data points are from \cite{Lourenco:1996wn,Gonin:1996wn,Abreu:1997jh}.}
 \end{figure*}

Indeed, broadening, $\Delta p_T^2$, is known to be proportional to the path length in the medium and its density.
The data show that broadening observed in nucleus-nucleus collisions is twice as large as in $\pA$ for the same total path length. This means that the cold nuclear matter is twice as dense in $\AA$ compared to $\pA$ collisions.
Correspondingly, the effective absorption cross section $\sigma_{abs}$ should be taken twice bigger when one makes predictions for $\AA$ collisions. Such a significant enhancement of absorption effects may explain the observed anomalous suppression.

This effect was successfully predicted in \cite{Hufner:1998hf} as a result of multiple $NN$ collisions and gluon radiation preceding their interaction with the $J/\psi$ (or $\bar cc$ dipole), as is illustrated in the left panel of Fig.~\ref{prompt}.

It was also shown that quantitatively this effect is able to explain the observed anomalous $J/\psi$ suppression,
as is depicted in the right panel of Fig.~\ref{prompt}. Simultaneously, the same effect of increased medium density leads to a significant increase of broadening in $\AA$ compared to $\pA$ collisions.

Suppression of $J/\psi$ caused by gluon radiation decreases as $1/\sqrt{s}$ and vanishes at high energies of RHIC and LHC. However, at the energies of LHC broadening of $J/\psi$ is expected to be highly enhanced in $\AA$ compared to $\pA$  collisions for another reason.

Notice that observation of an anomalously large broadening has an important advantage 
compared to anomalous suppression. While the latter may originate from either initial, 
or final state interactions, which can be easily mixed up, the former comes entirely 
from initial state multiple interactions. Indeed, a colorless $\bar cc$ dipole attenuates 
in a medium, but does not have any energy loss and does not change its momentum. 
Note that the partonic process may affect this result~\cite{Arleo:2010rb}.

\subsubsection{Transverse momentum broadening}

\begin{figure}[htb!]
\begin{center}
\includegraphics[width=0.9\columnwidth]{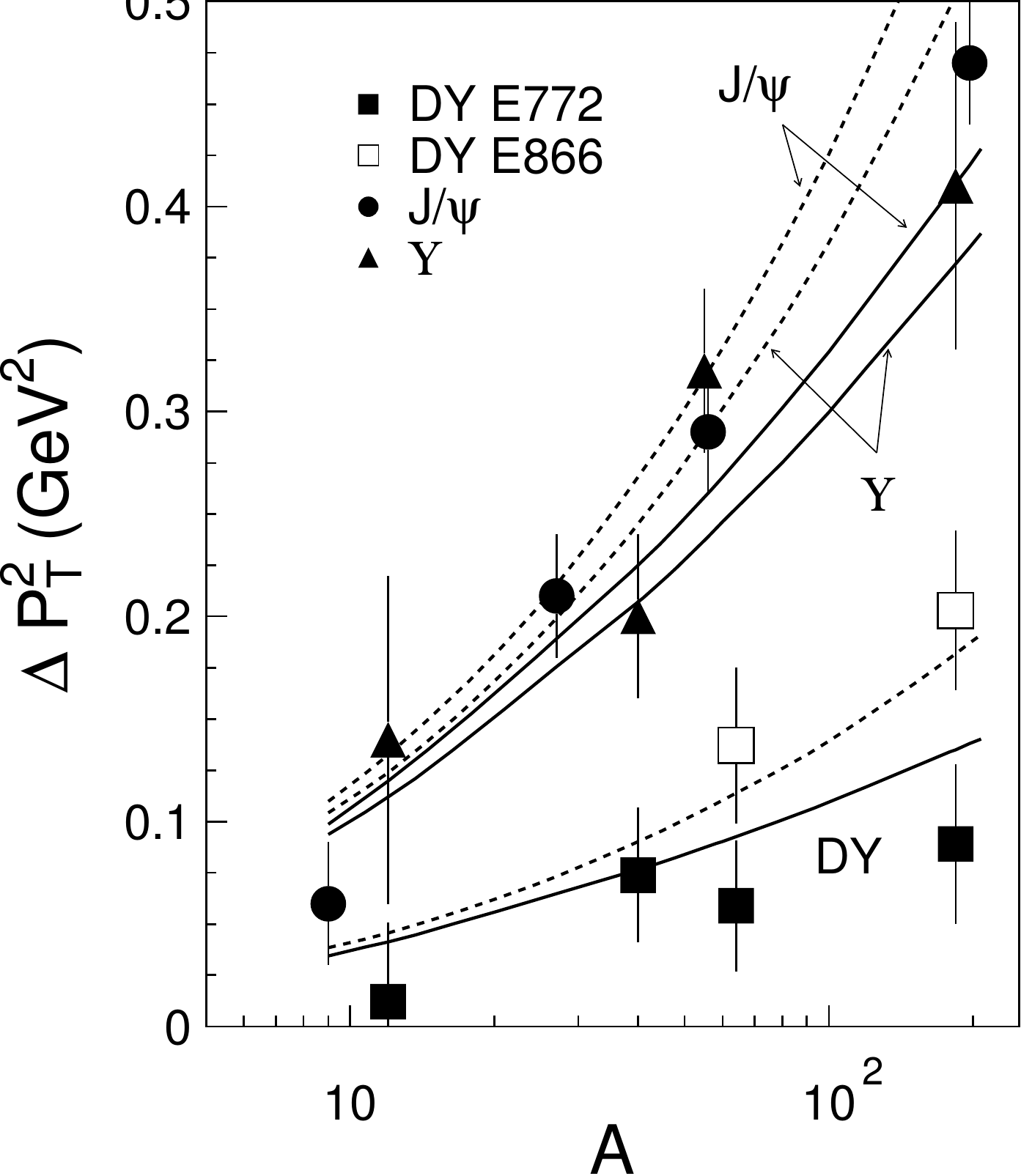}
\end{center}
\caption{\label{broad} Broadening for $J/\psi$	and  
$\Upsilon$ \cite{Alde:1991sw} is shown by circles and triangles respectively.
The dashed and solid curves correspond to the predictions based on Eq.~(\ref{2.100})
without and with the corrections for gluon shadowing respectively.}
 \end{figure}

The mean transverse momentum squared of heavy quarkonia increases in $\pA$ collisions compared to $pp$.
The magnitude of the effect, $\Delta p_T^2=\langle p_T^2\rangle_A - \langle p_T^2\rangle_p$ is usually called broadening. Broadening of a parton propagating through a nuclear medium can be calculated within the dipole approach as \cite{Dolejsi:1993iw,Johnson:2000dm},
\begin{equation}
\Delta p_T^2 = 2\,C(E) \int\limits_0^L dz\,\rho_A(z),
\label{2.100}
\end{equation}
where the factor $C(E)$ is related to the dipole-proton cross section $\sigma_{\bar qq}(r_T)$ known from phenomenology. For broadening of gluons
\begin{equation}
C_g=\left.\frac{9}{8}\,\vec\nabla_{r_T}^2\,\sigma_{\bar qq}(r_T)\right|_{r_T=0}.
\label{2.120}
\end{equation}
Using the dipole cross section in the saturated form fitted to DIS data \cite{GolecBiernat:1999qd,Kopeliovich:1999am} one can predict the factor $C_g(E)$ and broadening Eq.~(\ref{2.100}). The results depicted in Fig.~\ref{broad} by dashed curves well agree with data for $J/\psi$ and $\Upsilon$.
 Inclusion of gluon shadowing corrections reduces the amount of shadowing, but within the error bars.

\begin{figure}[htb!]
\begin{center}
\includegraphics[width=8cm]{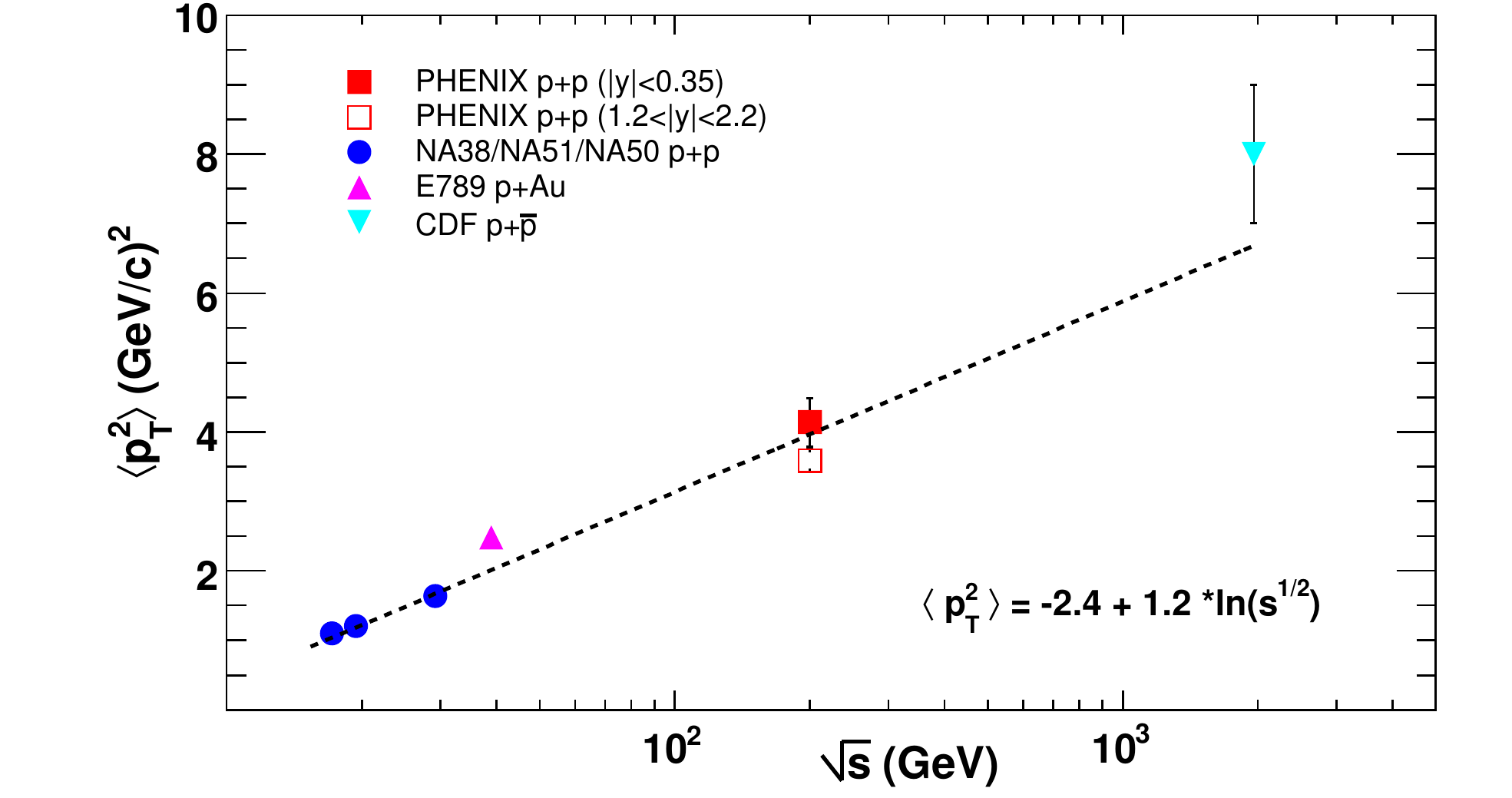}
\end{center}
\caption{\label{pt-e} The $J/\psi$ $\langle p_T^2\rangle$ in $pp$ collisions as function of energy. Data are from  \cite{Adare:2006kf}.}
 \end{figure}

One should also understand the $p_T$ dependence of the cross section and the energy dependence of the
mean value $\langle p_T^2\rangle$, depicted in Fig.~\ref{pt-e}.

\begin{figure}[htb!]
\begin{center}
\includegraphics[width=5.5cm]{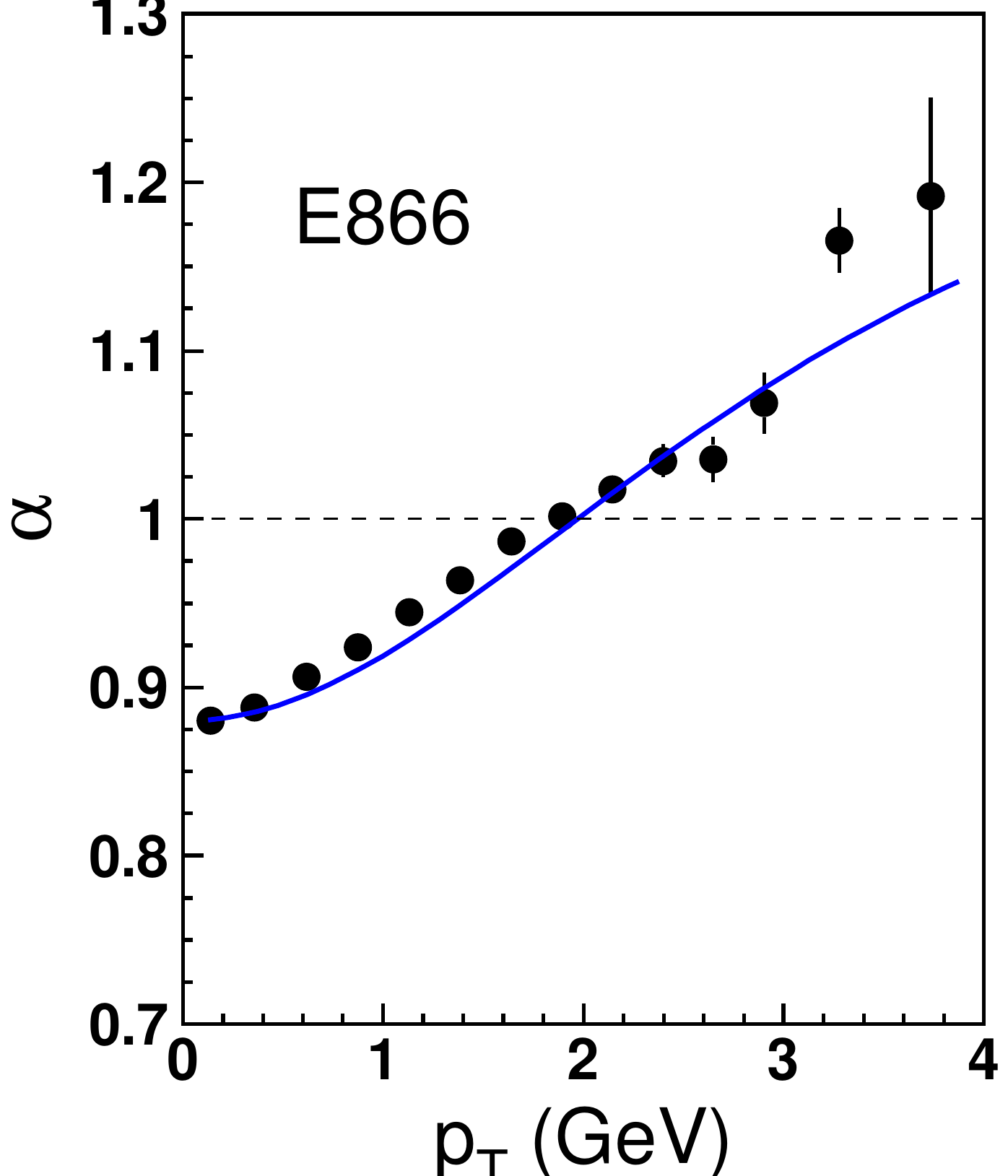}
\end{center}
\caption{\label{e866} The exponent $\alpha=1+\ln(R_{\pA})/\ln(A)$ as function of $p_T$ calculated with Eq.~(\ref{460}) 
in comparison with data from \cite{Leitch:2000zzz}.}
 \end{figure}

Remarkably,  data on  $pp$,
$\pA$ and even $\AA$ collisions, at the energies of fixed target
experiments and at RHIC \cite{Adare:2007gn}, are described
well by the simple parametrization,
\begin{equation} 
\frac{d\sigma}{dp_T^2} \propto
\left(1+\frac{p_T^2}{6\langle p_T^2\rangle}\right)^{-6}.
\label{460}
\end{equation}
If one considers that broadening does not alter the shape of the
$p_T$-distribution of produced $J/\psi$,
the simplest way to calculate the $p_T$-dependence of the $\pA$ cross section is just making a shift $\Delta p_T^2$  in the mean value $\langle p_T^2 \rangle $ for $\pA$ compared to $pp$. 
The resulting $A$-dependence of the $\pA$ over $\pp$ ratio, $R_{\pA}$, is presented in Fig.~\ref{e866} 
as function  of $p_T$ in comparison with data \cite{Leitch:2000zzz} at $\sqrt{s}=39$ GeV. 
The $A$-dependence is parametrized as $R_{\pA}=A^{\alpha-1}$, and $\alpha(p_T)$ is plotted in Fig.~\ref{e866}.

\subsubsection{Feynman $x_F$ dependence of nuclear effects}

\begin{figure}[htb!]
\begin{center}
\includegraphics[width=\columnwidth]{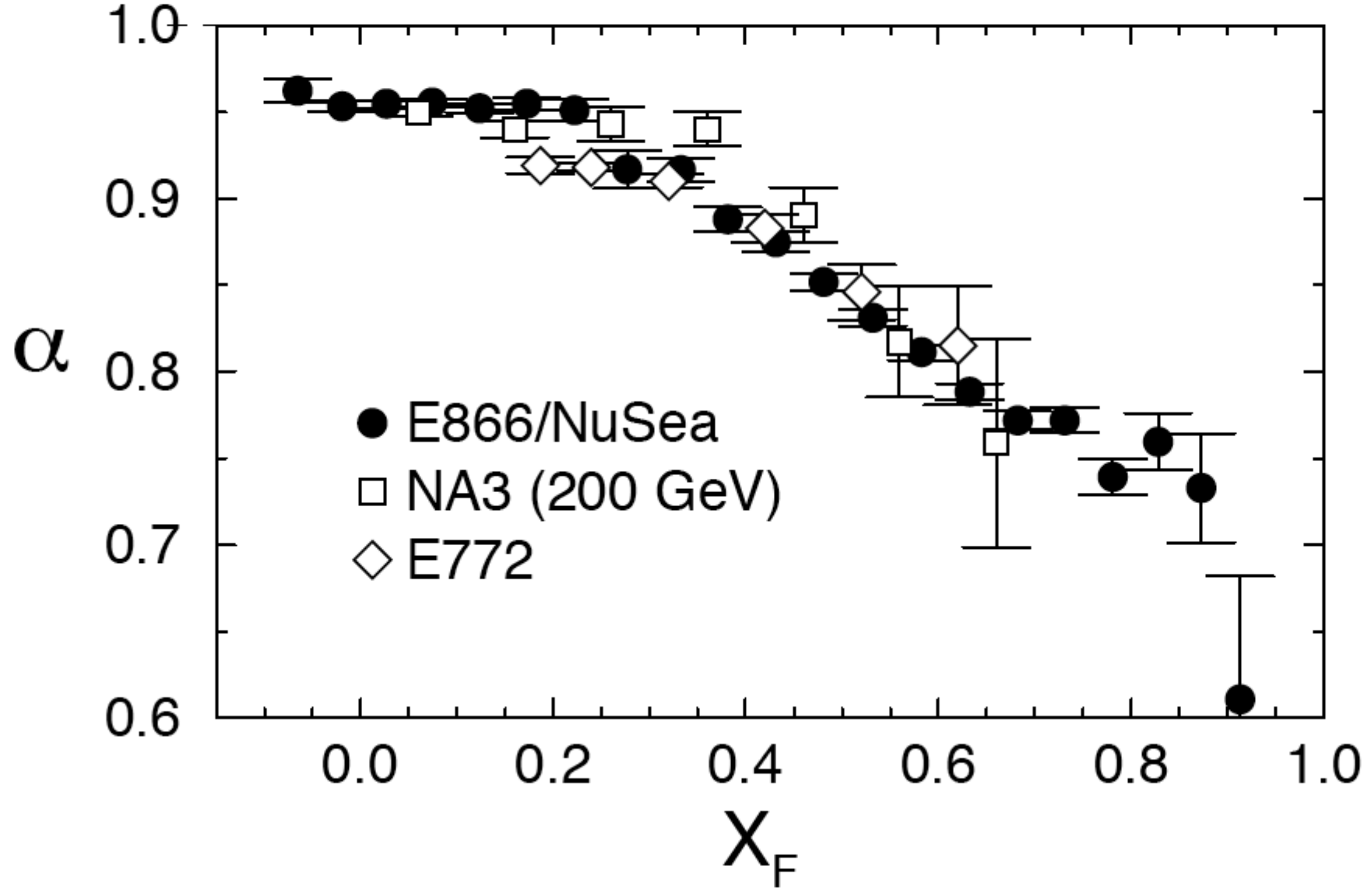}
\end{center}
\caption{\label{xf-dep} The exponent $\alpha(x_F)$ characterizing the $A$-dependence, 
$\sigma^{J/\psi}_{\pA}\propto A^{\alpha}$ at $\sqrt{s}=39$~GeV \cite{Leitch:2000zzz} an $\sqrt{s}=20$~GeV \cite{Badier:1983zzz}.}
 \end{figure}

Available data on $x_F$-dependence of nuclear effects in charmonium production  demonstrate increasing nuclear suppression towards large $x_F$, as is shown in Fig.~\ref{xf-dep}, for the exponent $\alpha(x_F)$ characterizing the $A$-dependence, $\sigma^{J/\psi}_{\pA}\propto A^{\alpha}$.

Although several mechanisms contributing to the $J/\psi$ suppression have been proposed, there is still 
no comprehensive description so far of the observed $x_F$-dependence of the nuclear effects.
The first mechanism proposed in~\cite{Kopeliovich:1984bf} explored initial state energy loss independent off the collision energy,
according to pQCD calculations, or string model. However, a finite energy loss becomes insignificant at high energies, while data in Fig.~\ref{xf-dep} demonstrate an approximate $x_F$ scaling. Apparently, the energy loss must be 
proportional to the initial energy, in order to provide an $x_F$-scaling. Such an energy loss was ad hoc introduced in~\cite{Gavin:1991qk}, and was criticized later in \cite{Brodsky:1992nq}, as contradicting perturbative QCD calculations.

The observed $x_F$-dependence of nuclear suppression was interpreted in~\cite{Kopeliovich:2005ym} in terms of the Fock state decomposition as the energy sharing problem enhanced by the nucleus towards the kinematic limit.
This mechanism also explains a similar suppression observed at large $x_F$ in many other reactions, Drell-Yan process, high and low $p_T$ production of different species of hadrons, etc.

Attempts to explain the observed suppression to gluons shadowing, as a dominant mechanism, was not successful. 
That would lead to $x_2$ scaling, which is severely broken in data~\cite{Alde:1991sw}.

\subsection{Existing measurements and future perspectives}

\subsubsection{Fixed-target experiments}

Since in charmonium production in $\pA$ a rather complicated interplay of various physical
processes occurs, the availability of accurate sets of data, spanning large intervals in
the incident proton energy, and covering large $x_{F}$ and $p_{T}$ regions, is 
essential for a thorough understanding of the involved mechanisms. At fixed target
energies, high-statistics \jpsi\ samples have been collected in recent years by the DESY 
experiment HERA-B~\cite{Abt:2008ya}, at 920 GeV incident energy, by E866~\cite{Leitch:1999ea} at FNAL 
at 800 GeV, by the CERN-SPS experiment NA50 at 400 and 450 GeV~\cite{Alessandro:2006jt} and by the NA60
experiment at 158 and 400 GeV~\cite{Arnaldi:2008er}. 
Usually, nuclear effects have been parametrized by fitting the $A$-dependence of the production cross section
with the simple power law $\sigma_{J/\psi}^{\pA} = \sigma_{J/\psi}^{pp}\cdot
A^{\alpha}$, and then studying the evolution of $\alpha$ with $x_{F}$ and $p_{T}$. 
Alternatively, nuclear effects have been expressed by fitting the data in the framework 
of the Glauber model~\cite{Glauber:1959zzz}, 
having as input parameters the inelastic nucleon-nucleon cross section
and the density distributions for the various nuclei~\cite{DeVries:1987zzz}. 
The model gives as an
output the so-called \jpsi\ absorption cross section $\sigma_{J/\psi}^{abs}$. 
Clearly, both $\alpha$ and $\sigma_{J/\psi}^{abs}$ represent effective quantities, 
including the contribution of the various
sources of nuclear effects sketched in the previous section.

\begin{figure}[htb!]
\begin{center}
\includegraphics[width=\columnwidth]{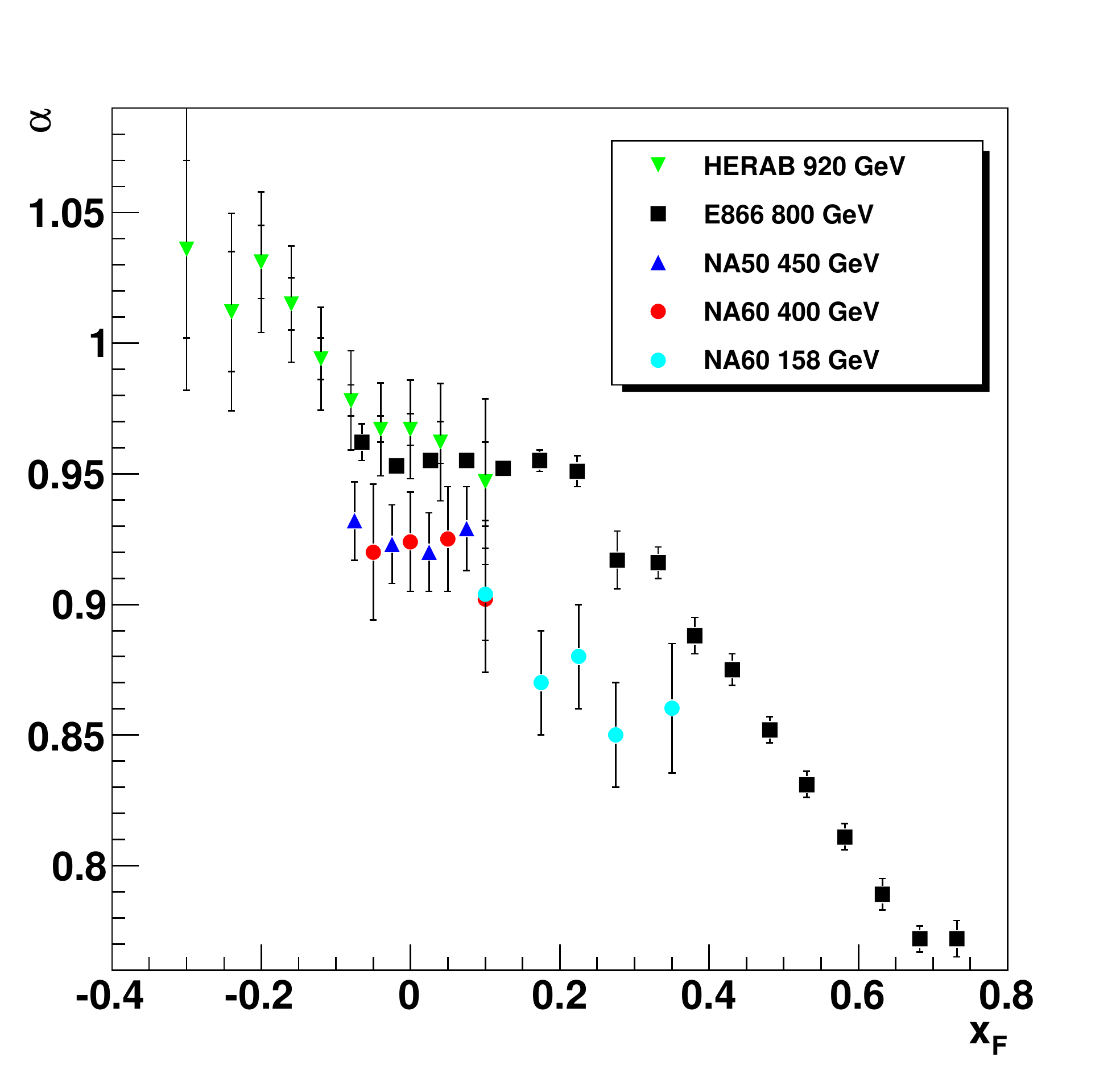}
\end{center}
\caption{\label{fig:alphavsxF}Compilation of experimental results on $\alpha$ vs $x_{F}$.}
 \end{figure}

The results for $\alpha$ as a function of $x_{F}$ are summarized in Fig.~\ref{fig:alphavsxF}, and
show various remarkable features.  First of all one can note a steady increase in the strength of
nuclear effects ($\alpha$ decreases) when increasing $x_{F}$. More in detail, the decrease of 
$\alpha$ at large $x_{F}$ has usually been attributed to energy loss of the projectile partons and indeed
various models have tried to explain, at least qualitatively, such an effect (see e.g.~\cite{Vogt:1999dw} and 
more recently~\cite{Arleo:2010rb}). On the other hand an important constraint on the size of parton energy loss comes
from experiments studying nuclear effects on Drell-Yan production. At this workshop~\cite{Woehri:QQ10}, it has been shown
that the initial-state parton energy loss needed to describe the NA3 data on Drell-Yan production~\cite{Badier:1984pm} 
is too small (by a factor 5 to 10) to explain at the same time the observed high-$x_{F}$ suppression for \jpsi. 
Therefore, a quantitative assessment of this effect is still lacking.
At the other extreme of the explored $x_{F}$ region (negative values), one observes a slight tendency towards 
enhancement of the \jpsi\ production ($\alpha >1$). In this region a model~\cite{Boreskov:2003ck}
formulated in the framework of reggeon phenomenology, indeed predicts an anti-screening effect, which cannot be reproduced
by more conventional approaches based on the commonly used shadowing parameterizations.

At midrapidity, several experiments have studied \jpsi\ production. The most remarkable feature is a tendency towards
stronger nuclear effects when $\sqrt{s}$ of the $\pA$ interaction decreases. In this region, the relevant processes 
affecting the production should be the final-state breakup of the $c{\overline c}$ pair and the parton shadowing in the nuclear target.
Both processes should approximately scale with $x_2$, i.e. one would expect nuclear effects to be the same at a certain $x_2$
independently of the initial proton energy. However, recent results by NA60~\cite{Arnaldi:2010ky}, comparing $\alpha$ as a function of 
$x_2$ for \jpsi\ production at 158 and 400 GeV, show that such a scaling does not hold, and that therefore other mechanisms
should be invoked to explain the $\sqrt{s}$-dependence of nuclear effects at midrapidity.

\begin{figure}[htb!]
\begin{center}
\includegraphics[width=\columnwidth]{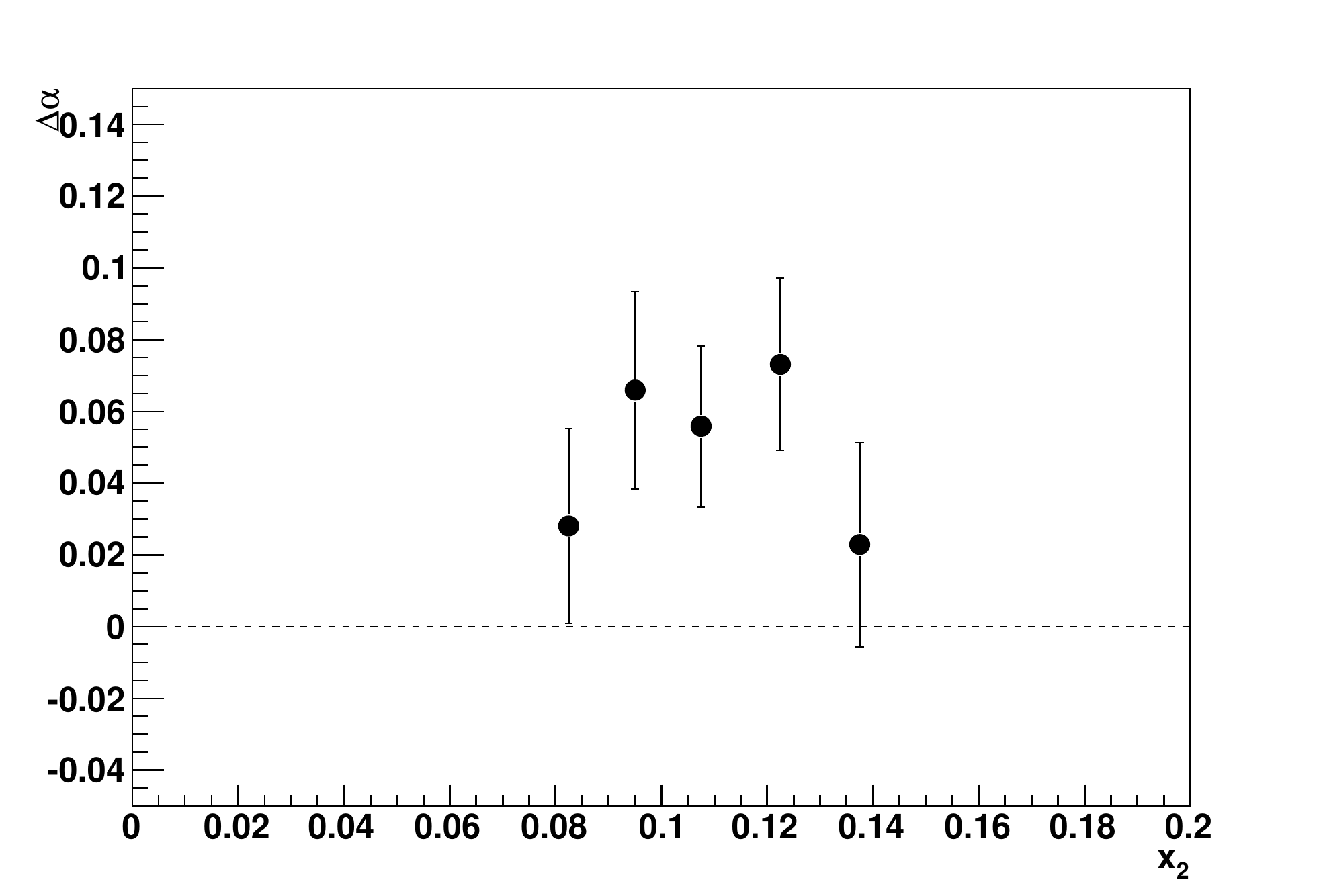}
\end{center}
\caption{\label{fig:alphavsx2} $\Delta\alpha$=$\alpha_{\rm 400 GeV}-\alpha_{\rm 158 GeV}$ vs $x_{2}$, as measured by NA60 in $\pA$ collisions at 158 and 400 GeV.}
 \end{figure}

To summarize, in spite of the rather extended set of measurements, which clearly define a trend for nuclear effects on
\jpsi\ production, the interpretation of fixed-target observations still remains unsatisfactory. 
For the future, further results
obtained at lower energies, at the SPS or at the future FAIR facility, may help produce a better characterization of the evolution
of nuclear effects vs $\sqrt{s}$. First ideas for fixed-target measurements at {\it{higher}} $\sqrt{s}$~\cite{Fle10}, 
using proton beams
extracted from the LHC, are also under consideration.

\subsubsection{Collider experiments}

The study of \jpsi\ production and propagation in cold nuclear matter has also been carried out at RHIC energies, through 
$\dAu$ collisions. The conclusions from the first round of measurements from the PHENIX experiment were not so sharp~\cite{Adare:2007gn}, 
because of the limited available statistics, but this problem has been recently overcome, with the availability of a first set of results corresponding 
to a much larger data sample (Run-6). One of the key features of the PHENIX results is the large rapidity coverage, including positive and
negative rapidities, corresponding to a large range of relative momenta between the $c{\overline c}$ pair and the medium.

More in detail, PHENIX has recently released $\dAu$ $\RCP$ data for \jpsi\ production
\cite{daSilva:2009yy} in nine rapidity bins 
over $|y| < 2.4$.  Systematic uncertainties associated with the 
beam luminosity, detector acceptance, trigger efficiency, and tracking 
efficiency cancel when \RCP, the ratio of central to peripheral events,
normalized to the number of $NN$ collisions, is formed. 
There is a remaining systematic uncertainty due to the centrality 
dependence of the tracking and particle identification efficiencies.
\begin{figure}[H]
  \centering
  \includegraphics[width=0.34\textwidth]{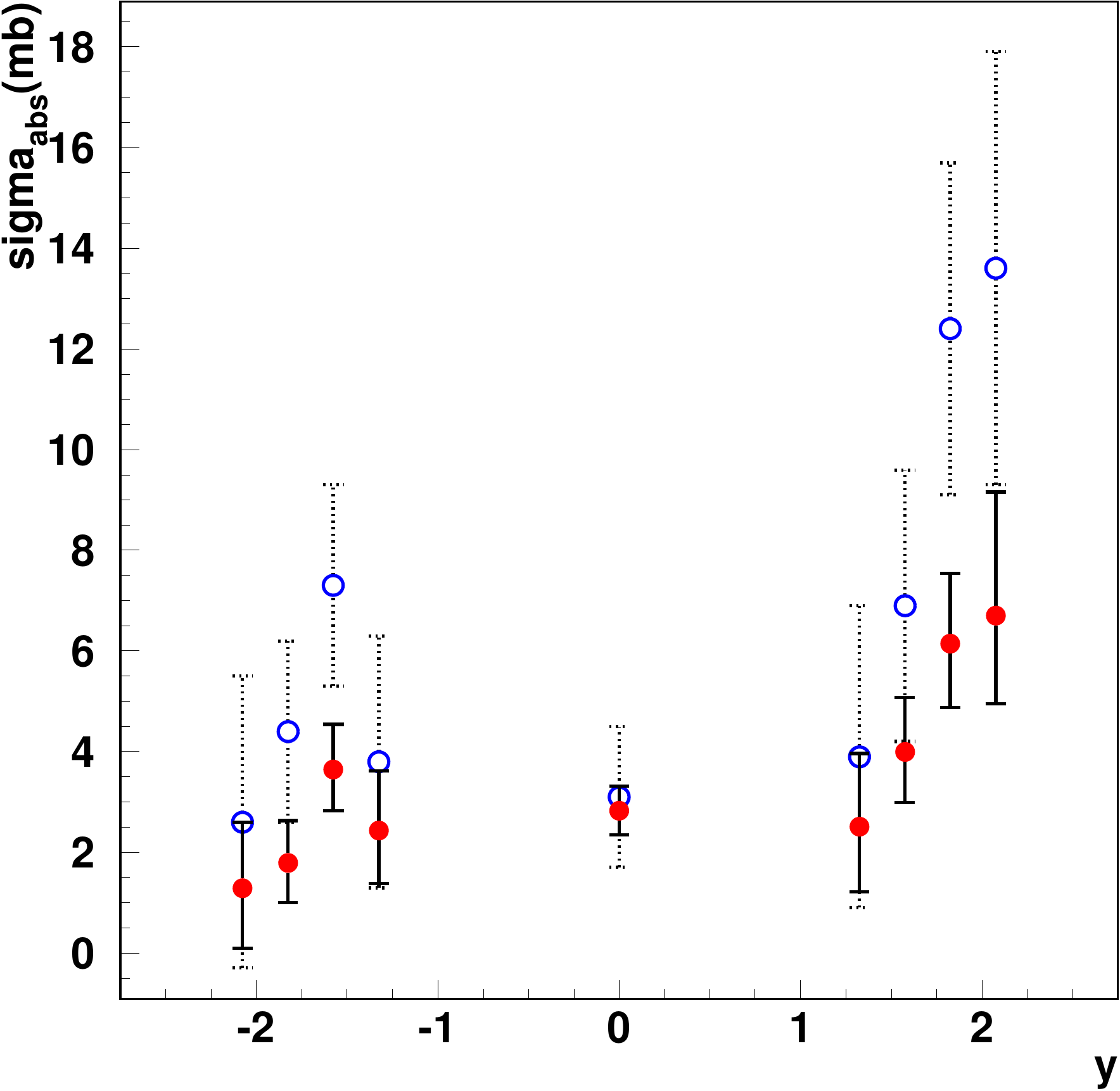}
  \includegraphics[width=0.34\textwidth]{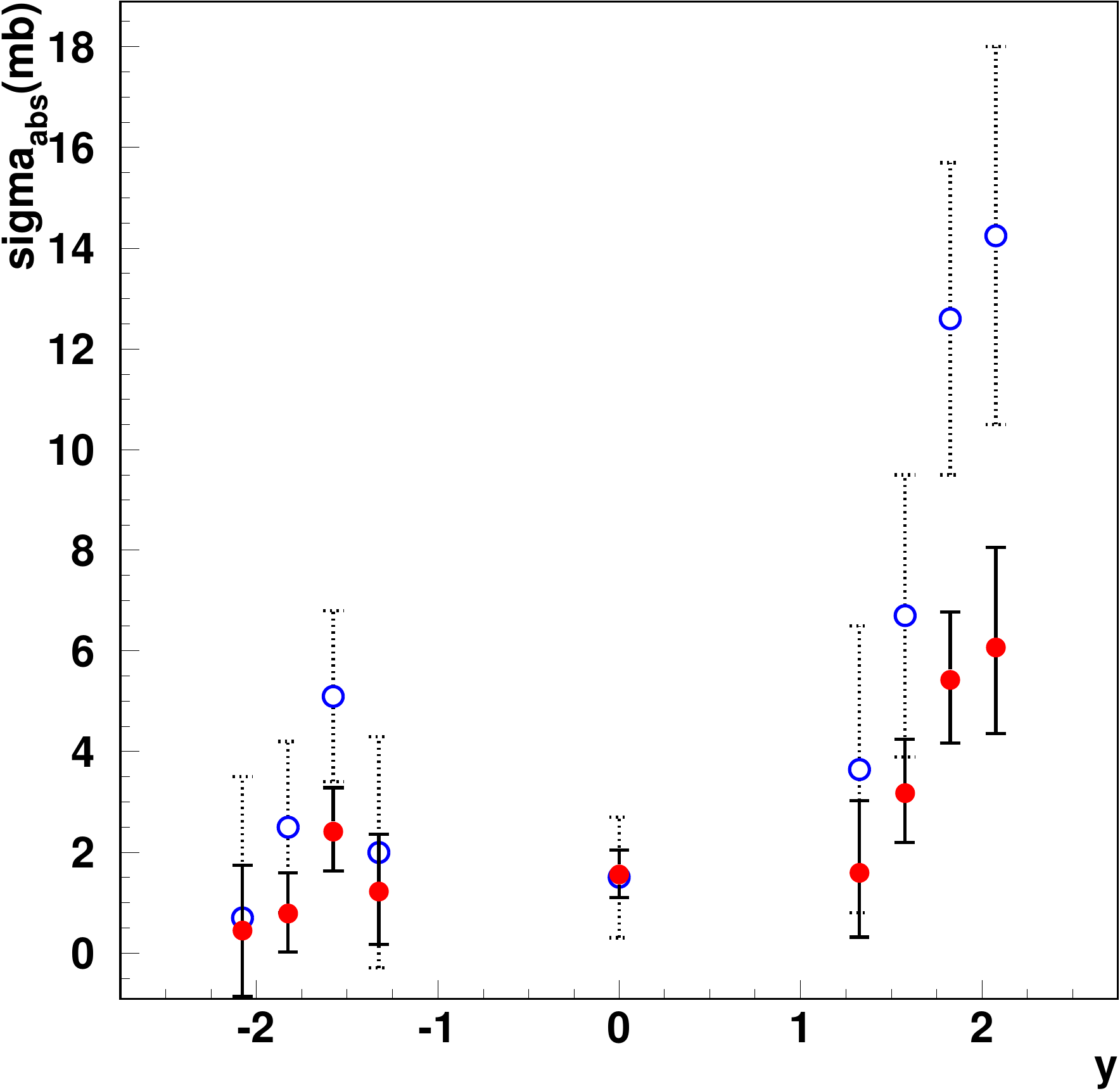}
  \includegraphics[width=0.34\textwidth]{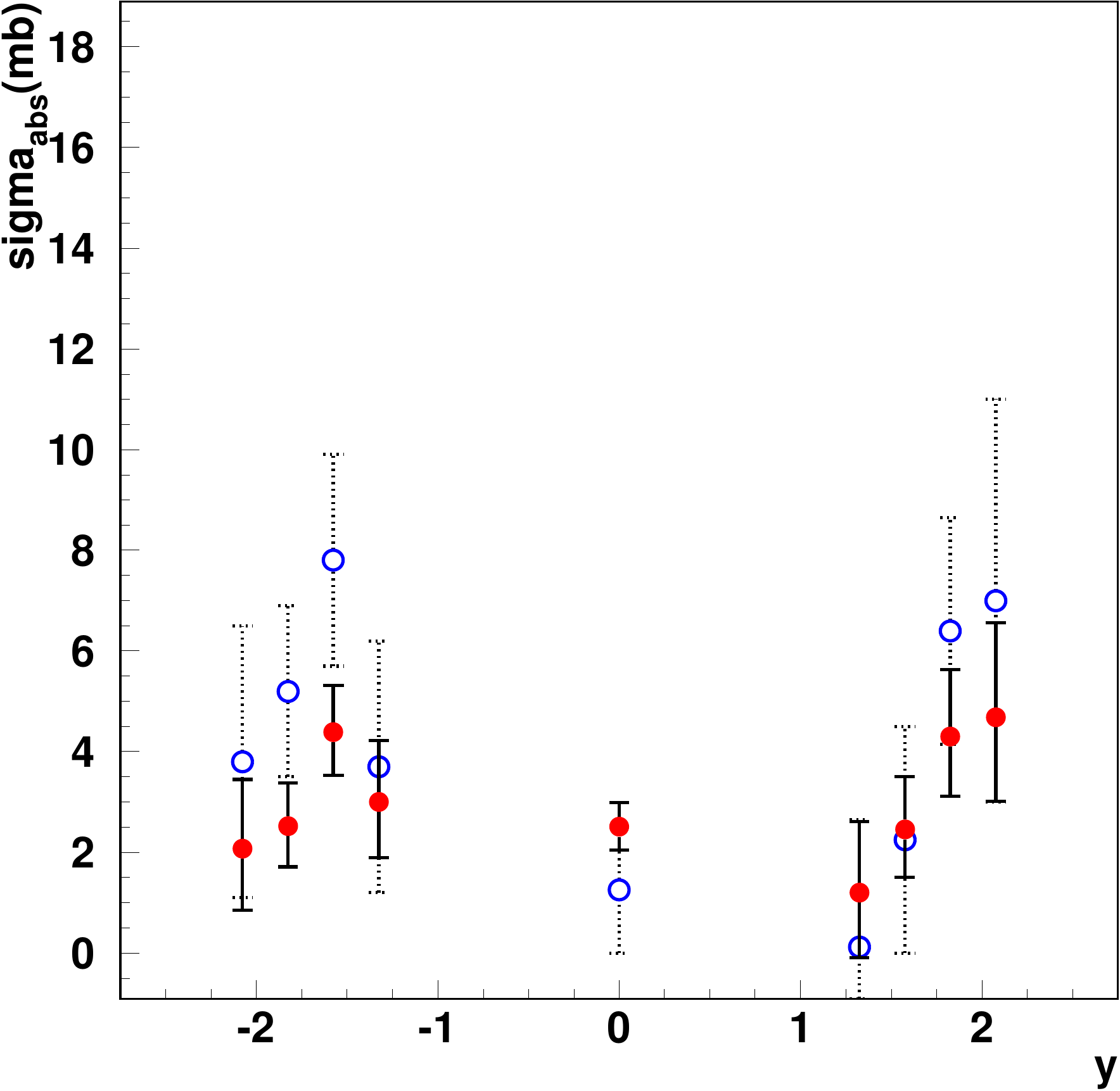}
  \caption{The effective absorption cross section as a function of rapidity
extracted from PHENIX $\dAu$ \RCP\ data in the extrinsic $2 \rightarrow 2$ scheme~\cite{Ferreiro:2009ur} (in red closed circle)
compared to the $2 \rightarrow 1$ scheme~\cite{tonyect} (in blue open circle) using a) EKS98, b) EPS08  and c) nDSg. 
From~\cite{Ferreiro:2009ur}.}
\label{fig:sigmaRCP}
\end{figure}
More importantly, there are significant systematic uncertainties in the centrality 
dependence of \RCP\ due to the use of a Glauber model to calculate the
average number of nucleon-nucleon collisions as a means of estimating the 
relative normalization between different centrality bins. The systematic 
uncertainty due to the Glauber calculation is independent of rapidity. 

The PHENIX $\dAu$ \RCP\ data have been independently fitted at each of the nine 
rapidities \cite{tonyect} employing a model including shadowing and 
\jpsi\ absorption.  The model calculations \cite{Vogt:2004dh} use the EKS98
and nDSg shadowing parameterizations with $0 \leq \sigma^{abs}_{J/\psi} \leq 15$~mb.
The best fit absorption cross section was determined at each rapidity, along 
with the $\pm 1\sigma$ uncertainties
associated with a) rapidity-dependent systematic uncertainties and b) 
rapidity-independent systematic uncertainties.  
The most notable feature is the stronger effective absorption cross section at 
forward rapidity, similar to the behavior observed at lower energies 
\cite{Leitch:1999ea}. In fact, it is striking that 
the extracted cross sections at forward rapidity are very similar for PHENIX 
($\sqrt{s_{_{NN}}} = 200$ GeV) and for the E866 collaboration ($\sqrt{s_{_{NN}}}
= 38.8$ GeV)~\cite{Lourenco:2008sk}, 
despite the large difference in center-of-mass energy.
One should note the large global systematic uncertainty in $\sigma_{\rm abs}$ extracted 
from the PHENIX \RCP\ data, dominated by the uncertainty in the 
Glauber estimate of the average number of collisions at each centrality. 
Although it does not affect the shape of the rapidity dependence of \sigabs, 
it results in considerable
uncertainty in the magnitude of the effective absorption cross section. 

It was recently suggested~\cite{Ferreiro:2008wc,Ferreiro:2009ur} that the large increase in 
effective absorption cross section at forward rapidity obtained from a LO CEM
calculation~\cite{tonyect},
may be significantly moderated if a more accurate $2 \rightarrow 2$ kinematics is used. 
Figure~\ref{fig:sigmaRCP} shows the results from the $2 \rightarrow 1$ kinematics~\cite{tonyect},
 compared to the ones from a $2 \rightarrow 2$ kinematics~\cite{Ferreiro:2009ur} using the $s$-channel cut~\cite{Haberzettl:2007kj}
as partonic model.  
The difference emphasizes the importance of
understanding the underlying production mechanisms.

PHENIX has very recently released~\cite{Adare:2010fn} final $R_{\dAu}$ and $R_{CP}$
data from the 2008 $\dAu$ RHIC run (where $R_{\dAu}$ is the \jpsi\ yield in $\dAu$
collisions normalized to the $\pp$ yield times the number of $NN$ collisions). 
The final $R_{CP}$ data are in very good agreement
with the preliminary results described above. A comparison of the $R_{\dAu}$ data 
with the $R_{CP}$ data shows that a simultaneous description of the two observables
will require a stronger than linear dependence of the \jpsi\ suppression on the
nuclear thickness function at forward rapidity. The dependence of the suppression on 
nuclear thickness is at least quadratic, and is likely higher. This result may have important 
implications for the understanding of forward-rapidity $\dAu$ physics, as well as for the 
estimate of cold nuclear matter effects in $\AuAu$ collisions.

\subsubsection{Cold nuclear matter effects at LHC energy}

The new energy regime recently opened up by the advent of the LHC offers new possibilities
for the study of quarkonium in cold nuclear matter. However, although the study of $\pA$
collisions is foreseen in the LHC physics program, the details of the acceleration scheme and
the choice of the nuclear beam still have to be finalized. Consequently, the study of the physics
observables accessible to the experiments is still at a rather preliminary level. 
Clearly, $\pA$ data at the LHC are essential to investigate various physics effects, and
``in primis'' nuclear shadowing in a still unexplored $x$-range. The final-state interaction
of charmonia with cold nuclear matter is another attractive topic. Due to the expected very short overlap 
time of the heavy-quark pair with nuclear matter, one could expect its influence on the still 
almost point-like $c\overline c$ to be very small. However, these considerations are
very qualitative and deeper theoretical studies are clearly necessary.

At this workshop, preliminary studies carried out in the frame of the ALICE experiment have been
discussed~\cite{Hadjidakis:QQ10}. In particular, the availability of a forward muon spectrometer (rapidity
coverage $2.5<y<4$) gives the opportunity for studying effects connected with gluon saturation via
the measurement of $R_{\pPb}$ for the \jpsi. A model for heavy-quark production in a 
Color-Glass Condensate environment~\cite{Fujii:2006ab} has been used to make predictions for $R_{\pPb}$ as a function
of $y$ and $p_{T}$ in the acceptance of the ALICE muon spectrometer. 
Such predictions quantitatively differ from the expectations related to 
a pure shadowing scenario, making such a measurement an interesting tool to investigate saturation-related
effects.

\section{Polarisation and feed-down in quarkonium production}\label{WG3}

Until recently, the numerous puzzles in the prediction of quarkonium-production 
rates at hadron colliders were attributed to 
non-perturbative effects associated with channels in which the heavy
quark and antiquark are produced in a colour-octet (CO) state~\cite{Brambilla:2004wf,Kramer:2001hh,Lansberg:2006dh,Lansberg:2008zm}.
It is now widely accepted that $\alpha^4_s$ and $\alpha^5_s$ corrections to the CSM~\cite{CSM_hadron} 
are fundamental for
understanding the $p_T$ spectrum of $J/\psi$ and $\Upsilon$ produced in
high-energy hadron  collisions~\cite{Campbell:2007ws,Artoisenet:2007xi,Gong:2008sn,Gong:2008hk,Artoisenet:2008fc,Lansberg:2008gk}.
The effect of QCD corrections is also manifest in the polarisation predictions. While the 
$J/\psi$ and $\Upsilon$  produced inclusively or in association with a photon are predicted to be 
transversally polarised at LO via a color singlet (CS), it has been recently emphasised that their polarisation at NLO is 
 increasingly longitudinal when $p_T$ gets larger \cite{Gong:2008sn,Artoisenet:2008fc,Li:2008ym,Lansberg:2009db}.

On the other hand, the LO NRQCD calculation (for which CO transitions dominante) 
predicts a sizable transverse polarisation rate for large $p_T$
${J/\psi}$ \cite{Beneke:1995yb,Beneke:1996yw,Braaten:1999qk,Kniehl:2000nn,Leibovich:1996pa} 
whereas the Tevatron CDF measurement at Fermilab \cite{Abulencia:2007us} 
displays a slight longitudinal polarisation at large $p_T$ for $J/\psi$ and $\Upsilon$ and a stronger
one for $\psi(2S)$.

To clarify the situation, more efforts on both experimental and theoretical aspects are 
expected from the forthcoming LHC. In the following, we review important aspects of quarkonium polarisation both 
from a theoretical and experimental perspective with particular emphasis on the $J=1$ states such as the $J/\psi$, $\psi(2S)$
 and $\Upsilon(nS)$ .

\subsection{Frames and polarizations}

The polarisation of the $J=1$ quarkonia is defined from their dilepton decay. In general, the angular distribution of
the dilepton is
\eqs{
  W(\vartheta, \varphi)   \,
   \propto \, \frac{1}{(3 + \lambda_{\vartheta})} \,
  (&1 + \lambda_{\vartheta} \cos^2 \vartheta
  + \lambda_{\varphi} \sin^2 \vartheta \cos 2 \varphi \\ +& \lambda_{\vartheta \varphi} \sin 2 \vartheta \cos \varphi ) ,
 \label{eq:ang_distr_general}
}
where $\vartheta$ and $\varphi$ are the (polar and azimuthal)
angles of the positive lepton with, respectively, the
polarisation axis $z$ and the production plane $xz$ (containing 
the colliding particles and the decaying meson).
Several polarisation frames have been used in the past.
In the helicity frame the polar axis coincides with the flight
direction of the meson in the centre-of-mass frame of the colliding
hadrons.  In the Collins-Soper frame, 
the polar axis reflects, on average, the direction of the relative
velocity of the colliding partons.

To clarify this puzzling situation, improved measurements are needed.
So far, most experiments have presented results based on a fraction of
the physical information derivable from the data: only one
polarisation frame is used and only the polar projection of the decay
angular distribution is studied.  These incomplete results prevent
model-independent physical conclusions.
Moreover, such partial descriptions of the observed physical processes
reduce the chances of detecting possible biases induced by
insufficiently controlled systematic effects.
In the forthcoming LHC measurements, 
it is important to approach the
polarisation measurement as a multidimensional problem, determining
the full angular distribution in more than one frame. 

In a series of recent works \cite{Faccioli:2008dx,Faccioli:2010ej,Faccioli:2010ji,Faccioli:2010kd},  a frame-invariant 
formalism was proposed to minimize the dependence of the measured result on
the experimental acceptance. It provides a much better control of the
systematic effects due to detector limitations and analysis biases. The overall idea
is to study both experimentally and theoretically the quantity
\begin{equation}
  \tilde{\lambda} 
  = \, \frac{\lambda_\vartheta + 3 \lambda_\varphi }{1 -
    \lambda_\varphi} \, , \label{eq:lambda_tilde}
\end{equation} 
instead of $\lambda_\vartheta$ and $\lambda_\varphi$.

\subsection{Charmonium polarization}

The NLO QCD corrections to ${J/\psi}$ polarisation in the CSM
at Tevatron and LHC have been calculated for the first time in~\cite{Gong:2008sn}. At
order $\alpha_S^4$, one should also take into account the contribution for 
$gg\to J/\psi+c \bar c$~\cite{Artoisenet:2007xi}. These are dominant at large $p_T$.
The results showed that the ${J/\psi}$ polarisation changes
drastically from strongly transverse  at LO
into strongly longitudinal  at NLO. While, this does not
completely agree with the CDF data, the trend is rather encouraging especially since
feed-down from  $P$-waves can alter the prompt yield polarisation.
The same trend is observed for the $\psi(2S)$~\cite{Lansberg:2008gk}, for which polarization measurements
are not affected by the $P$-waves. This is illustrated by \cf{fig:polar_psi2s}.

\begin{figure}[htb!]
\center{
\includegraphics[width=\columnwidth]{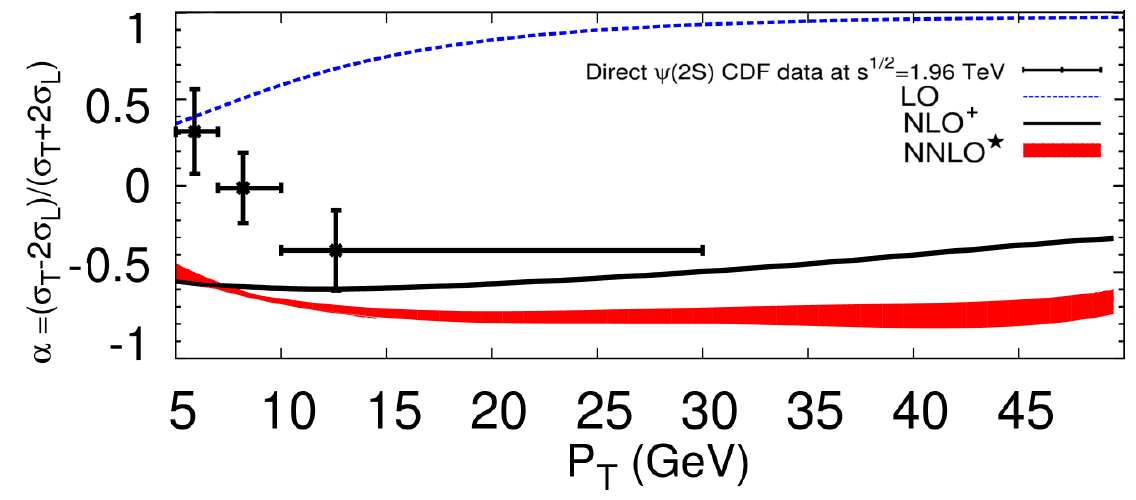}
\caption { \label{fig:polar_psi2s}
Polarisation of  $\psi(2S)$  directly 
produced as function of its transverse momentum $p_T$ at the Tevatron
as predicted by the CSM at LO, NLO$^+$ (including $gg\rightarrow J/\psi c\bar c$)
and at NNLO$^\star$ compared to the CDF measurement at $\sqrt{s}=1.96$ TeV.}}
\end{figure}

The first NLO QCD corrections of the ${J/\psi}$ 
production via CO states $\bigl[\bigl.^1S^{(8)}_0,\bigl.^3 S^{(8)}_1\bigr]$ at the Tevatron and LHC
were studied in~\cite{Gong:2008ft}.
Contrary to the CS case, these corrections only slightly change the 
 ${J/\psi}$ $p_T$ distributions and the polarisation compared to the LO NRQCD calculations. 
This result implies that the perturbative
QCD expansion quickly converges for ${J/\psi}$ production via the $S$-wave
CO state, in contrast to the CSM, where the NLO
contributions open new channel which show a leading $p_T$ behaviour with a 
different $J/\psi$ polarisation.

\begin{figure}[tbp]
\center{
\includegraphics*[scale=0.4]{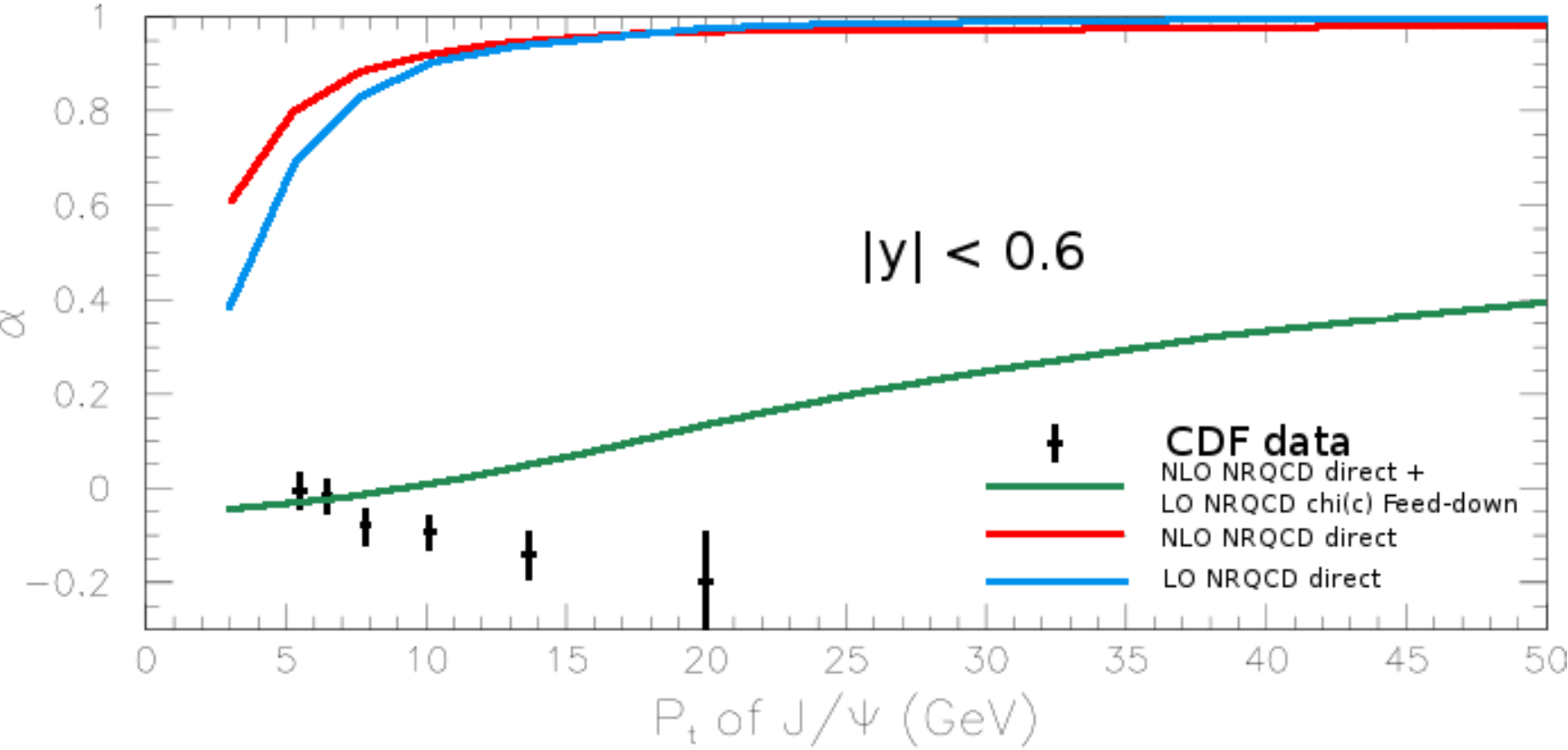}
\caption {\label{fig:polar}
Direct-$J/\psi$ yield polarisation $\alpha$ based on the work~\cite{Gong:2008ft} and the NRQCD matrix elements are from
the recent work~\cite{Ma:2010yw}. The experimental data is from the CDF measurement at the Tevatron~\cite{Abulencia:2007us} 
for the prompt $J/\psi$.}}\label{fig:pol-COM-NLO}
\end{figure}

As shown in \cf{fig:pol-COM-NLO}, an obvious gap  between the
theoretical results for $J/\psi$ polarisation calculated up to NLO including CO transition~\cite{Gong:2008ft}
and the experimental measurements at Tevatron  is observed for increasing $p_T$. There still remains only a
narrow window which might fill the gap if one follows the {\it a priori} rigourous theoretical framework of NRQCD, 
namely the NLO corrections to ${J/\psi}$ production via $P$-wave
CO states and ${J/\psi}$ production by feed down from $\chi_{cJ}$. It is worth noting however
that the latter possible solution does not apply for $\psi(2S)$ where a similar gap is observed --with a large magnitude even--, while
the former one would be at odds with the suppression of $C=+1$ CO transition expected from $e^+e^-$ studies.

The results given by two recent works~\cite{Ma:2010yw,Butenschoen:2010rq,Butenschoen:2010px} 
show that the $p_T$ distribution of the $P$-wave CO contribution at QCD NLO can be 
decomposed into the linear combination of the two $S$-wave COs. 
In Ref.~\cite{Ma:2010yw}, the $\chi_c$ feed down to prompt $J/\psi$ hadroproduction 
is taken into account, contributing at the level of $20-40 \%$. In this case, the NRQCD matrix elements 
${\langle\mathcal{O}^H_n\rangle}$ are determined to be ${\langle%
\mathcal{O}^\psi_8(\bigl.^3S_1)\rangle}=0.0005 \mathrm{~GeV}^3$
and ${\langle\mathcal{O}^\psi_8(\bigl.^1S_0)\rangle}=0.076
\mathrm{~GeV}^3$. It should be noted that some experimental data have been omitted since 
the fit of the $p_T$ distribution  starts at 7 GeV. The theoretical prediction for the polarisation of the direct $J/\psi$ 
hadroproduction is plotted in Fig.~\ref{fig:polar}, based on the calculation of Ref.~\cite{Gong:2008ft}. We also
note that such a value for ${\langle\mathcal{O}^\psi_8(\bigl.^1S_0)\rangle}$ does comply with the constraint of 
\ce{eq:constraint-CO-ee}. This either hints at an overestimation of this transition in these studies or at 
a breaking of NRQCD matrix-element universality for charmonia. In such a case, NRQCD would become nearly 
unpredictive as far as charmonium production is concerned.

Two aspects should be emphasized when comparing the theoretical prediction and the existing experimental measurements.
It is impossible to give the polarisation of the prompt $J/\psi$ hadroproduction without 
a calculation of the polarisation for the $\chi_c$ feed-down contribution if it really contributes  $20-40 \%$
to the $p_T$ distribution. In addition, the polarisation prediction in Fig.~\ref{fig:polar} for direct 
$J/\psi$ production assumes that the CO $^3P_J^{[8]}(J/\psi)$ NRQCD matrix element
can be neglected. If not, its contribution to $J/\psi$ polarisation  must calculated
at NLO accuracy.  At LO, the $\chi_c$ and the $P$-wave CO contributions were both taken into account in the first
 NRQCD-based predictions~\cite{Braaten:1999qk}. 

\subsection{Bottomonium polarization}

As regards the polarisation of $\Upsilon$ produced in $pp$ collisions, the NLO CSM correction was first presented 
in~\cite{Gong:2008hk}, and followed by an analysis including the real-emissions at NNLO~\cite{Artoisenet:2008fc} which we discuss later on. 
The polarization is quite similar to that of $J/\psi$, \ie the polarisation changes drastically  from 
strongly transverse at LO into strongly longitudinal at NLO. The NLO results are also rather similar to those
obtained in the $k_T$ factorisation approach at LO~\cite{Baranov:2007ay,Baranov:2008yk}.

\begin{figure}[hbt!]
\center{
\includegraphics*[scale=0.45]{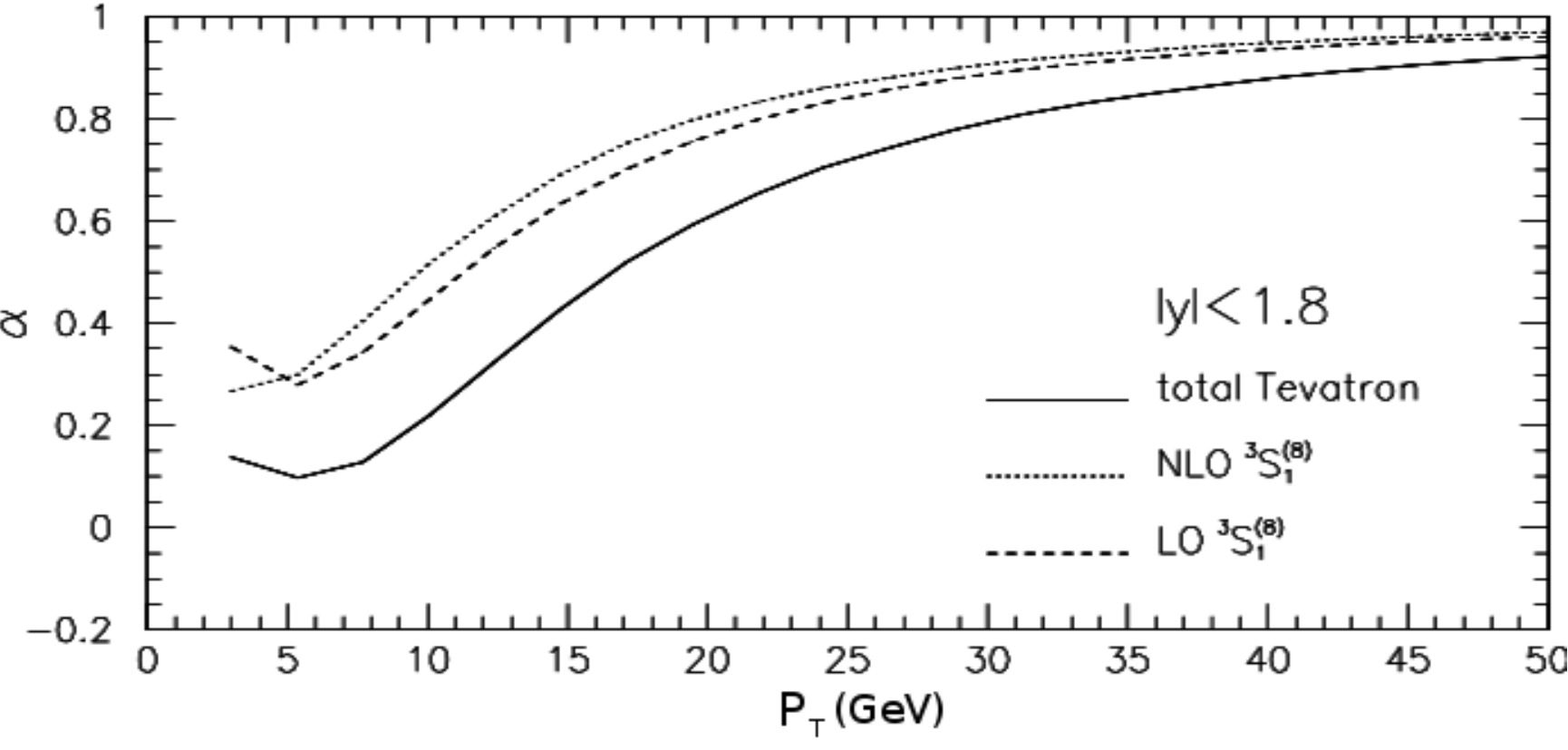}\\
\caption {\label{fig:polar_dir}Transverse momentum distribution of polarization parameter
$\alpha$ for direct $\Upsilon$ production at the Tevatron.}}
\end{figure}

The NLO QCD corrections of the $\Upsilon$ production via CO ${\bigl[\bigl.^1S^{(8)}_0,\bigl.^3S^{(8)}_1\bigr]}$ 
at the Tevatron and the LHC were studied for the first time in~\cite{Gong:2010bk}. 
The $\Upsilon$ polarization parameter, $\alpha$, from CO ${\bigl[^3S^{(8)}_1\bigr]}$
as function of $p_T$  is shown in \cf{fig:polar_dir}. Slight changes can be observed when the
 NLO corrections are taken into account. For any $p_T$, the direct $\Upsilon$ yield from CO is 
transversally polarized, with $\alpha > 0.5$ as soon as $p_T > 10$ GeV. 
 The predictions for the polarization of $\Upsilon$ production when the CSM NLO yield is added 
are also presented in the figure as a ``total'' result.

\begin{figure*}[hbt!]
\center{
  \includegraphics[width=1.75\columnwidth]{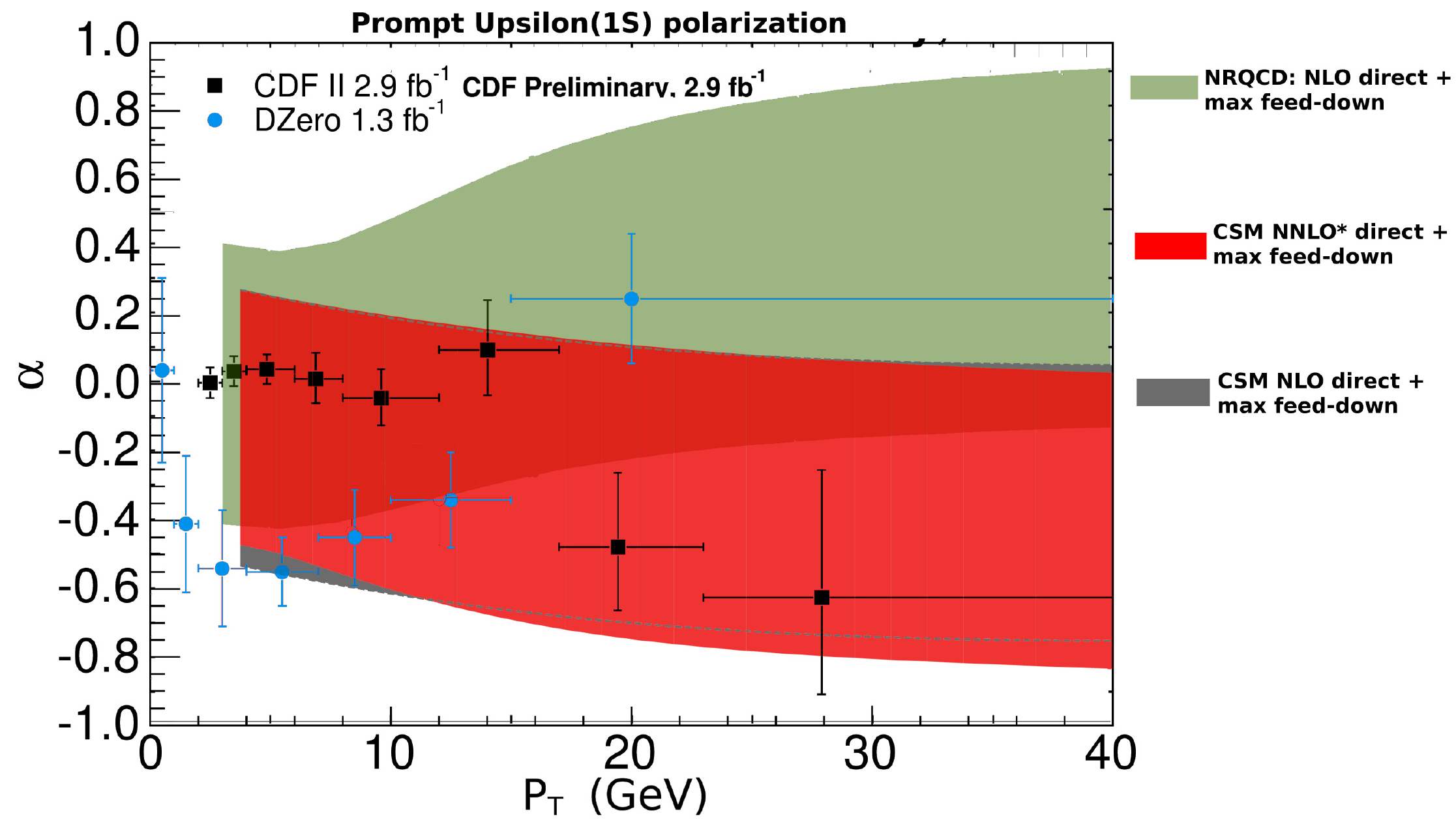}
\caption {\label{fig:polar_inc} Extrapolated prompt $\Upsilon(1S)$ polarization as a function of its transverse momentum at the Tevatron
as predicted by NRQCD (including CO transitions)~\cite{Gong:2010bk}, CSM at NLO~\cite{Artoisenet:2008fc} and CSM at 
NNLO$^\star$~\cite{Artoisenet:2008fc}, compared to the  D0~\cite{Abazov:2008za} and CDF~\cite{cdf-note-9966} data.}}
\end{figure*}

In \cf{fig:polar_inc}, 
the polarization of inclusive (\ie prompt) $\Upsilon$ production at the 
Tevatron is shown. As the polarization of $\Upsilon$ from the feed-down of $\chi_b(nP)$ 
is not available yet, a huge band is obtained only by imposing that the 
polarization of this part is between -1 to 1. 
The experimental data from D0 and CDF is also shown in the same figure. Since both data sets clearly disagree, 
it is difficult to draw any strong conclusion from such a comparison. Yet, if one follows the CDF analysis based on a higher integrated
luminosity and where the 3 $\Upsilon$ states are clearly separated out, one observes the same gap opening with $p_T$
as for the $\psi$'s, even with such a large theoretical uncertainty. For the sake of completeness, one should
underline that the direct yield from CO transitions is always predicted to be transverse. The CO yield can only become
slightly longitudinal thanks to the (unknown) feed-down. This is a clear motivation for the CDF and D0 to measure
the polarisation of direct $\Upsilon(1S)$ or that of $\Upsilon(3S)$, the latter being {\it de facto} direct.

In addition to the evaluation of the NLO CSM contributions to $\Upsilon$ production, 
a subset of the  NNLO contributions were evaluated in~\cite{Artoisenet:2008fc}, these are 
the real-emission contribution at order $\alpha_S^5$ believed to provide the leading contribution at large $p_T$. 
It was indeed found that these contributions, included in the NNLO$^\star$ yield, could be large and 
bring an agreement between a CSM-based evaluation and the CDF data. It also occured that the NNLO$^\star$ yield showed 
the same polarisation than the NLO yield, excepted for the --important-- fact that the yield was correctly described. 
The polarization of directly-produced $\Upsilon$ predicted likewise does not agree with the experimental 
data since it is strongly longitudinally polarized. Nevertheless, as we have discussed 
above, the experimental data includes contributions from $P$-wave decays, up to roughly 40 \%, and these can decay into 
transversally polarised $\Upsilon(1S)$. Even if the feed-down yield is unpolarized, the prompt polarization 
would only be $0.6$ times the direct polarization. Clearly when this is taken into account, the CSM NNLO$^\star$ and 
the NLO agree with the CDF data, as depicted by the gray (labeled NLO) and red (labeled NNLO$^\star$) band of \cf{fig:polar_inc}. 

In this partial QCD NNLO calculation, some infrared divergences are not automatically regulated 
since the loop corrections at $\alpha_s^5$  are missing. This imposed the introduction of an infrared cut-off, whose effect 
is believed to decrease as $p_T$ increases, provided that the real-emissions dominates at large $p_T$. This may not be exactly realised 
if double $t$-channel gluon exchanges have an important impact. In this case, the result could be overestimated. However, the agreement 
with the $k_T$ factorisation results gives us the hope that the infrared divergences are for a large part under control.

This partial QCD NNLO calculation was also applied to the $J/\psi$ and the $\psi(2S)$
case~\cite{Lansberg:2008gk} and the contribution was also found to be 
important. Yet, contrary to the $\Upsilon$ case, there is still a small gap opening at large $p_T$. At the leading
order in $v$, the prediction for the $J/\psi$ and the $\psi(2S)$ are identical. The $\psi(2S)$
polarisation  as function of  $p_T$ is depicted by the red band (labeled NNLO$^\star$) band of \cf{fig:polar_psi2s}. 
It is similar to that of NLO.

\subsection{Polarization in photoproduction and at RHIC}

The $J/\psi$ polarization in photoproduction at HERA was also studied up to NLO in~\cite{Artoisenet:2009xh,Chang:2009uj}. The results showed that the transverse momentum, $p_T$, and energy fraction, $z$, distributions of $J/\psi$ 
production do not agree well with the observations. The theoretical uncertainties on the $z$ distributions of the
$J/\psi$ polarization parameters in the target frame for various choices of the renormalization and factorization
scales are too large to provide a definite prediction relative to the experimental data~\cite{Jungst:2008ip}. It 
is quite easy to understand that the uncertainty in the QCD effects could be rather large since the $p_T$ of the $J/\psi$ is 
quite small at HERA, $1 < p_T < 5$ GeV. Thus the theoretical prediction is not expected to agree 
well with the  measurements.

The NLO CSM polarisation of  the $J/\psi$  produced in proton-proton collisions at RHIC at $\sqrt{s}=200$ GeV 
has been studied~\cite{Lansberg:2010vq}. The results show that 
the polarisation pattern at NLO in the helicity frame is in good agreement with the PHENIX  data both 
in the central~\cite{Adare:2009js} and the forward~\cite{Atomssa:2008dn} regions
when extra contribution from $cg$ fusion and a data-driven range for the $\chi_c$ contribution to the $J/\psi$ 
polarisation are considered. For the time being, nothing is known about $P$-wave feed downs at RHIC especially at large $p_T$
where $x_T$ starts to be large. This prevents any strong conclusion based on the comparison with the data regarding the status of the CO
contributions and a possible overestimation by the partial NNLO contribution.

\subsection{Perspectives on polarization studies}

To clarify the $J/\psi$ polarisation puzzle, other observables such as the measurements 
of $J/\psi$ ($\Upsilon$) hadroproduction in association with a photon at the LHC can
 be considered.  The NLO CS contribution was studied~\cite{Li:2008ym} and the 
results show that the  $J/\psi$ polarization changes from transverse
 at leading-order to longitudinal  at NLO.  It is known that the
CO contribution to the inclusive $J/\psi$ hadroproduction may be one order 
of magnitude larger than the CS contribution, and the 
polarization distribution may be
dominated by the CO  at NLO. In contrast, the CS contribution for
$J/\psi+\gamma$ production is at least of the
same order as the CO contribution. The polarization
should then  arise from both the CS and CO.
Therefore, measurements of $J/\psi$ production associated with a
direct photon at hadron colliders could be an important
test of the theoretical treatment of 
heavy-quarkonium production. The partial NNLO  
result~\cite{Lansberg:2009db}, NNLO$^\star$, is about one order of magnitude larger than the NLO 
$J/\psi$ $p_T$ distribution. The predicted polarization  are almost the same. 
Therefore, it will be reality check on the partial NNLO treatment. 

Moreover, to compare the polarisation measurements in different frames, we need theoretical predictions 
based on different polarisation frames such as the helicity frame (HX) and  Collins-Soper frame. Along these lines, we would like
to mention a recent pioner work on the polarization studies of $\Upsilon+ \gamma$ in the $k_T$ factorization approach
using different polarization frames~\cite{baranov-these-proc}.

Theoretical predictions of the new, proposed, frame-independent polarisation parameter $\tilde{\lambda}$ are needed. 
Of course, all the prediction should at NLO accuracy.
Furthermore, the comparison between experimental data and theory must consider
the experimental acceptance and efficiency.  Experiments measure the
net polarisation of the specific cocktail of quarkonium events
accepted by the detector, trigger and analysis cuts.  
It will be very useful to have an event generator where the quarkonium decay into lepton pairs
for inclusive production at NLO which could be embedded into Monte Carlo simulations.

\section{Open heavy-flavor production in $\pp$ and $\pA$ collisions}\label{WG4}
Open heavy-flavor production in $\pp$ and $\pA$ collisions is tightly
connected to the process of quarkonium (hidden flavor) production.
Nevertheless, it is in itself one of the most interesting processes to
study from theoretical and experimental points of view (see, e.g.,
Refs.~\cite{Jung:2009eq,Lipka:2010xr} for a review on recent results).  In fact,
the production of heavy quarks allows us to test fundamental concepts,
such as perturbative QCD, factorization and non perturbative power
corrections. It also represents a mandatory benchmark for quarkonium
production and parameters derived in open heavy-flavor production
serve as the basis for calculations of quarkonium production.

Thanks to the massiveness of heavy quarks, the production cross
section can be calculated analytically in perturbative QCD down to
$p_T = 0$ at the partonic level \cite{Nason:1987xz}.  However, differential
distributions and event shapes exhibit large logarithmic
contributions, typically corresponding to soft or collinear parton
radiation, which need to be resummed to all orders~\cite{Bonciani:1998vc,Czakon:2009zw,Cacciari:2005rk}.  
Furthermore, dead-cone effects suppress
gluon radiation around the heavy-quark direction, which has relevant
phenomenological implications on both parton- and hadron-level
spectra.  As far as heavy-hadron production is concerned, one can
model non perturbative effects by means of a fragmentation function,
depending on a few tunable parameters, or by including power
corrections in a `frozen' or `effective' strong coupling 
constant~\cite{Dokshitzer:1995ev,Dokshitzer:1995qm,Aglietti:2006yf}. Reliable hadronization models 
should also be capable of distinguishing the spin of the produced hadrons, e.g.
separating the $D^*$ from the $D$, and describing both baryons and
mesons and the multiplicity ratios $\Lambda_c/D$ or $\Lambda_b/B$.
Typically, hadronization models, such as the cluster \cite{Webber:1984zzz} or
string~\cite{Andersson:1983ia} models, are implemented within the framework
of Monte Carlo generators, such as HERWIG~\cite{Corcella:2000bw}, PYTHIA~\cite{Sjostrand:2006za} or MC@NLO~\cite{Frixione:2002ik} and are tuned to heavy-hadron
data from $e^+e^-$ experiments, e.g. LEP and SLD data (see, for
example, \cite{Corcella:2005dk} for a tuning to $b$-quark fragmentation). 
Multi-purpose
generators have been available for several years for lepton/hadron
collisions in the vacuum and have been lately modified to include the
effects of dense matter \cite{Armesto:2009fj,Armesto:2009ab} (see also \cite{Lokhtin:2005px,Zapp:2008gi,Renk:2008pp,Zapp:2008af,Schenke:2009gb} for generators specific for heavy-ion collisions).

A profound knowledge of open heavy-quark production is crucial for the
understanding of quarkonium phenomenology. The production,
e.g., of the $J/\psi$ in Non relativistic QCD (NRQCD) is
described as the convolution of the perturbative $c\bar{c}$ cross
section and a non perturbative operator, possibly depending on the
color of the $c\bar c$ state, according to the
color-singlet and color-octet mechanisms
(see e.g. \cite{kn}).

In order to promote the formalism used to describe open heavy flavors
to $\pA$ and ultimately $\AA$ collisions, one has to introduce
medium-modification effects, taking into account the existing
differences between light and heavy quarks (hadrons) \cite{Armesto:2003jh}.
In particular, one of the most striking observations of heavy-ion
collisions is the jet-quenching phenomenon, namely the suppression of
hadron multiplicity at large transverse momentum ($p_T$) with respect
to $\pp$ processes \cite{StarWhitePaper,PhobosWhitePaper,BrahmsWhitePaper,PhenixWhitePaper}.

\begin{figure}[tbh]
    \includegraphics[width=0.98\columnwidth]{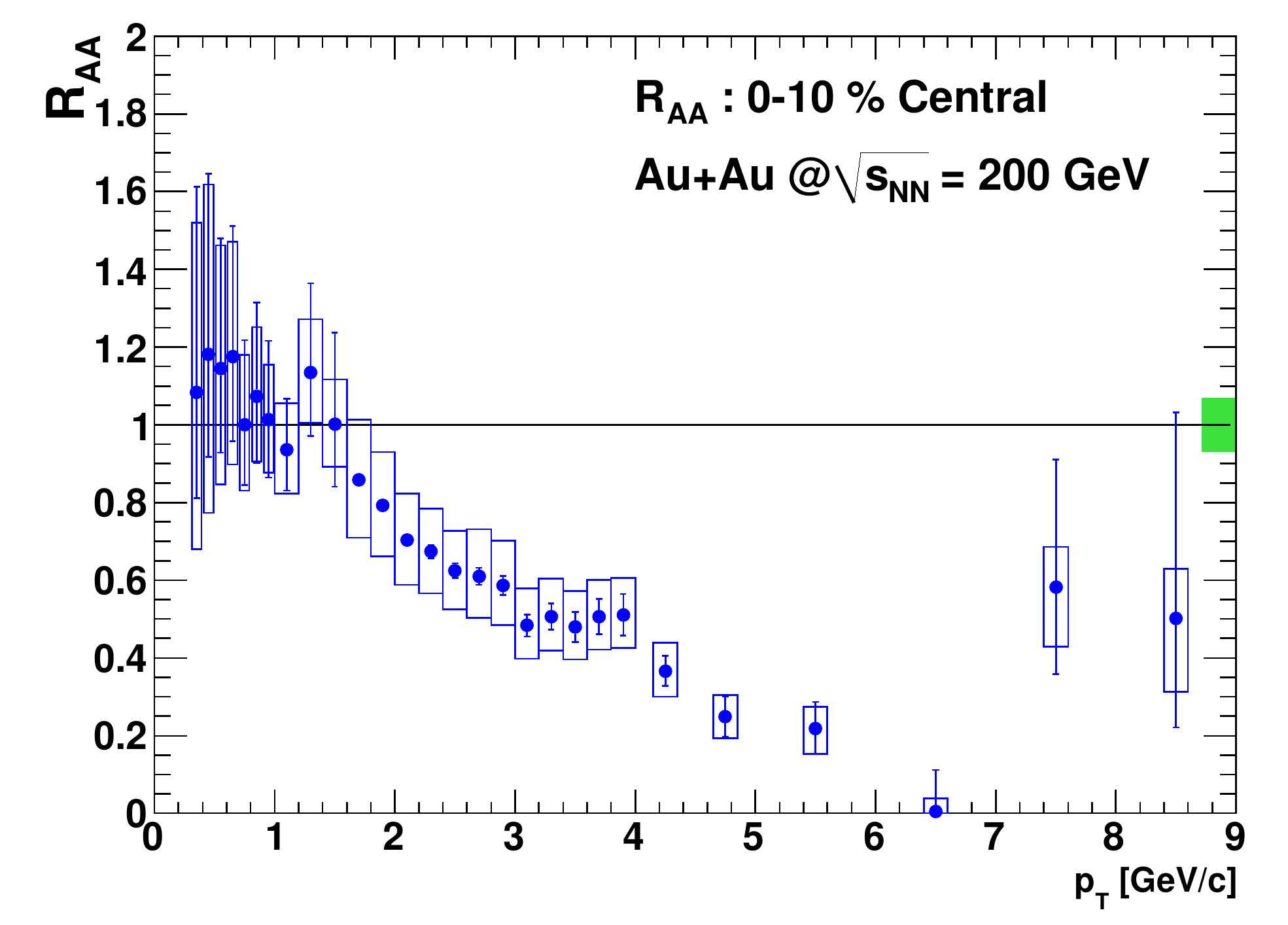}
    \caption{Nuclear modification factor ($R_{AA}$) for non-photonic electrons 
originating from semileptonic decays of heavy-flavored mesons in the $0$-$10\%$ 
most central \AuAu\ collisions at RHIC measured by PHENIX~\cite{Adare:2010de}. 
This corresponds to the ratio of the invariant yields in \AuAu\ and $\pp$ collisions 
normalized by the number of binary nucleon collisions. This measurement indicates a 
strong suppression of heavy-flavored mesons in central heavy-ion collisions.}
    \label{fig:npePhenix}
\end{figure}

In the case of heavy hadrons, the measured suppression is expected to
be lower than that of light-flavor mesons, due to the 
dead-cone effect discussed above.
This effect becomes less important in the
high-$p_T$ limit where $p_T\gg m_Q$. So far, suppression of 
heavy-flavored mesons in $\pA$ or $\AA$ collisions has only been 
measured through leptons, i.e., the non-photonic electrons and muons from semi leptonic
decays of heavy-flavored mesons~\cite{Adare:2010de,Garishvili:2009ei,Abelev:2006db}.
Despite being a rather complex study, non-photonic electrons and muons have
various experimental advantages, the most important one being the
possibility to deploy the fast triggers, in such a
way to allow the experiments to
collect large data samples. On the other hand, they
originate from both charm and bottom mesons with only few indirect
experimental handles available to disentangle the two contributions.

Suprisingly, the energy loss observed through non-photonic
electrons is substantial already at moderate $p_T$ of $\sim$3 GeV/$c$.
Figure \ref{fig:npePhenix} shows this suppression via the nuclear modification factor
$R_{AA}$ defined as the ratio of invariant yields in \AuAu\ over that in $\pp$
collisions times the number of binary nucleon collisions in the \AuAu\
system at a given centrality. In the absence of any nuclear or dense
matter effects $R_{AA}$ is unity.  Several mechanisms to describe
heavy-flavor energy loss in the dense medium have been so far
proposed. Most prominent are collisional and radiative energy loss,
typically calculated in the weak-coupling regime.  
However, the magnitude of the
observed suppression at RHIC is hard to accommodate in these models.
Recent attempts, assuming strong coupling between the heavy-flavor
mesons and the dense medium, and based on the AdS/CFT correspondence,
lead to higher 
suppression factors, but further studies and experimental data are
needed before any conclusions can be drawn.

It is still an open question to what extent the energy-loss mechanisms,
observed in open heavy-flavor production, affect the quarkonium yields in
$\AA$ collisions.  After including quark-antiquark recombination effects,
the quenching of open heavy-flavored hadrons will also serve as a
benchmark for quarkonia in heavy-ion collisions.

Another crucial issue in heavy-flavor phenomenology concerns
heavy-quark parton distributions in protons and nuclei.  In fact,
processes with the production of heavy quarks are fundamental
to constrain the PDFs.  Higher-order corrections to the production of
heavy quarks, with the implementation of soft/collinear resummation to
improve the differential distributions, will also be helpful to
acquire information on heavy quark densities. The possible inclusion
of medium effects in the hard-scattering cross section, along with
nuclear parton distribution functions, will eventually lead to a
prediction taking fully into account the modifications induced by the
dense matter. However, unlike collisions in the vacuum, there is no
guarantee that factorization should also work in a medium.

In the following, we shall review the main results discussed within
the Open Heavy-Flavor Working Group in the Quarkonium 2010 workshop.
We shall discuss higher-order calculations of heavy-flavor production
spectra at RHIC, taking particular care about the role played by the
resummation on transverse-momentum distributions. We shall then
present a few selected topics on specific features of heavy quarks and
dead-cone effects, as well as novel developments on the modelling of
energy loss from heavy quarks in dense matter and its possible
relation with the jet-quenching observation.  Later on, recent results
on the production of photon plus heavy quarks, along with its
relevance to constrain charm/bottom PDFs in proton and heavy ions,
will be investigated.

\subsection{Open heavy-flavor production at RHIC and FONLL calculations}

As our understanding of heavy-ion collisions becomes more mature, it
becomes increasingly clear that heavy-flavor studies are a mandatory
tool to further our understanding of the quark-gluon-plasma formed in
high energy nuclear collisions at RHIC and now at LHC.  Having
reliable measurements of all aspects of heavy-flavor production
is compelling.

\begin{figure}[tbh]
  \includegraphics[width=0.98\columnwidth]{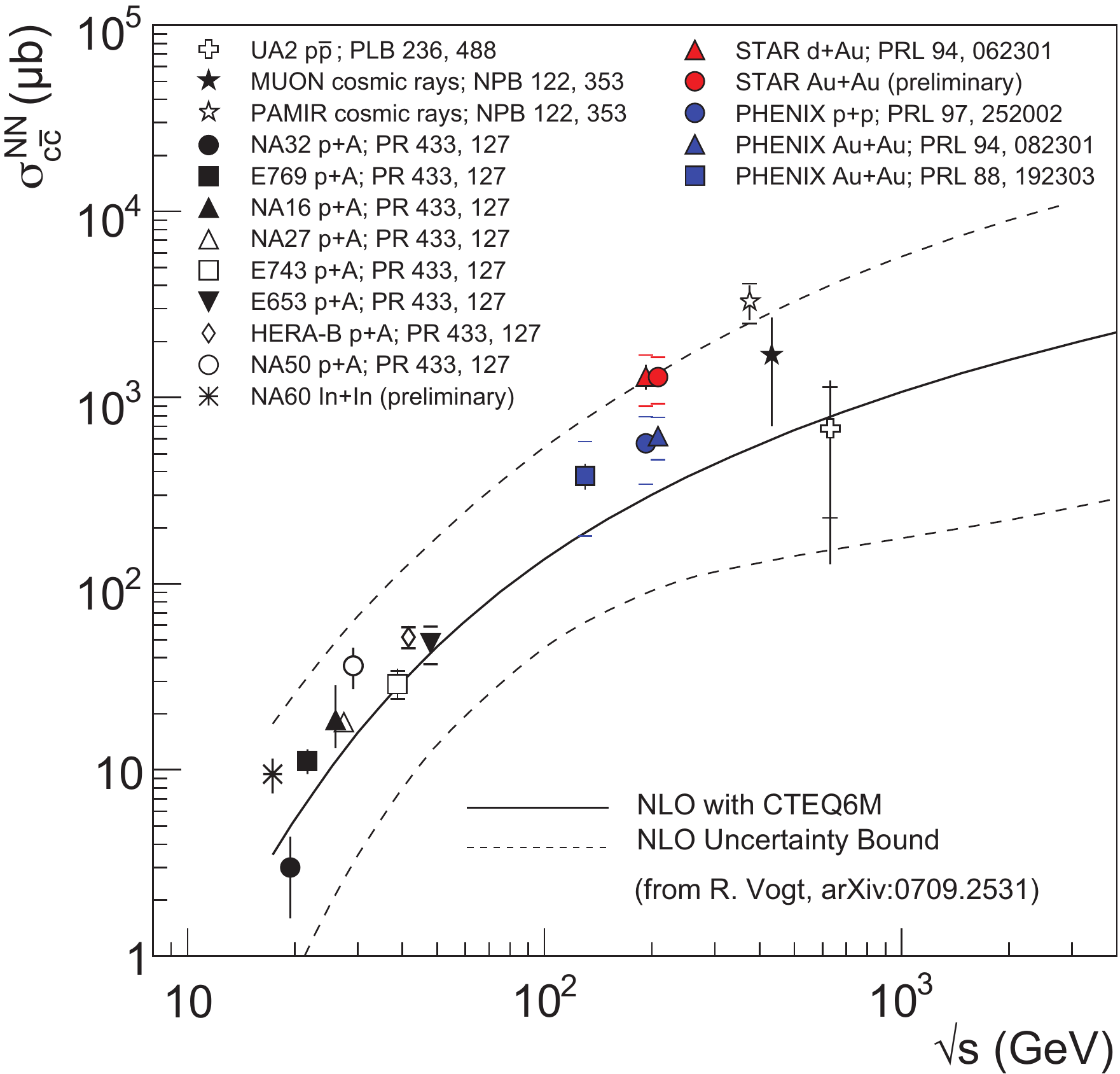}
    \caption{Comparison of total charm cross section measurements. The STAR and 
PHENIX results are given as cross section per binary collisions. Vertical lines 
reflect the statistical errors, horizontal bars indicate the systematic uncertainties 
(where available). From \cite{Frawley:2008kk}.}
    \label{fig:xsec}
\end{figure}

Experimentally, the total charm cross section is not well constrained.
Measurements of charm meson production at the Tevatron could be
conducted only at large transverse momenta ($p_T > 5$ GeV/$c$) while
measurements at lower energies exhibit
rather large uncertainties. At RHIC, measurements of $D$ mesons in conjunction with low-$p_T$
non-photonic electrons and muons in \dAu\ collisions \cite{Abelev:2008hja} (STAR) indicate 
a rather large total charm cross section of $\sim$1.3 mb. 
On the other hand, the non-photonic lepton measurements in pp collisions~\cite{Adare:2010de} (PHENIX) suggest a considerably smaller cross section of $\sim$ 0.6 mb. 
In all cases assumptions about rapidity
distributions and contributions from unmeasured charm mesons and
baryons (e.g. $\Lambda_c$) have to be made, adding further
contributions to the already large
statistical and systematic uncertainties. Upgrades
to the existing RHIC detectors (for an overview see
\cite{Frawley:2008kk}) will help to overcome these shortcomings. A
summary of  current total charm cross sections ($\sigma_{c\bar{c}}$) is
presented in Fig.~\ref{fig:xsec}.

On the theory side, the state of the art for charm-quark production is
NLO with the inclusion of $n_f=3$ active flavors \cite{Nason:1987xz}.  A few
ingredients of the NNLO corrections are available
\cite{Czakon:2007ej,Czakon:2007wk}, but the full NNLO calculation has not been
completed yet.  Another possible approach to compute the cross section
is to integrate the transverse momentum distribution computed
in the so-called FONLL approximation, i.e. soft/collinear resummation
to next-to-leading logarithmic accuracy (NLL), matched to the
NLO result. The total cross section is still NLO, but the charm quark
is treated as an active flavor, and therefore $n_f=4$ in the
calculation, which leads to a difference in the result \cite{Cacciari:2005rk}.

The FONLL calculation can of course be employed to predict
differential distributions, in particular the transverse momentum one,
wherein contributions $\sim\alpha_S^k\ln^{k-2}(p_T/m_Q)$ (LL) and
$\sim\alpha_S^k\ln^{k-3}(p_T/m_Q)$ (NLL), large for $p_T\gg m_Q$, are
summed up to all orders.  The spectra of $c$-flavored hadrons are then
obtained by convoluting the charm distribution with the appropriate
non perturbative fragmentation function, fitted, e.g., to LEP data. 
When doing the fit, the calculation of heavy-quark production in
$e^+e^-$ annihilation must be carried out in the same perturbative
approximation, i.e. NLO+NLL, and the scales are to be consistently set
\cite{Cacciari:2001cw,Cacciari:2002pa}.

The main sources of uncertainty on such predictions are the
renormalization and factorization scales, the quark masses, entering
in the perturbative calculation, parton distribution functions, such
as the gluon density about the charm/bottom mass scale.  As for the
quark mass, it is also unclear whether in the calculation one should
use the pole, the $\overline{\mathrm{MS}}$ or even the hadron mass.
For the purpose of the PDFs, the gluon distribution exhibits large
uncertainties at small $x$, especially a strong dependence on the
factorization scale.

\begin{figure}[tbh]
    \includegraphics[width=0.98\columnwidth]{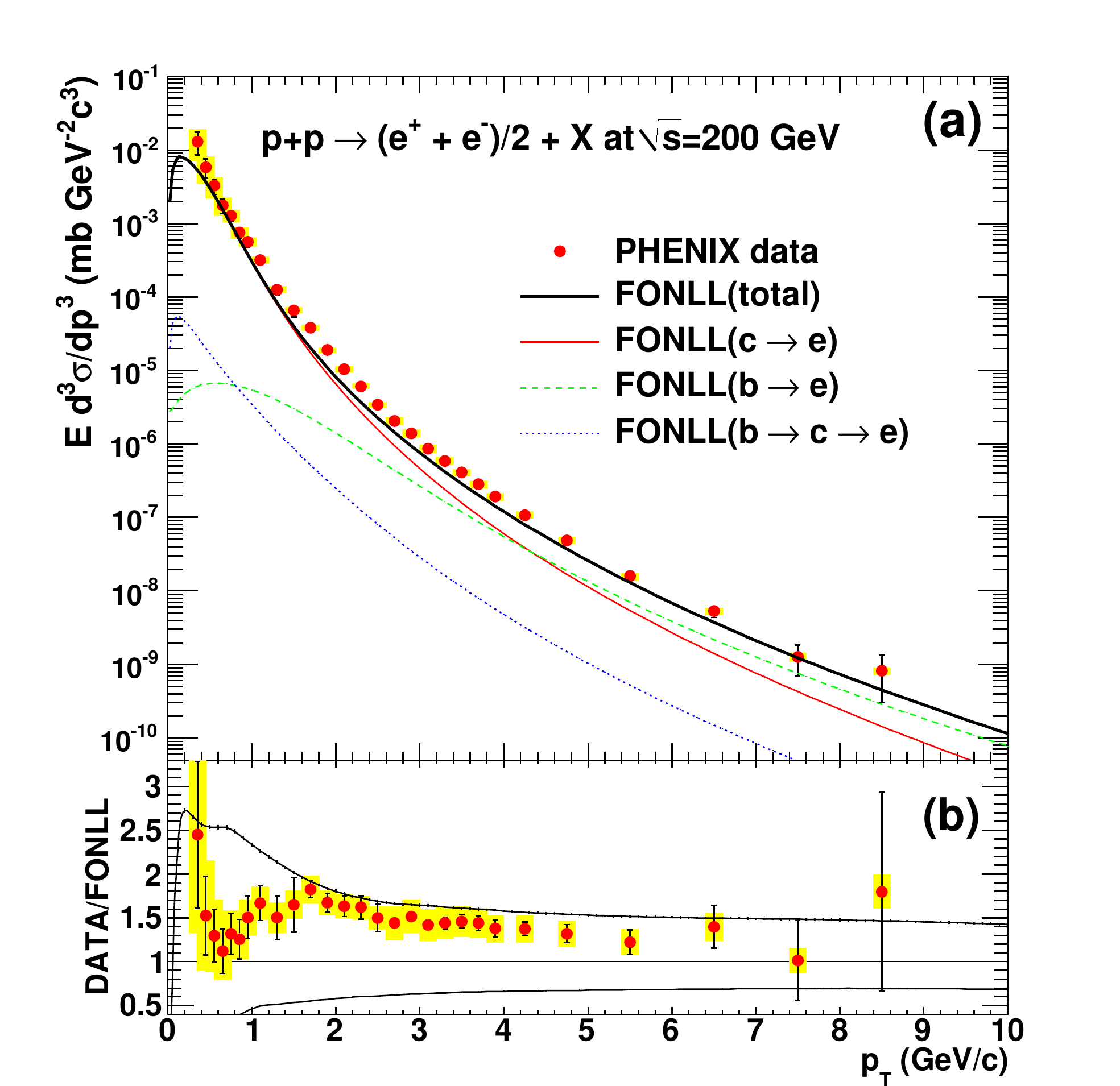}
    \caption{Non-photonic electron measurements at RHIC (PHENIX) in
        $\pp$ collisions at $\sqrt{s}=200$ GeV.
        (a) Invariant differential cross section of single electrons
        as a function of $p_T$.  (b) The ratio of FONLL/Data as a
        function of $p_T$.  The upper (lower) curve shows the
        theoretical upper (lower) limit of the FONLL calculation.
        From \cite{Adare:2010de}.
        }
    \label{fig:fonllcomp}
\end{figure}

From the $D$- and $B$-hadron distribution one can derive the non-photonic
electron spectrum at FONLL and compare it with RHIC data. 
Both the new PHENIX (shown in Fig.~\ref{fig:fonllcomp}) and STAR $\pp$ data 
are compatible with the 
FONLL computation,
within the theoretical and experimental uncertainties. 
However, the data favors large cross sections, about 1.5 times the nominal FONLL
calculation, just on the upper edge of the uncertainty band (Fig.~\ref{fig:fonllcomp}b).
Non-photonic electron spectra at low $p_T$ are notoriously difficult to analyze
due to the increased photonic backgrounds, which increase the uncertainties for
$p_T < 2$ GeV/$c$ considerably and make 
the evaluation of $\sigma_{c\bar{c}}$ from electrons alone problematic.

\begin{figure}[tbh]
    \includegraphics[width=0.98\columnwidth]{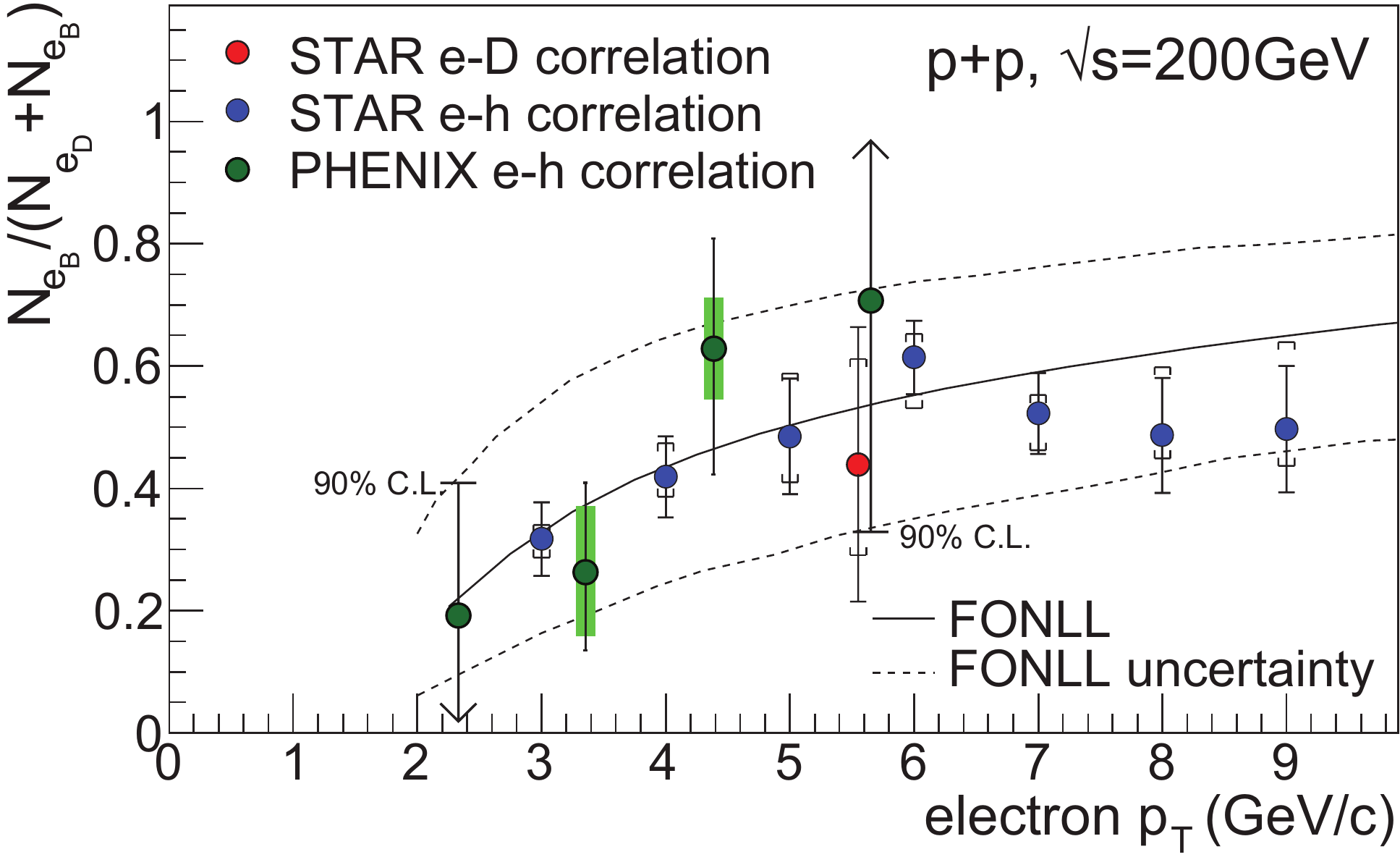}
    \caption{Transverse momentum dependence of the relative
        contribution from B mesons to the non-photonic electron yields
        as measured by PHENIX \cite{Adare:2009ic} and STAR
        \cite{Aggarwal:2010xp}.  Error bars are statistical and
        brackets/boxes are systematic uncertainties.  The solid line
        is a FONLL prediction and the dotted lines represent the
        uncertainty on this FONLL prediction.}
    \label{fig:candb}
\end{figure}

Furthermore, much work has been carried out towards the separation
of charm from bottom contributions to the lepton spectra in $\pp$ and $\pA$
collisions.
Experimental progress has been made using $e$-$h$ and $e$-$D$ correlations 
\cite{Adare:2009ic,Aggarwal:2010xp} shown in Fig.~\ref{fig:candb}. 
Again, results 
are consistent with the FONLL calculation, although experimental errors are 
currently still too large to allow an accurate separation of bottom and 
charm production in $\pp$ processes.

The final answer has to come from direct measurements of charm through
$D$ mesons (if able to subtract the $B \rightarrow D$ contribution looking e.g. at the impact parameter distribution)
 and that of bottom through $B \rightarrow J/\psi + X$ and/or $b$-tagged jets or leptonic decays.
These measurements will require high precision vertexing that will become available
for both RHIC experiments in the near future \cite{Frawley:2008kk}.

\subsection{Properties of gluon radiation off heavy quarks}

A heavy quark $Q$ is characterized by the fact that its mass is much
larger than the typical hadronization scale, say $m_Q\gg
\Lambda_{\mathrm{QCD}}$, e.g., in the $\overline{\mathrm{MS}}$
renormalization scheme.  The fact that the quark mass acts as a
regulator of the collinear singularity and that the strong coupling constant is
evaluated at relatively large scales, depending on the observable one
is looking at, makes perturbative QCD applicable.

When dealing with multiple emissions from heavy quarks, an essential
difference with respect to light quarks is that gluon radiation 
is suppressed inside the forward angular
cone with an opening angle $m_Q/E$, the so-called dead cone,
with $E$ being the heavy-quark energy~\cite{JPhysG17.1602}.  The dead cone implies the decrease of the overall energy
loss and thus to the leading-particle effect, i.e. heavy-hadron energy
distributions are peaked at large $x$, with $x$ being the energy
fraction in the centre-of-mass frame \cite{Dokshitzer:1995ev}.  In other words,
after multi-parton radiation and hadronization, a heavy hadron still
carries a significant fraction of the parent-quark energy.  Also, for
$x\simeq 1-\Lambda_{\mathrm{QCD}}/m_Q$, one starts to be sensitive to
non perturbative effects which can be included, e.g., by means of a
fragmentation function or a frozen/effective coupling constant.
Another relevant result, which can be related to the dead cone is that,
unlike the naive parton-model expectation, the multiplicity difference
between heavy and light hadrons is roughly independent of the
centre-of-mass energy \cite{Dokshitzer:2005ri}.

More generally, soft radiation obeys the Low--Barnett--Kroll 
(LBK) theorem~\cite{PhysRev110.974,PhysRevLett20.86}, according to which soft gluons are classical and therefore
incapable of modifying the quantum numbers, such as the color, of a
given system. Also, due to its classical nature, soft-parton emission
is independent of the underlying hard-scattering process and of the
quantum properties of the emitter.  This implies that, if a process is
forbidden by a given symmetry, such a veto cannot be relaxed by
emitting extra soft photons/gluons.

This property of soft radiation can be connected to the longstanding
puzzle of $J/\psi$ production, whose measured rate at large $p_T$ at
the Tevatron was about 50 times larger than the theoretical
prediction.  In fact, the color-octet mechanism was formulated in order to 
explain such a discrepancy.
In other words, such a model predicts, at one loop in
gluon-gluon fusion, the production of a $J/\psi$ and a gluon, with the
$J/\psi$ being in a color-octet state.

However, as discussed above, in order to change the $J/\psi$ color,
the emitted gluon cannot be soft: according to the LBK theorem, the
price to pay for the emission of a hard gluon is $\sim
(\Lambda_{\mathrm{QCD}}/m_c)^2$.

Therefore, it is probably worthwhile investigating alternative
mechanisms which, besides the color-evaporation model and the color-octet
model, can increase
the yield of large-$p_T$ $J/\psi$ production in hadron collisions and
hence recover the agreement with the Tevatron data. In fact, the
production of quarkonia at large transverse momentum has a very small
cross section and therefore it is a rare configuration, wherein
fluctuations are expected to play a role \cite{Dokshitzer:QQ10}.
 
\subsection{Heavy quarks and energy loss in dense matter}

A crucial issue in heavy-quark phenomenology is the description of
energy loss in a dense medium and its relation to the quarkonium
spectrum in matter. It is commonly assumed that partons at
large transverse momentum traversing the medium lose energy through 
incoherent radiation of gluons as well as collisional energy loss.
The approach presented in \cite{Gossiaux:2008jv} assumes that a fast parton in a medium
undergoes multiple collisions, characterized by a moderate momentum
transfer and large values of the strong coupling constant.  As for the
coupling, non perturbative effects are taken into account by using a
frozen coupling constant \cite{Dokshitzer:1995ev,Dokshitzer:1995qm}.  Indeed, such a model manages
to reproduce quite well the $R_{AA}$ ratio measured by the PHENIX
collaboration, for \AuAu\ collisions at 200 GeV over the full transverse-momentum range, up to a normalization $K$-factor of about 2.

As for the radiative energy loss, for light quarks one usually relies
on the Baier--Dokshitzer--Mueller--Peign\'e--Schiff (BDMPS)
approximation \cite{Baier:1996sk} which assumes static scattering centres,
independent soft emissions and hadronization outside the medium. For heavy quarks one can rely to the
Gunion--Bertsch approach \cite{PhysRevD25.746}: the medium-induced gluon-radiation spectrum is corrected
by means of terms depending on the heavy-quark mass and on the fictitious gluon
thermal mass. Moreover, gluons radiated by massive quarks
are resolved in a smaller time with respect to light quarks, which
implies a milder dependence on coherence effects and evolution
variables for multiple radiation. Therefore, as a first
approximation, one can even think of neglecting coherence effects when
dealing with emissions from heavy quarks.  The result is that if one
adds both collisional and radiative energy loss contributions, a
scaling factor of 0.6 is enough to reproduce the $R_{AA}$ data at RHIC.
Future data from the LHC will possibly shed light on the issue of the
energy loss; to this goal, it will be very useful having more
exclusive probes, such as azimuthal correlations or tagged $c$- or
$b$-flavored jets. Another open issue concerns the gluon thermal mass
and whether it can be theoretically predicted or it should rather be
fitted to experimental data, like a non perturbative parameter.

\subsection{Heavy quark plus photon production as a probe of the heavy quark PDF}

The production of a direct photon accompanying a heavy quark at hadron
colliders is an important process, also relevant to constrain the
heavy-quark density. As it escapes confinement, a photon can be
exploited to tag the hard scattering; also, its transverse momentum
distribution gives meaningful information on the heavy-quark
production process and PDF.

At leading order, i.e. ${\cal O}(\alpha\alpha_S)$, there is only one
hard scattering process, namely $gQ\to Q\gamma$. At NLO in the strong
coupling constant, i.e.  ${\cal O}(\alpha\alpha_S^2)$, several
subprocesses contribute to $Q\gamma$ production, with gluons, heavy
and light quarks in the initial state \cite{Stavreva:2009vi}. Considering,
e.g., charm production, such processes will help to shed light on the
charm distribution and whether it is radiatively generated by means of
gluon emissions, in which case $c(x,Q^2) \simeq g(x,Q^2)$, or
there is an intrinsic charm contribution to the proton.  In fact,
there exist light cone models, wherein intrinsic charm is relevant at
large $x$ \cite{PhysLettB93.451}, and sea-like models, with $\bar c(x) \simeq \bar
u(x)+\bar d(x)$.  As for the comparison between the photon spectra
produced in association with charm and bottom quarks, there are
remarkable differences both at LO and at NLO. At large transverse
momenta, the discrepancy between LO and NLO gets larger
\cite{Stavreva:2009vi}.

The NLO calculation has been compared with D0 data \cite{Abazov:2008er} on the
photon transverse momentum in $Q\gamma$ events: the agreement is
pretty good in the case of $b\gamma$ production, whereas discrepancies
are present for charm production, with the NLO computation yielding a
lower cross section for moderate and large values of $p_T$.  Such disagreement can be traced back to 
the possible existence of intrinsic
charm in the proton.  Using the Brodsky--Hoyer--Peterson--Sakai (BHPS)
PDF set \cite{PhysLettB93.451} slightly improves the comparison, although
meaningful differences are still present at very large $p_T$. The
discrepancy is instead milder when comparing the ratio of charm and
bottom cross sections. At the LHC, one can still investigate the
presence of the intrinsic charm. However, since the probed
values of $x$ are on average smaller than at the Tevatron, one is
basically sensitive to the sea-like model. Also, the forward-rapidity
region will be particularly suitable to discriminate between radiative
and intrinsic contributions to the charm PDF.

The LO/NLO calculations for $b/c+\gamma$ production can be extended to
$\pA$ collisions, e.g. proton-lead processes at the LHC. A first
approximation consists in using the same amplitudes as in the vacuum
and convoluting them with nuclear parton densities, such as the EPS
\cite{Eskola:2008ca}, HKN \cite{Hirai:2007sx} and nCTEQ \cite{Schienbein:2009kk} sets. Unlike the
proton case, the gluon distribution in a nucleus is poorly known, and
it is one of the main differences among the nuclear PDF sets. In
fact, possible measurements at the LHC of $\gamma+c/b$ final states
will help to investigate both gluon and charm/bottom distributions
in dense matter.

Likewise, similar studies can be carried out for deuterium-gold collisions at RHIC.  
Since the Bjorken-$x$ regions probed at RHIC and LHC are complementary, these measurements will help to discriminate among the nuclear parton distribution sets. 
As for the hard scattering, how to implement medium modifications, for both
light and heavy quark production, is still an open issue; moreover,
collinear factorization is an assumption that has not yet been proved. 
The BDMPS and Gunion--Bertsch approximations discussed above
work well only for soft/collinear emissions. Indeed, there are
prescriptions for hard/large-angle radiation, but a thorough
implementation and phenomenological analysis on medium-modified hard
scatterings is still missing. Convoluting possible medium-induced
hard matrix elements with nuclear parton densities will 
ultimately lead to cross sections fully including dense-matter effects.

\section{Quarkonia as a Tool}\label{WG5}

\subsection{Proton-Proton Collision Studies}

Since the start of LHC operation, new experimental data on quarkonium production
has revived the interest in understanding the production mechanisms of various 
quarkonium states in hadronic collisions. All the models describing various sources
of onia in hadronic collisions share the common 
basis: the factorisation between the hard collison subprocess and the
parton-parton collision luminosity, calculated as a convolution
of the parton distribution functions (PDFs).  

Clearly, the range of accessible $x_{1,2}$ depends on the rapidity
interval covered by the experiments.
For charmonium at the Tevatron, with a partonic center-of-mass energy $\sqrt{\hat s} \simeq 3.5$~GeV, $\sqrt{s}\simeq 2$ TeV, 
the CDF range is :
\begin{eqnarray}
y\simeq 0 & \Rightarrow & { x_{1,2}\simeq 1.8 \times 10^{-3}} \nonumber
\end{eqnarray}
while for D0:
\begin{eqnarray}
-1.6 \lesssim y \lesssim 1.6 &\Rightarrow& { x_{1,2}\simeq (0.36 - 8.9)\times 10^{-3}} \nonumber
\end{eqnarray}
Their values of $x_{1,2}$ lie within the range where the PDFs are
well-established. Hence the uncertainties on the
parton luminosities at the Tevatron are fairly small.

At the LHC, not only is the energy higher by a factor of 3.5, but
the rapidity coverage of the various experiments is also significantly wider, especially
if one combines the results from ATLAS and CMS with those of
LHCb and ALICE. One get for ATLAS and CMS:
\begin{eqnarray}
-2.4\lesssim  y\lesssim 2.4 &\Rightarrow& x_{1}\simeq (0.04 - 6.0) \times 10^{-3}  \nonumber\\
& & x_{2}\simeq (6.0 - 0.04) \times 10^{-3}  \nonumber
\end{eqnarray}
while for LHCb:
\begin{eqnarray}
2 \lesssim y \lesssim 5 &\Rightarrow&  x_{1}\simeq (4 - 80) \cdot 10^{-3}  \nonumber\\
& & x_{2}\simeq (0.07 - 0.003) \times 10^{-3} \nonumber
\end{eqnarray}
ALICE has complementary coverages ($|y|<1$ \& $2.5.<y<4.0$).
For the kinematics of LHC experiments, the accessible $x$ range 
of charmonium production measurements is up to three 
orders of magnitude below the ``safe'' value of 10$^{-3}$. 
Processes usually used for determining gluon PDFs, such as
jets, prompt photons or open charm pairs, are for various reasons unlikely to be useful 
at these extremely low $x$, leaving charmonium as the only source of information
on gluon  PDFs in this area.
  
The  smallest $\sqrt{\hat s}$ value for bottomonium is about 10~GeV,  three times larger than for charmonium. This restricts 
the ``reach'' of $\Upsilon$ studies to $x_{1,2}$ values three times larger than those
contributing to $J/\psi$. However,  this is still significantly smaller than
the currently accessible range, and thus should be useful as a  consistency check and maybe also to test
the $Q^2$ evolution (providing data with a higher $Q^2$ for similar $x_B$, which may be helpful to
 differentiate between a BFKL- and  DGLAP-like evolution).

The main experimental difficulties include a reduced acceptance for $J/\psi$ 
when both $p_T$ and $|y|$ are small, and the necessity of suppressing
non-prompt charmonia (which come from larger $\hat s$).
Theoretical complications include proper and consistent treatment of 
intrinsic parton $p_T$, initial-state radiation, and higher-order
perturbative effects, as well as the possibility of gluon recombination at high
densities. So concerted efforts from theorists and experimentalists will be needed
to extend our measurements of the PDFs to much smaller  
$x$ than prior to the LHC. This is  necessary 
 for understanding the dynamics of quarkonium hadronic production.

\subsection{Nuclear Collision Studies}

From statistical QCD we know that strongly interacting matter undergoes 
a deconfinement transition to a new state, the quark-gluon plasma (QGP). 
The objective of high energy nuclear collisions is to produce and study 
the QGP under controlled conditions in the laboratory. How then 
can we probe the QGP - what phenomena provide us with information about 
its thermodynamic state? Here the ultimate aim must be to carry out 
{\sl ab initio} calculations of the in-medium behaviour of the probe 
in finite temperature QCD. Quarkonia may well provide the best tool for 
this presently known, but in nuclear collisions other phenomena will 
also lead to modified quarkonium production. As a consequence, 
parton distribution changes, parton energy loss and cold nuclear matter 
effects (absorption) on quarkonium production must be accounted for 
before any QGP studies become meaningful. For the sake of definiteness, 
in this section we shall refer to the non-QGP effects as normal, in contrast to the 
anomalous suppression (or enhancement) we want to look for.

Any modifications observed when comparing \J~production in $\AA$ collisions
to that in $pp$ interactions thus have two distinct origins. Of primary
interest is obviously the effect of the secondary medium produced in the
collision - this is the candidate for the QGP we want to study. In 
addition, however, the presence of the ``normal'' initial and final
state effects will presumably also affect the production process and 
the measured rates. This ambiguity in the origin of any observed 
\J~suppression will thus have to be resolved.

A second empirical feature to be noted is that the observed \J~production 
consists of directly produced $1S$ states as well as of decay products 
from \X(1P) and \P(2S) production. The presence of a hot QGP affects the 
higher excited quarkonium states much sooner (at lower temperatures) 
than the ground states. This results in another ambiguity in observed 
\J~production - are only the higher excited states affected, or do all 
states suffer?

An ideal solution of these problems would be to measure separately
\J, \X~and \P~production first in$\pA$ (or $\dA$) collisions, to determine
the effects of cold nuclear matter, and then measure, again separately,
the production of the different states in $\AA$ collisions as function of
centrality at different collision energies.

Until such studies become available, we resort to a more
operational approach, whose basic features are:
\begin{itemize}
\vspace*{-0.2cm}
\item{We assume that the \J~feed-down rates in $\pA$ and $\AA$ are the same 
as in $pp$, i.e., 60 \% direct \J~, 30 \% decay of $\x(1P)$, 
and 10 \% decay of $\p(2S)$.}
\vspace*{-0.2cm}
\item{We specify the effects due to cold nuclear matter by a Glauber 
analysis of $\pA$ or $\dA$ experiments in terms of $\sigma_{\rm{abs}}^i$ for 
$i=$\J, \X, \P. This $\sigma_{\rm{abs}}^i$ is not meant to be a real cross section
for charmonium absorption by nucleons in a nucleus; it is rather used to
parametrize all initial and final state nuclear effects, including
shadowing/antishadowing, parton energy loss and pre-resonance/resonance 
absorption.}
\vspace*{-0.2cm}
\item{In the analysis of $\AA$ collisions, we then use $\sigma_{\rm{abs}}^i$ 
in a Glauber analysis to obtain the predicted form of {\sl normal} 
\J~suppression. This allows us to identify {\sl anomalous} \J~suppression
as the difference between the observed production distribution and that
expected from only normal suppression. We parametrize the anomalous
suppression through the survival probability
\be
S_i = {(dN_i/dy)_{\rm exp} \over (dN_i/dy)_{\rm Glauber}}
\ee
for each quarkonium state $i$.}
\end{itemize}

\begin{figure}[htb]
\centerline{\includegraphics[width=6cm]{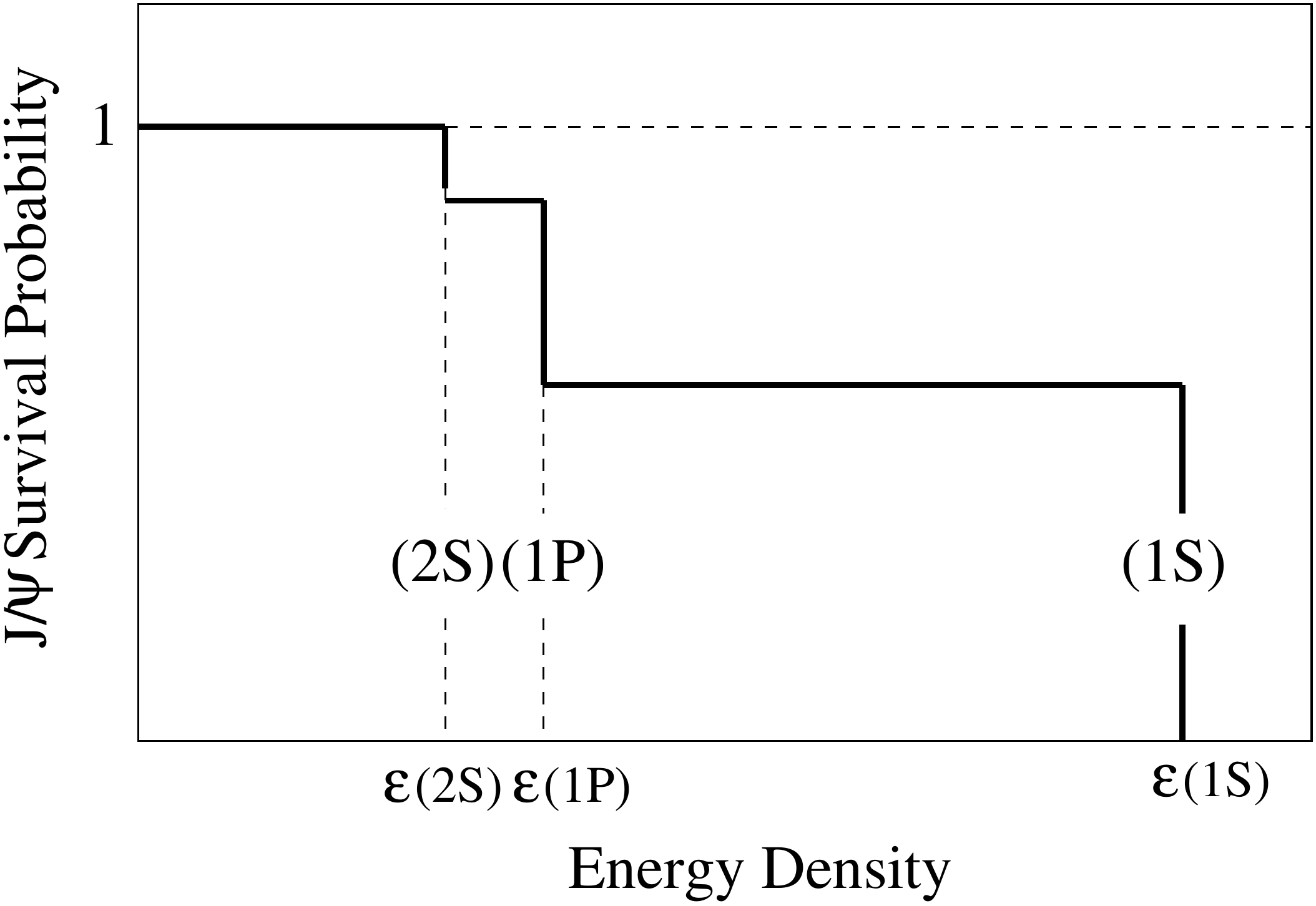}}
\caption{Illustration of sequential \J~suppression.}
\label{sequential}
\end{figure}

With the effects of cold nuclear matter thus accounted for, what form do
we expect for anomalous \J~suppression? If $\AA$ collisions indeed produce
a fully equilibrated QGP, we should observe a sequential suppression pattern 
for \J~and \U, with thresholds predicted (quantitatively in temperature or
energy density) by finite temperature QCD \cite{Karsch:1990wi,Gupta:1992cd,Digal:2001ue,Karsch:2005nk}. The 
resulting pattern for the \J~is illustrated in Fig.\ \ref{sequential}.

Its consequences are quite clear. If, as present statistical QCD studies
indicate, the direct \J~ survives up to about $2~T_c$ and hence to
$\e \geq 25$ GeV/fm$^3$, then all anomalous suppression observed at SPS 
and RHIC must be due to the dissociation of the higher excited states 
\X~and \P. The suppression onset for these is predicted to lie around 
$\e \simeq 1$ GeV/fm$^3$, and once they are gone, only the unaffected  
\J~ production remains. Hence the \J~survival probability for 
central $\AuAu$ collisions at RHIC should be the same as for central 
$\PbPb$ collisions at the SPS.

A further check to verify that the observed \J~production in central
collisions is indeed due to the unmodified survival of the directly
produced $1S$ state is provided by its transverse momentum behaviour.
Initial state parton scattering causes a broadening of the $p_T$
distributions of charmonia \cite{Karsch:1988ri,Gavin:1988tw,Hufner:1988wz}: the gluon from the proton projectile 
in $\pA$ collisions can scatter a number of times in the target nucleus
before fusing with a target gluon to produce a $\C$. Assuming the
gluon from the proton to undergo a random walk through the target leads to
\be 
\langle p_T^2 \rangle_{pA} = \langle p_T^2 \rangle_{pp} + N_c^A \delta_0
\ee
for the average squared transverse momentum of the observed \J. Here
$N_c^A$ specifies the number of collisions of the gluon before the parton 
fusion to $\C$, and $\delta_0$ the kick it receives at each collision.
The collision number $N_c^A$ can be calculated in the Glauber formalism;
here $\sigma_{\rm{abs}}$ has to be included to take the presence of cold nuclear 
matter into account, which through a reduction of \J~production shifts
the effective fusion point further ``down-stream'' \cite{Kharzeev:1997ry}.

In $\AA$ collisions, initial state parton scattering occurs in both
target and projectile and the corresponding random walk form becomes
\be
\langle p_T^2 \rangle_{AA} = \langle p_T^2 \rangle_{pp} + N_c^{AA} \delta_0,
\ee
where now $N_c^{AA}$ denotes the sum of the number of collisions in the 
target and the projectile prior to parton fusion.  It can again be
calculated in the Glauber scheme including $\sigma_{\rm{abs}}$. The crucial
point now is that if the observed \J's in central $\AA$ collisions are
due to undisturbed $1S$ production, then the centrality dependence of 
the $p_T$ broadening is fully predicted by such initial state parton 
scattering \cite{Karsch:2005nk}. In contrast, any onset of anomalous
suppression of the \J~ would lead to a modification of the random
walk form \cite{Kharzeev:1997ry}.

\begin{figure}[htb]
\centerline{\includegraphics[width=6cm]{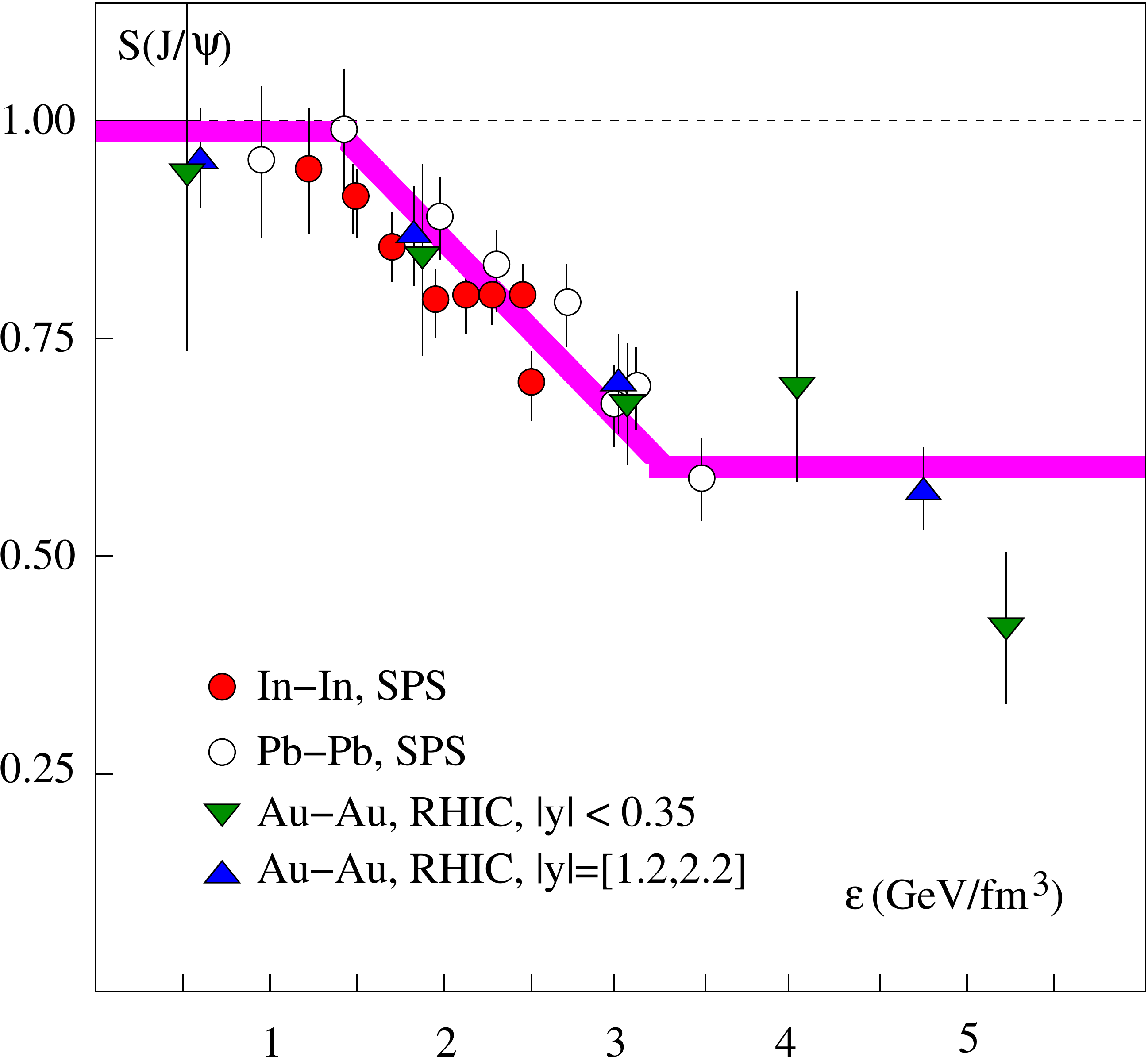}}
\centerline{\includegraphics[width=6cm]{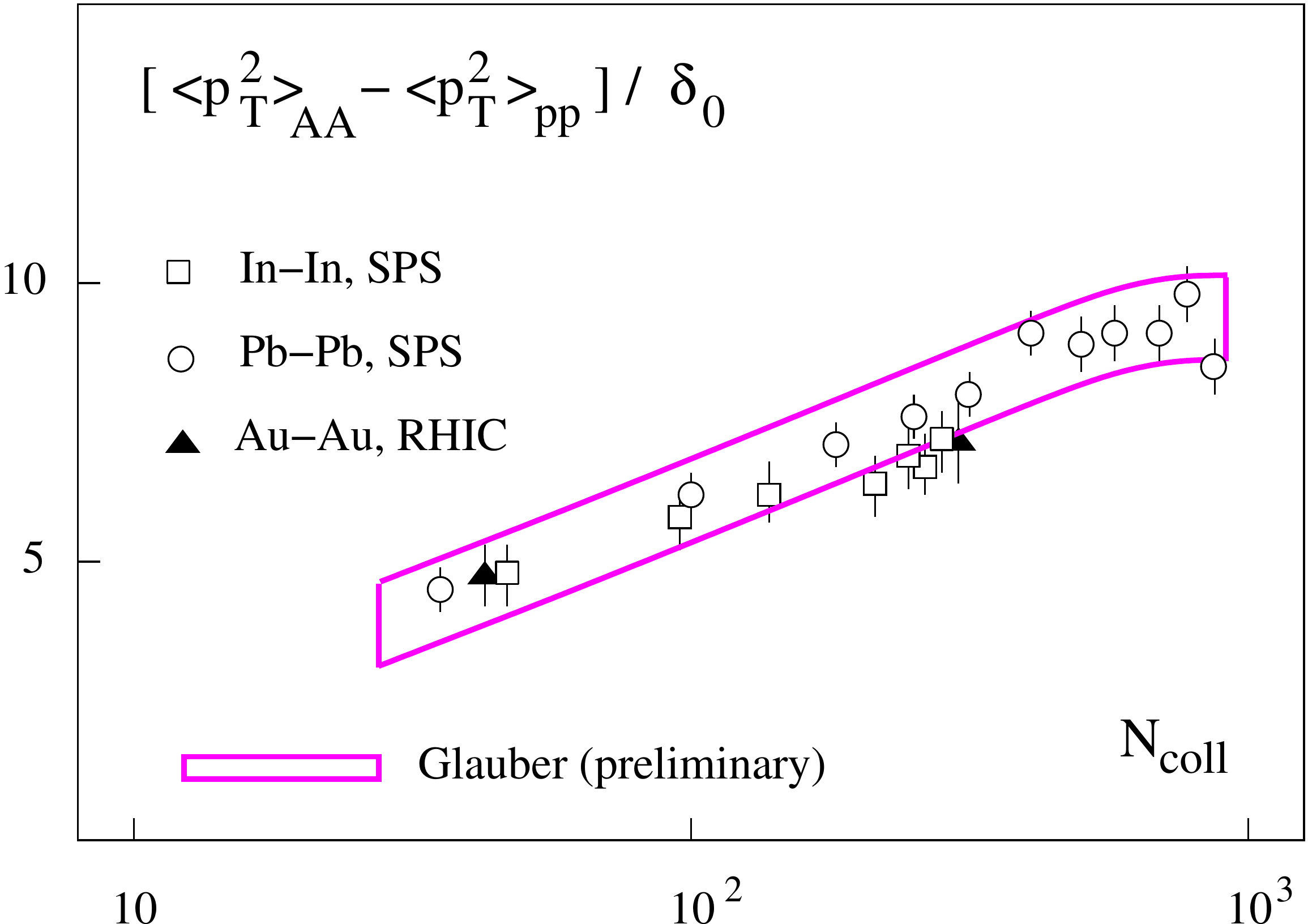}}
\caption{\J~survival and transverse momentum at SPS and RHIC.}
\label{data}
\end{figure}

In Fig.\ \ref{data} we summarize the predictions for \J~survival and 
transverse momentum behaviour in $\AA$ collisions at SPS and RHIC, as 
they emerge from our present state of knowledge of statistical QCD. 
Included are some preliminary and some final data; for a discussion 
of the data analysis and selection, see Ref.\ \cite{Karsch:2005nk}. 

We conclude that the present experimental results are compatible with
the present information from statistical QCD. This was not the case
previously, and such a conclusion can be drawn today because of 
several changes in our theoretical and experimental understanding. 
\begin{itemize}
\vspace*{-0.2cm}
\item{Finite $T$ lattice QCD suggests direct \J~suppression
at energy densities beyond the RHIC range; previous onset values were 
much lower (see, e.g., Ref.\ \cite{Digal:2001iu}).}
\vspace*{-0.2cm}
\item{SPS \InIn\ data suggest an onset of anomalous suppression at 
$\e \simeq$ 1 GeV/fm$^3$; previous onset values from $\PbPb$ and $\SU$
interactions were considerably higher, with $\e \simeq 2 - 2.5$ GeV/fm$^3$
(see, e.g., Ref.\ \cite{Abreu:2002qs}).}
\vspace*{-0.2cm}
\item{Within statistics, there is no further drop of the \J~survival rate 
below 50 - 60 \%, neither at RHIC nor at the SPS; a second drop in very
central SPS $\PbPb$ data (see, e.g., Ref.\ \cite{Abreu:2002qs}) is no longer 
considered viable.}
\vspace*{-0.2cm}
\end{itemize}

\subsubsection{ \J~Enhancement by Regeneration}

In this section we want to consider an alternative approach which can so 
far also account for the available data. The basic idea here is that the 
medium produced in nuclear collisions is not a QGP in full equilibrium 
but rather one which is oversaturated in its charm content.

\begin{figure}[htb]
\centerline{\includegraphics[width=6cm]{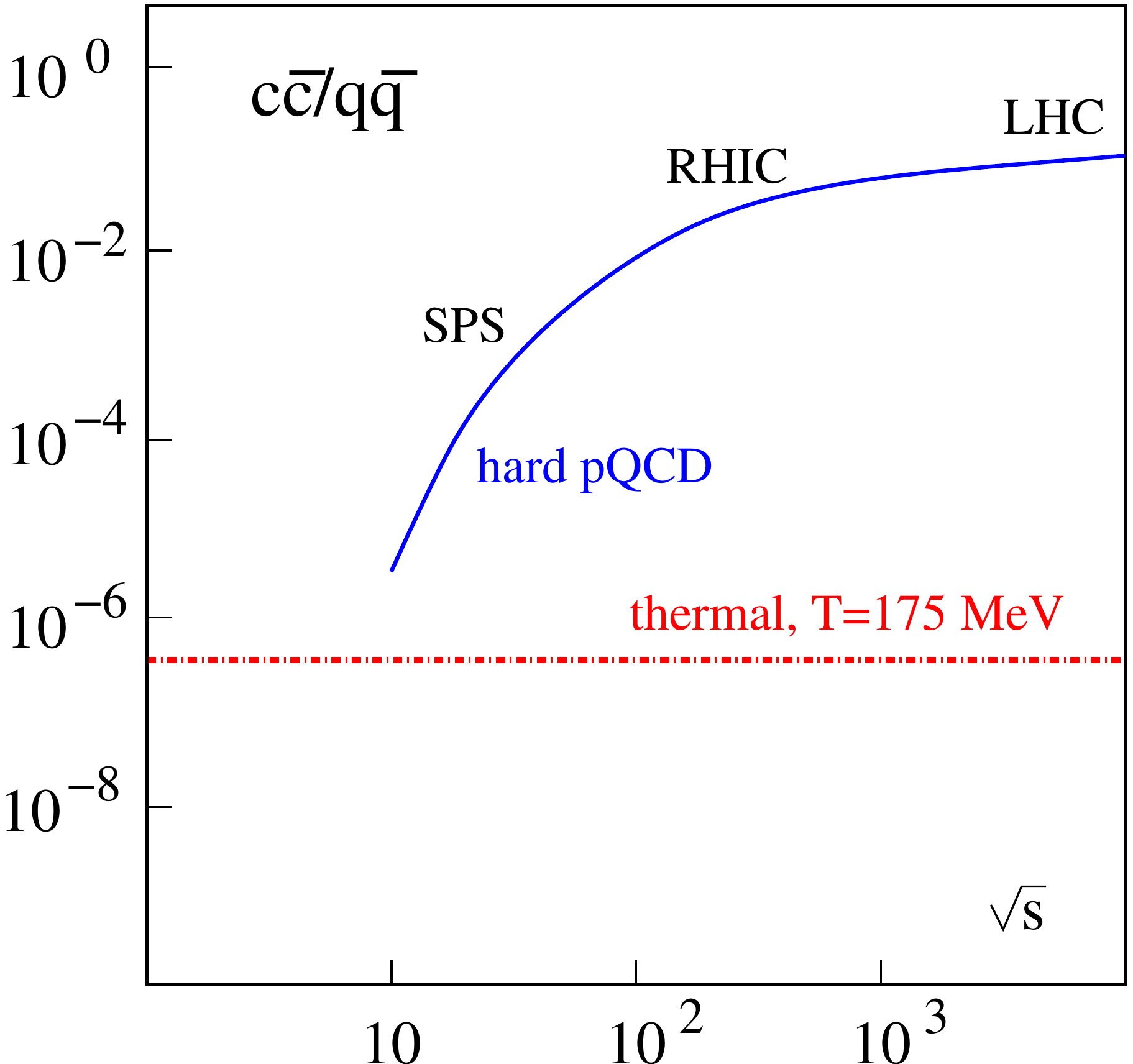}}
\caption{Thermal vs.\ hard charm production.}
\label{initial}
\end{figure}

A crucial aspect in the QGP argumentation of the previous sections was 
that charmonia, once dissociated, cannot be recreated at the hadronization 
stage since the abundance of charm quarks in an equilibrium QGP is far 
too low to allow this. The thermal charm quark production rate, relative 
to that of light quarks, is 
\be
{c\bar c \over q \bar q} \simeq \exp\{-m_c/T_c\} \simeq 6 \times 10^{-4},
\ee
with $m_c = 1.3$ GeV for the charm quark mass and $T_c=0.175$ GeV for the
transition temperature. The initial charm production in high energy
nuclear interactions, however, is a hard non-thermal process, and the 
resulting $c/\bar c$ production rates grow with the number ($N_{\rm{coll}}$) 
of nucleon-nucleon collisions. In contrast, the light quark production
rate grows (at least in the present energy regime) essentially as the
number ($N_{\rm{part}}$) of participant nucleons, i.e., much slower. The
initial charm abundance in $\AA$ collisions is thus much higher than the
thermal value; we illustrate this in Fig.\ \ref{initial} for $A=200$
as function of the collision energy $\sqrt s$. What happens to this 
excess in the course of the collision evolution?

The basic assumption of the regeneration approach \cite{BraunMunzinger:2000ep,Thews:2000rj,Grandchamp:2002wp}
is that the initial charm excess is maintained throughout the subsequent
evolution, i.e., that the initial chemical non equilibrium will persist
up to the hadronization point. If that is the case, a $c$ from a given
nucleon-nucleon collision can, at hadronization, combine with a $\bar c$
from a different collision (``off-diagonal'' pairs) to recreate a \J.
This pairing provides a new secondary statistical charmonium 
production mechanism, in which the $c$ and the $\bar c$ in a charmonium 
state have different parents, in contrast to the primary dynamical production 
in a \pp\ collision. At sufficiently high energy, this mechanism will lead 
to enhanced \J~production in $\AA$ collisions in comparison to the scaled 
\pp\ rates. When should this enhancement set in?

In recent work \cite{BraunMunzinger:2000ep,Thews:2000rj,Grandchamp:2002wp}, it is generally assumed that
the direct \J~production is strongly suppressed for $\e \geq 3$ GeV/fm$^3$.
This is evidently in contrast to the statistical QCD results discussed in
the previous sections; however, we recall the caveat that the temperature 
dependence of the charmonium widths is so far not known. Moreover, it is
of course always possible that the medium produced in nuclear collisions
is quite different from the quark-gluon plasma of statistical QCD.

Next, it is assumed that the regeneration rate is determined by 
statistical combination in a QGP in {\sl kinetic} equilibrium, with or 
without wave function corrections. In other words, if a $c$ and a 
$\bar c$ meet under the right kinematic conditions, they are taken 
to form a \J. An evolution towards a QGP in {\sl chemical} equilibrium 
would also allow annihilation at this point.

\begin{figure}[htb]
\centerline{\includegraphics[width=6cm]{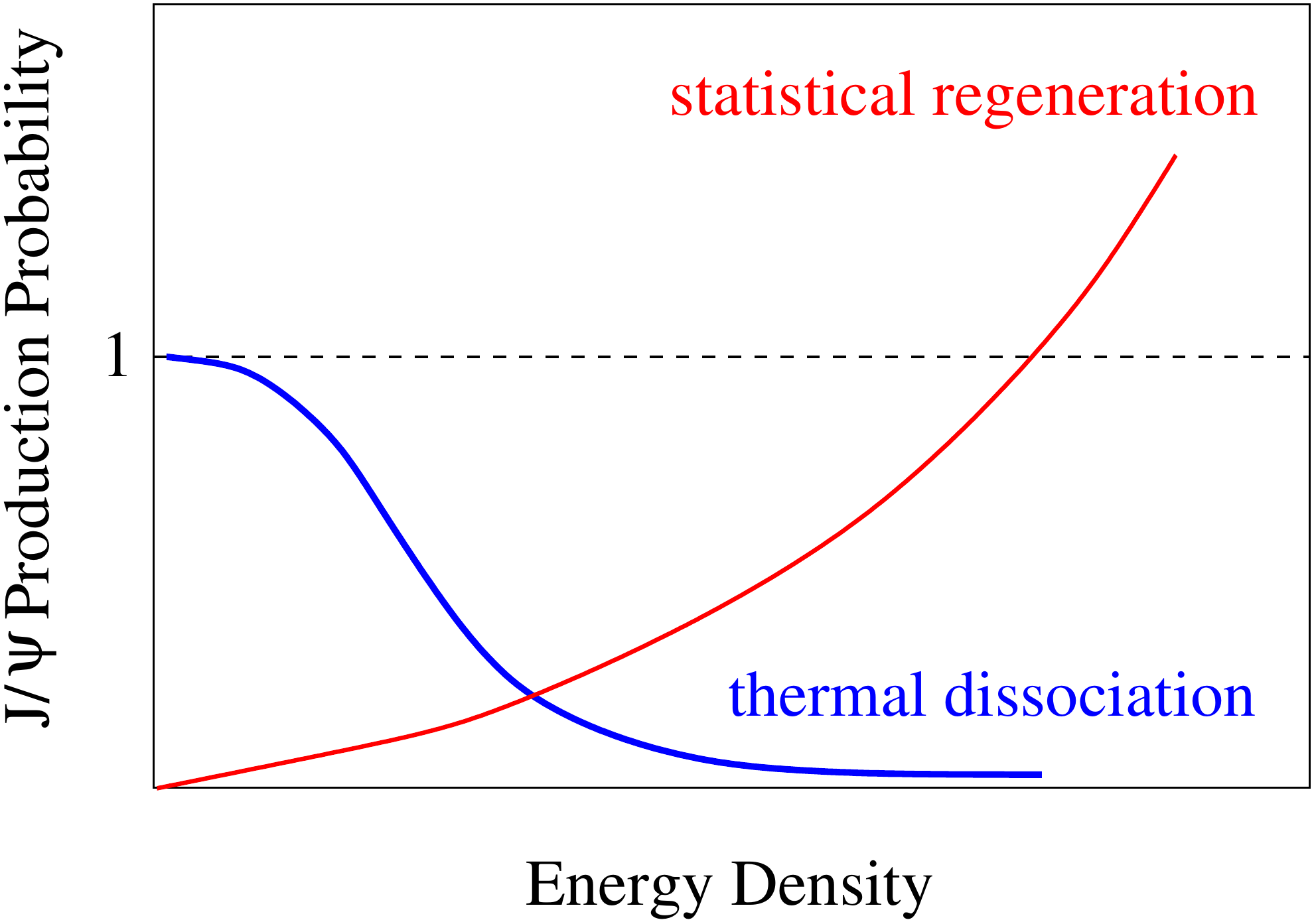}}
\caption{Illustration of regeneration of \J~production.}
\label{recom}
\end{figure}

To account for the \J~production rates observed at RHIC, it is then
assumed that the new statistical production just compensates the proposed
decrease of the direct primary $1S$ production, as illustrated in
Fig.\ \ref{recom}. At the LHC, with much higher energy densities, one 
should then observe a \J~enhancement relative to the rates expected
from scaled $pp$ results.

We thus have to find ways of distinguishing between the two scenarios
discussed here: the sequential suppression predicted by an equilibrium 
QGP or a stronger direct suppression followed by a \J~regeneration in a
medium with excess charm. Fortunately the basic production patterns in
the two cases are very different, so that one may hope for an eventual
resolution.

\begin{figure}[htb]
\centerline{\includegraphics[width=6cm]{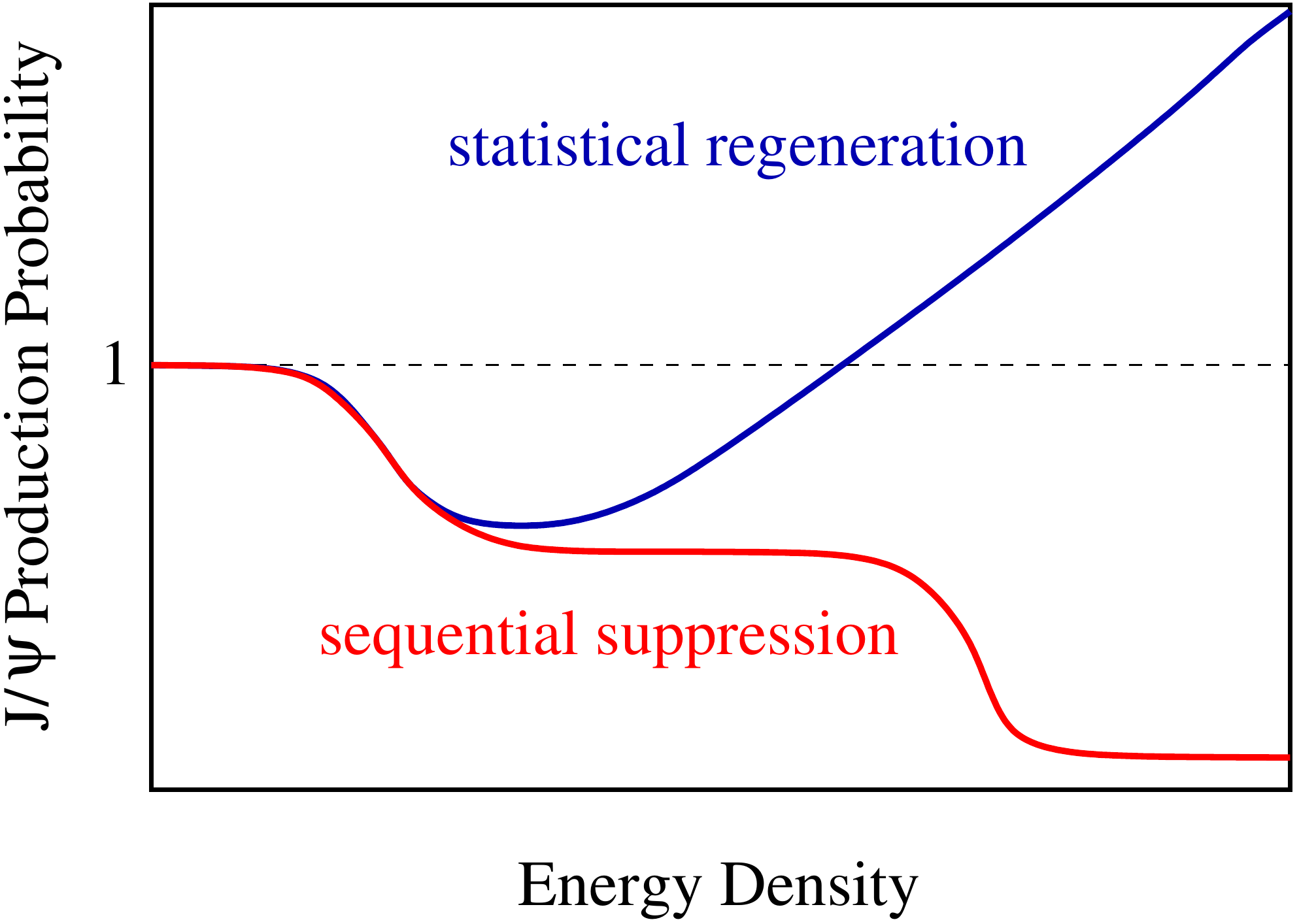}}
\caption{\J~survival: sequential suppression vs.\ regeneration.}
\label{surv}
\end{figure}

The overall \J~survival probability in the two cases is illustrated in
Fig.\ \ref{surv}. Sequential suppression provides a step-wise reduction:
first the higher excited charmonium states are dissociated and thus their 
feed-down contribution disappears; at much higher temperature, the \J~  
itself is suppressed. Both onsets are in principle predicted by lattice QCD
calculations. In the regeneration scenario, the thermal dissociation of 
all ``diagonal'' \J~production is obtained by extrapolating SPS data to
higher energy densities. The main prediction of the approach is therefore
the increase of \J~production with increasing energy density. Ideally, the
predictions for the LHC are opposite extremes, providing of course that 
here the feed-down from $B$-decay is properly accounted for.

\begin{figure}[htb]
\centerline{\includegraphics[width=6cm]{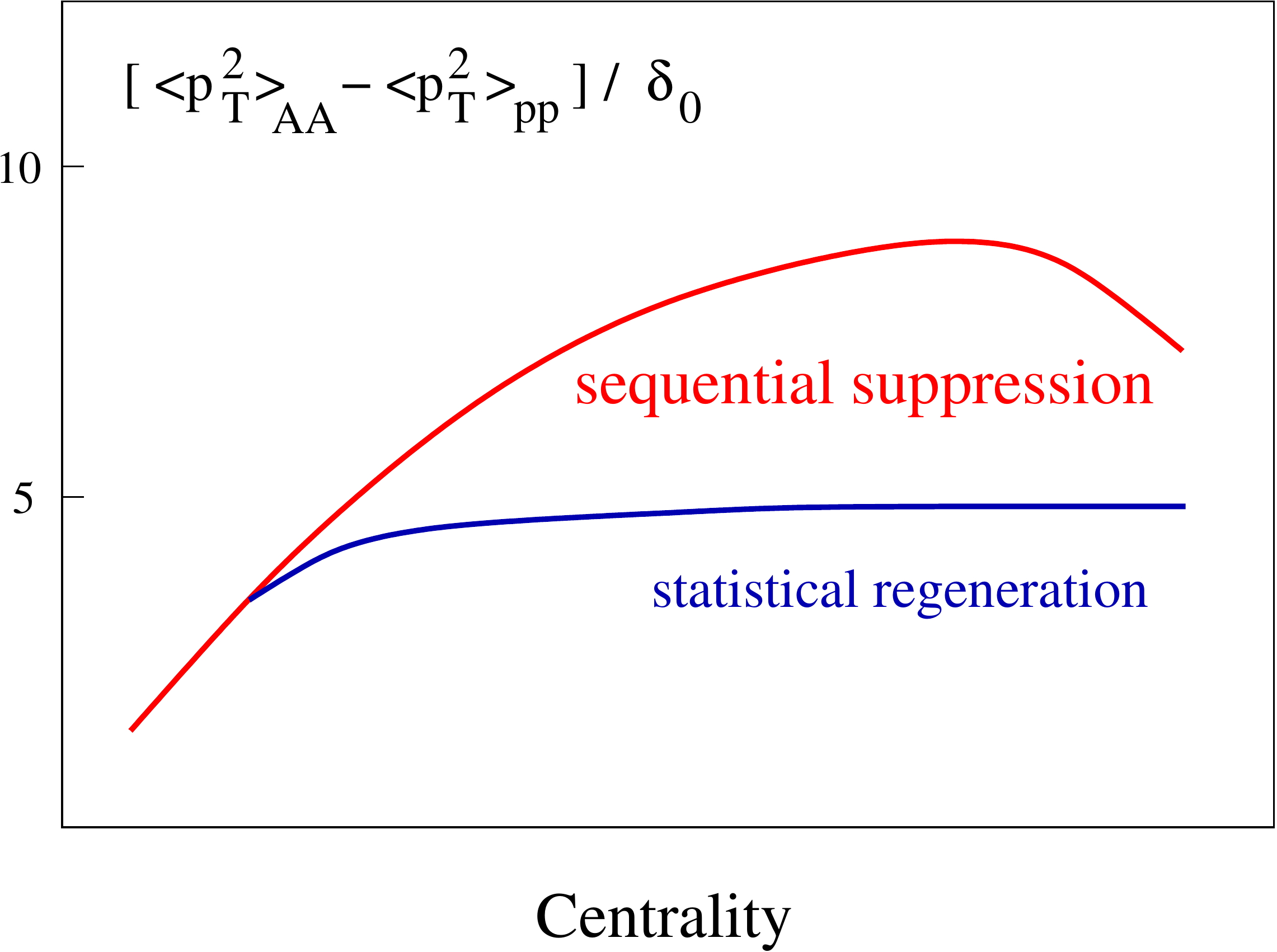}}
\caption{$p_T$-behaviour: sequential suppression vs.\ regeneration.}
\label{trans}
\end{figure}

The expected transverse momentum behaviour in the two cases is also
quite different. In the region of full \J~ survival, sequential
suppression predicts the normal random walk pattern specified through
$\pA$ studies; the eventual dissociation of direct \J~ states then
leads to an anomalous suppression also in the average $p_T^2$ \cite{Kharzeev:1997ry}.   
Regeneration alone basically removes the centrality dependence since
the different partners come from different collisions. It is possible
to introduce some small centrality dependence \cite{Thews:2005vj}, but the random 
walk increase is essentially removed. The resulting behaviour is 
schematically illustrated in Fig.\ \ref{trans}. More generally, the 
quarkonium momentum distributions, whether transverse or longitudinal, 
should in the regeneration scenario be simply a convolution of the 
corresponding open charm distributions; this provides a further 
check \cite{Thews:2005vj}. 

\subsubsection{Conclusions}
~
\begin{itemize}
\vspace*{-0.2cm}
\item{In statistical QCD, the spectral analysis of quarkonia provides  
a well-defined way to determine temperature and energy density of the
QGP.}
\vspace*{-0.2cm}
\item{If nuclear collisions produce a quark-gluon plasma in equilibrium, 
the study of quarkonium production can provide a direct way to connect 
experiment and statistical QCD.}
\vspace*{-0.2cm}
\item{For a QGP with surviving charm excess, off-diagonal quarkonium 
formation by statistical combination may destroy this connection and
instead result in enhanced \J~production.} 
\end{itemize}

\section{New observables in quarkonium production}\label{WG6}

In order to better understand
the mechanisms responsible for heavy quarkonium production, it is now essential to introduce
 new observables. The theoretical uncertainties on the predicted yields
 are too large to draw any strong conclusions on production models. Moreover, polarization 
studies are known to be experimentally challenging
and not always possible over the full phase-space region even if the yield is otherwise measured. The LHC will certainly
provide new information on the yields and polarizations. 
In addition, the LHC is well positioned to make unique new measurements previously unavailable 
due to unsufficient luminosity, center-of-mass energy, detector resolution or even  manpower.

\subsection{Hadronic activity around quarkonium}

The first  observable that we   discuss,  hadronic activity
around  quarkonium~\cite{Kraan:2008hb},  is, in fact,
not completely new. Indeed, UA1 compared their
charged-track distributions with Monte Carlo simulations of a $J/\psi$ coming
from a $B$ decay and a $J/\psi$ coming from a
$\chi_c$ decay~\cite{Albajar:1987ke,Albajar:1990hf}. In the early nineties --before the Tevatron results--  
$\chi_c$ feed-down was still expected to be the major source of prompt $J/\psi$ production.
However, if we assume either that  production proceeds through CO transitions or through CS transitions at
higher-orders, we now expect  more complex distributions,
even for the prompt yield. It is therefore not clear how accurately such methods --  recently revived by STAR
in~\cite{Abelev:2009qaa} --
can determine the $B$-feeddown to $J/\psi$, other than with measurements
of displaced vertices typical of $B$ decays. Instead, the hadronic activity 
around  quarkonium could be a good discriminant between CO and CS contributions
to quarkonium production since gluon emission during CO transitions would increase
the hadronic activity. This
 study is, however,  difficult to carry out in practice~\cite{Kraan:2008hb}.

\subsection{Associated production}

We therefore urgently need more observables easier to predict that
test the many production models
available~\cite{Brambilla:2004wf,Lansberg:2006dh}. We argue  that the
study of associated production channels, $\psi + c \bar c$ and $\Upsilon + b \bar b$, 
 first in $\pp$ collisions, then in $\pA$ and $\AA$, fills both these requirements.
These reactions are potentially very interesting since the LO description already
shows  leading $p_T$ behaviour. We expect that higher-QCD corrections will be less
important and the cross section will thus be less sensitive to the renormalisation scale,
allowing  for more precise predictions.

Further motivation for such studies is  the amazingly large fraction of $J/\psi$ produced
in association with  another $c \bar c$ pair at $B$-factories.
For instance, the Belle collaboration first found \cite{Abe:2002rb} 
\eqs{\frac{\sigma\,(e^+ e^-
\to J/\psi +c \bar c)}{\sigma\,(e^+ e^- \to J/\psi +X)}=0.59^{-0.13}_{+0.15}\pm 0.12.} 
More recently, the Belle collaboration reported new
measurements~\cite{Belle:2009nj}:
\eqs{
\sigma\, (J/\psi+ X) &= 1.17 \pm 0.02 \pm 0.07 \hbox{ pb ,} \\
\sigma\, (J/\psi+ c \bar c)&= 0.74 \pm 0.08^{+0.09}_{-0.08}  \hbox{ pb ,} \\
\sigma\, (J/\psi+ X_{{\rm non} \ c \bar c})&=0.43\pm 0.09\pm 0.09)  \hbox{ pb} ,}
which  confirm that associated $J/\psi+ c \bar c$ production is the 
dominant production mechanism at $B$ factories.

Until now, it is  unknown whether such a high associated fraction also holds for hadroproduction. 
Analyses at the Tevatron (CDF and $D0$) and at RHIC (PHENIX and STAR) are feasible, in addition to the LHC.
The predicted Tevatron Run II integrated cross sections at $\sqrt{s}=1.96$ TeV 
are significant~\cite{Artoisenet:2007xi} :

\eqs{\sigma(J/\psi +c \bar c) \times Br (\ell^+\ell^-)& \simeq 1~
\hbox{nb},\\
\sigma(\Upsilon +b \bar b)\times Br (\ell^+\ell^-)& \simeq  1~
\hbox{pb}.}

Neglecting the likely reduction of the CO matrix elements needed to comply
with the $e^+e^-$ results and induced by higher-order QCD corrections to the color-singlet contributions
(see section \ref{WG1}),
the integrated cross sections for associated $J/\psi+ c \bar c$ production were found~\cite{Artoisenet:2008tc} to be
dominated by the CS part. This is also the case for the $p_T$ distribution 
up to at least 5~GeV for $\psi$ and 10~GeV for $\Upsilon$.
This clearly means that such observables directly probe the CS
mechanism. For most of the other observables, it is not as clear whether the CO contribution dominante over
the CS one and they cannot be regarderd as a clean probe of CS constributions due
to the inherent uncertainty from the CO yield.

If the effect of CO transitions is indeed negligible for  {\it inclusive}  $\Upsilon$ production,  
$\Upsilon$'s produced in association with a $b \bar b$ pair would  be unpolarised at the LHC,
independent of $p_T$.

Besides discriminating between the CO and the CS
transitions, the $\psi$  yield in association with a $c\bar c$ should show
an {\it a priori} completely different sensitivity to the $\chi_c$ feed-down
than the inclusive yield. The same holds true for $\Upsilon+b\bar b$ relative to
 $\chi_b$ feed-down. The CS  $P$-state yield is
expected to be smaller than the $S$-state one since they are suppressed
by powers of the relative velocity $v$. Contrary to the inclusive case, 
no additional gluon is needed to attach
to the heavy quark loop to produce the $\psi$
(or $\Upsilon$) compared to $P$-states.

For the CO transitions, associated production of $\chi_c+c\bar c$ can occur via
 $gg\to gg$  where both final-state gluons split into
$c \bar c$ pairs. One of them then hadronizes into a  $\chi_c$ via a CO transitions.
This contribution is certainly suppressed up to $p_T \simeq 20$ GeV. At larger
$p_T$, a dedicated calculation is needed. However, this mechanism could be 
easily disentangled from the CS contributions since both $c$ quarks are
necessarily emitted back to back from the  $\chi_c$ and thus to the $J/\psi$.

The non-prompt signal  would originate as usual from $gg\to b\bar b$, where 
one $b$ quark decays into a $\psi+X$. 
Usually, hadronisation of the $b$ produces the $\psi$ with light quarks only. This means 
that we have a single $c$ quark in the event, produced from the decay 
of the recoiling $b$ quark and therefore back to back from the $\psi$. The 
non-prompt signal could then be simply suppressed down by searching for a $D$ meson 
near the $\psi$. It is possible  that  hadronisation of the $b$ produces 
both the $\psi$ and a $D$ meson. In this case, kinematic cuts would not help 
suppress the non-prompt yield. Fortunately, this final state is {\it a priori} suppressed 
compared to the first case and even further suppressed relative to the direct yield. Indeed, 
there is no gain in the $p_T$ dependence since both the  $gg\to b\bar b$ and 
$gg\to \psi +c \bar c$ cross sections scale like $p_T^{-4}$. A cross check 
sizing up the non-prompt yield with a displaced vertex measurement would be 
instructive.

Associated production has also been studied in 
direct $\gamma\gamma$ collisions in ultraperipheral collisions 
(UPC)~\cite{Klasen:2008mh}. At least for direct 
$\gamma\gamma$ collisions, associated production is the dominant 
contribution to the inclusive rate for $p_T \geq 2$~GeV/$c$. 

To conclude,  studies can be carried on by
detecting either the ``near'' or ``away'' heavy-quark with respect to
the quarkonia. There are, of course, different ways to detect the $D$,
$B$, or $b$-jet, ranging from a displaced vertices
to  detection of  semileptonic decays to $e$ or $\mu$. 

As mentionned in  section~\ref{WG3}, associated production of $J/\psi$ or $\Upsilon$
with a prompt photon may also be investigated at the LHC.  Interestingly, contrary to the inclusive case, 
the  $C=+1$ CO should impact the large $p_T$ region, while  $C=-1$ CO the small and mid $p_T$ region. The NLO CS contribution 
was recently studied~\cite{Li:2008ym} and, contrary to the inclusive case, it was shown that for this 
process the CO and CS induced yield should be similar, even with the largest
possible CO matrix elements. At NNLO, the CS yield is expected to be further enhanced~\cite{Lansberg:2009db}
and to dominate over the CO yield. The measurements of $J/\psi$ and $\Upsilon$ production associated with a
direct photon at hadron colliders could thus be an important test of the theoretical treatment of 
heavy-quarkonium production. 

The background to these processes should also be taken into account.
The Quarkonium event generator Madonia~\cite{Artoisenet:2007qm}, imbedded in Madgraph~\cite{Alwall:2007st},  will surely
be of a great help in achieving this.
In any case, we hope that such measurements would provide clear
information on the mechanisms at work in quarkonium production.

\section{Conclusion} 

Quarkonium production measurements in $pp$ collisions have motivated many theoretical investigations. 
However, despite advances, there is still no clear picture of the quarkonium
hadroproduction mechanism. Such a mechanism should be able to explain both the
cross section and the polarization measurements at the Tevatron and RHIC, as well as
comparable measurements at $B$ factories and lepton-hadron colliders. 
Presently, we have  several different approaches at our disposal, such as
the Color Singlet Model (CSM), in which the $Q \bar{Q}$ is produced color neutral at short distances, 
and Non-Relativistic QCD (NRQCD), where quarkonium production can 
also proceed via creation of color-octet $Q \bar{Q}$ pairs, via the Color Octet Mechanism (COM).
These approaches have both advantages and disadvantages.
 
NRQCD has successfully explained 
the excess in charmonium production reported by the CDF
Collaboration, orders of magnitude larger than the LO CSM. 
However so far it fails to provide a consistent description of production and polarization in $\pp$ collisions
and the rates at $B$ factories.

Higher order corrections in $\alpha_s$  have only recently been
calculated. The CS corrections at high $p_T$ are very large given that the 
corrections open new production channels which are relevant at high $p_T$.
On the contrary, CO calculations show that the cross
sections do not increase much when these corrections are taken into account 
since NLO CO contributions do not open any new channels. 
The contributions of the NLO CS corrections reduce the discrepancy
between the CSM cross sections and the CDF data, but still fall
too steeply at high $p_T$ to describe this region successfully.  The NNLO$^\star$ contribution
may be able to fill the gap between calculations and high $p_T$ data.
At RHIC, the high $p_T$ STAR data seems to favor the COM over the CSM, while at low
$p_T$ the PHENIX data are well reproduced by both approaches. It is, however,
worth noting that little is known about feed down contributions at RHIC, while
there does not exist any NLO theoretical predictions for the $\chi_c$ yield, in either approach.

In addition, the LO NRQCD calculation predicts a sizable transverse polarization rate for large $p_T$
${J/\psi}$ whereas the Tevatron CDF measurement at Fermilab
displays a slight longitudinal polarization at large $p_T$ for the prompt yield. This problem is still 
present in calculations at NLO, since
these corrections only slightly change the $p_T$ dependence 
 of ${J/\psi}$
production rate and polarization with respect to the LO calculations. 
However, the CSM results 
drastically change from transverse-polarization dominance at LO
to longitudinal-polarization dominance at NLO, nearly in agreement with the data on $J/\psi$ polarization 
produced in $pp$ collisions both at the Tevatron and RHIC, if the polarization of the $J/\psi$ from $\chi_c$ feed down is assumed.

In order to solve this puzzle, improved measurements are needed.
So far, most experiments have presented results based on a fraction of
the physical information derivable from the data: only one
polarization frame is used and only the polar projection of the decay
angular distribution is studied.  These incomplete results prevent
model-independent physical conclusions. In the forthcoming LHC measurements, 
it is important to approach the
polarization measurement as a multidimensional problem, determining
the full angular distribution in more than one frame.

The feed downs from
more massive states, such as higher-mass quarkonium states, cannot yet be  taken into account in most of the calculations.
While there are CDF measurements of direct production, the
$\chi_c$ feed down can significantly influence  the polarization of the prompt
$J/\psi$ yield. 
A better way to avoid the feed-down problem is to study  $\psi^\prime$ and $\Upsilon(3S)$ where 
no important feed down from higher excited states exist. 


The behavior of quarkonium production in $\pA$ collisions due to  Cold Nuclear Matter (CNM) effects is not yet understood.
Puzzling features in proton-nucleus data put forward new aspects of charmonium physics in nuclear reactions,
namely the role of cold nuclear matter effects (CNM). Two 
CNM effects are considered to be of particular importance: 
  the modification of the initial parton distributions (PDFs)
due to the nuclear environment, an initial-state effect known as shadowing; and the breakup of 
$c\overline c$ pairs after multiple scatterings with the remnants of target nucleus, referred to as the nuclear absorption.

Parton shadowing in the target nucleus may suppress (or enhance, in case of antishadowing) 
the probability of producing a \jpsi. Its effect depends on the kinematics of the $c\overline c$ pair production. 
The theoretical expectations for gluon shadowing are quite diverse. 
A weak shadowing effect was predicted by the dipole approach and in some analysis of DIS data such as 
the one developed by deFlorian and Sassot at leading and
next-to-leading order -- the nDS and nDSg parameterizations --. However, 
strong gluon shadowing and antishadowing was obtained in 
the EKS98, EPS08, and EPS09
parameterizations by Eskola and collaborators.
We note that some $\pA$ data were included in the EPS08 and EPS09 analyses.
The nDSg and EKS98 parameterizations are compatible for $x<10^{-3}$.


Secondly, there is the effect of the nuclear absorption. The strength of the interaction of the evolving $c\overline c$ pair 
with the target nucleons can break up the pair and consequently
suppress the \jpsi\ yield. The intensity of this effect may also depend on the quantum states 
of the $c\overline c$ pair at the production level 
(color octet or color singlet), and on the kinematic variables of the pair. Indeed, a colorless 
$c\overline c$ pair created in a hard reaction is not yet  charmonium since it does not have a fixed mass. 
It takes time to disentangle ground state charmonium from its excitations, $t_f \sim 1/(m_{\psi'}^2-m_{J/\psi}^2)$.
This time scale is interpreted as the formation time of the charmonium wave function.
 At low energies, this time is shorter than the mean nucleon spacing in a nucleus so that the formation process can be treated
as instantaneous. 
At much higher energies, the formation time is long and the initial size of the $c\overline c$ 
dipole is "frozen" by Lorentz time dilation while propagating through the nucleus. 
%

According to the above arguments,
at high energy, the heavy state in the projectile should 
undergo  coherent scattering off the nucleons of the target nucleus, 
in contrast to the incoherent, longitudinally ordered scattering that takes place at low energies.
This should lead to a decrease of \sigabs\ with increasing energy $\sqrt{s_{NN}}$. Compilation and systematic study of many experimental data indicate that \sigabs\ appears either constant or decreasing with energy, following the most recent theoretical expectations.


The  energy regime recently opened up by the advent of the LHC offers new possibilities. 
Clearly $\pA$ data at the LHC are essential to investigate various physics effects, in particular 
nuclear shadowing in a still unexplored $x$-range. The final-state interaction
of quarkonia with CNM is another attractive topic. Due to the expected very short overlap 
time of the heavy-quark pair with nuclear matter, one could expect its influence on the still 
almost point-like $c\overline c$ to be rather limited. However, these considerations are still
very qualitative and deeper theoretical studies are clearly necessary.



The study of open heavy flavour production in $\pp$ and $\pA$ collisions is tightly connected to the 
process of quarkonia (hidden flavor) production. It offers a good basis to establish fundamental 
parameters for quarkonium calculations, together with an outstanding framework to test fundamental concepts,
such as perturbative QCD, factorization and non-perturbative power
corrections.
Because heavy quarks are massive, the production cross
section can be calculated  in perturbative QCD down to
$p_T = 0$ at the parton level. 

However, differential distributions and event shapes exhibit large logarithmic 
contributions, typically corresponding to soft or collinear parton 
radiation, which need to be resummed to all orders. 
Furthermore, dead-cone effects suppress 
gluon radiation around the heavy-quark direction, which has relevant 
phenomenological implications on both parton- and hadron-level 
spectra. As far as heavy-hadron production is concerned, one can 
model non perturbative effects by means of 
hadronization models such as the Lund-based string models, 
which are implemented within the framework 
of Monte Carlo generators HERWIG, PYTHIA, MC@NLO, etc.
Multi-purpose generators have been available for several years for lepton/hadron 
collisions in the vacuum and some have been lately modified to include the 
effects of dense matter.

In order to promote the formalism used to describe open heavy flavors 
to $p$A and ultimately $\AA$ collisions, one has to introduce 
medium-modification effects. 
In particular, one of the most striking observations of heavy-ion 
collisions is the jet-quenching phenomenon, namely the suppression of 
hadron multiplicity at large transverse momentum with respect 
to $pp$ processes. In the case of heavy hadrons, the measured suppression is expected to 
be lower than that of light-flavor mesons, due to the 
dead-cone effect. 
So far, suppression of heavy-flavored mesons in $\pA$ or $\AA$ collisions has only been 
measured through leptons (non-photonic electrons and single muons), i.e., the leptons from semi leptonic 
decays of heavy-flavored hadrons. Surprisingly, the energy loss observed through 
heavy-flavor leptons is substantial already at moderate $p_T$ of $\sim$3 GeV/$c$. 
Several mechanisms to describe 
heavy-flavor energy loss in the dense medium have been 
proposed. Most prominent are collisional and radiative energy loss, 
typically calculated in the weak-coupling regime. 
However, the magnitude of the 
observed suppression at RHIC is hard to accommodate in these models. 
Recent attempts, assuming strong coupling between the heavy-flavor 
mesons and the dense medium and based on  AdS/CFT correspondence 
lead to higher suppression factors. Further studies as well as experimental data are 
needed before any conclusions can be drawn. 
Future data from the LHC will possibly shed light on the issue of the 
energy loss. It will be very useful to have more 
exclusive probes such as azimuthal correlations or tagged $c$- or 
$b$-flavored jets. 

Another crucial issue in heavy-flavor phenomenology concerns
heavy-quark parton distributions in protons and nuclei.  In fact,
processes with the production of heavy quarks are fundamental
to constrain the PDFs.  
In particular, the production of a direct photon accompanying a heavy quark at hadron
colliders is an important process, also relevant to constrain the
heavy-quark density. As it escapes confinement, a photon can be
exploited to tag the hard scattering; also, its transverse momentum
distribution gives meaningful information on the heavy-quark
production process and PDF.

The LO/NLO calculations for $\gamma+c/b$ production can be extended to
$\pA$ collisions, e.g. proton-lead processes at the LHC. 
To a first approximation, the process  can be calculated  by convoluting
the vacuum matrix elements with nuclear PDFs. While the gluon distribution
in the proton is rather well constrained, it is poorly known in the nucleus, 
hence large discrepancy between gluon shadowing parametrizations.
In fact, possible measurements at the LHC of $\gamma+c/b$ final states 
will help to investigate both gluon and charm/bottom distributions 
in dense matter. Since the Bjorken-$x$ regions probed at RHIC and LHC are complementary, these measurements will help to discriminate among the nuclear parton distribution sets.

Lattice QCD calculations predict that, at sufficiently large energy
densities, hadronic matter undergoes a phase transition to a plasma of
deconfined quarks and gluons (QGP). 
Considerable efforts have been invested
 in the study of high-energy heavy-ion collisions to reveal the
existence of this phase transition and to study the properties of strongly
interacting matter in the new phase, in view of \eg~improving our understanding of
confinement, a crucial feature of QCD. The study of quarkonium production and
suppression is among the most interesting investigations in this field.
Calculations indicate that the QCD with binding potential is screened in the QGP
phase, the screening level increasing with  energy density.
Given the existence of several quarkonium states of different binding
energies, it is expected that they will be consecutively dissolved (into open
charm or bottom mesons) above certain energy density thresholds. 

One of the objectives of high energy nuclear collisions is to produce and study
the QGP under controlled conditions in the laboratory. How then
can we probe the QGP - what phenomena provide us with information about
its thermodynamic state? Here the ultimate aim must be to carry out 
{\sl ab initio} calculations of the in-medium behaviour of the probe 
in finite temperature QCD. Quarkonia may well provide the best tool presently known for
this. However, quarkonium production will also be modified in nuclear collisions by other phenomena. 
Parton distribution modifications, parton energy loss and cold nuclear matter
effects on quarkonium production must be accounted for
before any QGP studies become meaningful.

Any modifications observed when comparing \J~production in $\AA$ collisions
to that in $pp$ interactions thus have two distinct origins. Of primary
interest is obviously the effect of the medium produced in the
collision - this is the candidate for the QGP we want to study. In 
addition, however, the presence of the ``normal'' initial and final
state effects will presumably also affect the production process and 
the measured rates. This ambiguity in the origin of any observed 
\J~suppression will thus have to be resolved.

A second empirical feature to be noted is that the observed \J~production 
consists of directly produced $1S$ states as well as of decay products 
from \X(1P) and \P(2S) production. The presence of a hot QGP affects the 
higher excited quarkonium states much sooner (at lower temperatures) 
than the ground states. This results in another ambiguity in observed 
\J~production - are only the higher excited states affected, or do all 
states suffer?

Ideally, one would measure  
\J, \X~and \P~production separately first in $pA$ (or $dA$) collisions, to determine
the effects of cold nuclear matter, and then, again separately,
in $\AA$ collisions as a function of
collision centrality at different center-of-mass energies.

To deepen our understanding of heavy-quarkonium production mechanisms in $\pp$ and $\pA$,
 we must now begin to study  new  observables such as associated production. 
As we have discussed above, theoretical uncertainties on the yields
 are too large to draw any strong conclusions in favour or disfavour of one model
or the other. In addition, polarization studies are in practice challenging 
and not always possible in the complete phase-space region where one can measure the yield.
Much is expected from the LHC on the latter two observables. Nonetheless, one should also use the LHC 
for experimental investigations of new measurements, not possible previously.
These observables can test the many production models available, 
in particular the CSM, the COM, the CEM and the $k_T$ factorization. 
The study of associated production,  $\psi + c \bar c$, $\Upsilon + b \bar b$, $\psi + \gamma$, \dots,  
first in $\pp$ collisions, then in $\pA$ and $\AA$, should be very fruitful.

{\small
\section*{Acknowledgments}
The authors appreciate and acknowledge support for this workshop
and for the work on this document provided, in part or in whole, by
\begin{itemize}
\item the French IN2P3/CNRS;
\item the ReteQuarkonii Networking of the EU
I3 Hadron Physics 2 program;
\item Xunta de Galicia under contract (2008/012);
\item the Spanish Ministerio de Ciencia under contract FPA2008-03961-E;
\item the U.S. Department of Energy under the
contracts DE-AC52-07NA27344 (R. V.), DE-AC02-07CH11359 (V. P.);
\item the U.S. National Science Foundation under grant NSF PHY-0555660 (R. V.);  
\item the German Research Foundation (DFG) under grant PI182/3-1 (B. K.);
\item the Fondecyt (Chile) under grant 1090291;
\item the Conicyt-DFG under grant No. 084-2009;
\item the Georgian National Science Foundation grant GNSF/ST08/4-421.
\end{itemize}
}


\begin{thebibliography}{99}




\bibitem{Abe:1997jz} F.~Abe {\it et al.}  [CDF], Phys.\ Rev.\ Lett.\  {\bf 79}  (1997) 572.

\bibitem{Abe:1997yz} F.~Abe {\it et al.}  [CDF], Phys.\ Rev.\ Lett.\  {\bf 79}  (1997) 578.

\bibitem{CSM_hadron} C-H. Chang, {Nucl. Phys. } B {\bf 172} (1980) 425; R. Baier, R. R\"uckl, {Phys. Lett. } B {\bf 102} (1981) 364; E.~L.~Berger, D.~L.~Jones,   Phys.\ Rev.\  D {\bf 23} (1981) 1521.

\bibitem{Bodwin:1994jh} G.~T.~Bodwin, E.~Braaten, G.~P.~Lepage, Phys.\ Rev.\  D {\bf 51}  (1995) 1125 [Erratum-ibid.\  D {\bf 55} (1997) 5853].

\bibitem{Abachi:1996jq} S.~Abachi {\it et al.}  [D0], Phys.\ Lett.\  B {\bf 370}  (1996) 239.

\bibitem{Affolder:2000nn} A.~A.~Affolder {\it et al.}  [CDF], Phys.\ Rev.\ Lett.\  {\bf 85} (2000) 2886.

\bibitem{Acosta:2004yw} D.~Acosta {\it et al.}  [CDF], Phys.\ Rev.\  D {\bf 71}  (2005) 032001.

\bibitem{Abulencia:2007us} A.~Abulencia {\it et al.}  [CDF], Phys.\ Rev.\ Lett.\  {\bf 99}  (2007) 132001.

\bibitem{Aaltonen:2009dm} T.~Aaltonen {\it et al.}  [CDF], Phys.\ Rev.\  D {\bf 80}  (2009) 031103.

\bibitem{Adare:2006kf} A.~Adare {\it et al.}  [PHENIX], Phys.\ Rev.\ Lett.\  {\bf 98}  (2007) 232002.

\bibitem{Adler:2003qs} S.~S.~Adler {\it et al.}  [PHENIX], Phys.\ Rev.\ Lett.\  {\bf 92}  (2004) 051802.

\bibitem{Atomssa:2008dn} E.~T.~Atomssa  [PHENIX], Eur.\ Phys.\ J.\  C {\bf 61} (2009) 683.

\bibitem{Abelev:2009qaa} B.~I.~Abelev {\it et al.}  [STAR], Phys.\ Rev.\  C {\bf 80} (2009) 041902.

\bibitem{daSilva:2009yy} C.~L.~da Silva [PHENIX], Nucl. Phys. A {\bf 830} (2009) 227c.

\bibitem{Adare:2009js}  A.~Adare  [PHENIX], arXiv:0912.2082 [hep-ex].

\bibitem{Butenschoen:QQ10}  M.~Butenschoen, these proceedings [arXiv:1011.3670 [hep-ph]].

\bibitem{Wang:QQ10} B. Gong, Z.G. He, R. Li, J.X. Wang, these proceedings.

\bibitem{Artoisenet:2008fc} P.~Artoisenet, J.~M.~Campbell, J.~P.~Lansberg, F.~Maltoni, F.~Tramontano, Phys.\ Rev.\ Lett.\  {\bf 101} (2008) 152001 .  

\bibitem{Affolder:1999wm} T.~Affolder {\it et al.}  [CDF], Phys.\ Rev.\ Lett.\  {\bf 84} (2000) 2094.

\bibitem{Acosta:2001gv} D.~Acosta {\it et al.}  [CDF], Phys.\ Rev.\ Lett.\  {\bf 88} (2002) 161802.

\bibitem{Abazov:2005yc} V.~M.~Abazov {\it et al.}  [D0], Phys.\ Rev.\ Lett.\  {\bf 94}, 232001 (2005), 
[Erratum-ibid.\  {\bf 100}, 049902 (2008)].

\bibitem{Abazov:2008za} V.~M. Abazov et~al. [D0] Phys.\ Rev.\ Lett. {\bf 101} (2008) 182004.

\bibitem{Artoisenet:2007xi} P.~Artoisenet, J.~P.~Lansberg, F.~Maltoni, Phys.\ Lett.\ B {\bf 653} (2007) 60.  

\bibitem{Campbell:2007ws} J.~M.~Campbell, F.~Maltoni, F.~Tramontano, Phys.\ Rev.\ Lett.\ {\bf 98} (2007) 252002.

\bibitem{Gong:2008sn} B.~Gong, J.~X.~Wang, Phys.\ Rev.\ Lett.\ {\bf 100} (2008) 232001.

\bibitem{Gong:2008hk} B.~Gong, J.~X.~Wang, Phys.\ Rev.\  D {\bf 78}  (2008) 074011 .

\bibitem{Gong:2008ft} B.~Gong, X.~Q.~Li,  J.~X.~Wang, Phys.\ Lett.\  B {\bf 673} (2009)  197.
[Erratum-ibid.\  {\bf 693}, 612 (2010)].

\bibitem{Brodsky:2009cf} S.~Brodsky, J.~P.~Lansberg, Phys.\ Rev.\  D {\bf 81} (2010) 051502.   

\bibitem{Lansberg:2010cn}  J.~P.~Lansberg, PoS  {\bf ICHEP 2010} (2010) 206,  arXiv:1012.2815 [hep-ph].

\bibitem{Kramer:1995nb} M.~Kramer, Nucl.\ Phys.\  B {\bf 459} (1996) 3.

\bibitem{Brambilla:2010cs} N. Brambilla {\it et al.}, Eur. Phys. J. C{\bf 71} (2011) 1534.

\bibitem{Lansberg:2008gk} J.P.~Lansberg, Eur. Phys. J. C {\bf 60} (2009) 693.


\bibitem{Gong:2010bk} B.~Gong, J.~X.~Wang, H.~F.~Zhang, arXiv:1009.3839 [hep-ph].

\bibitem{Ma:2010yw} Y.~Q.~Ma, K.~Wang, K.~T.~Chao,  Phys.\ Rev.\ Lett.\  {\bf 106} (2011) 042002.

\bibitem{Butenschoen:2010rq} M.~Butenschoen, B.~A.~Kniehl, arXiv:1009.5662 [hep-ph].


\bibitem{Zhang:2009ym}  Y.~J.~Zhang, Y.~Q.~Ma, K.~Wang,  K.~T.~Chao,  Phys.\ Rev.\  D {\bf 81} (2010) 034015.

\bibitem{Belle:2009nj}
  P.~Pakhlov {\it et al.}  [Belle],   Phys.\ Rev.\  D {\bf 79} (2009) 071101.


\bibitem{Nayak:2003jp} G.~C.~Nayak, M.~X.~Liu,  F.~Cooper, Phys.\ Rev.\ D {\bf 68} (2003) 034003.

\bibitem{Chung:2009xr} H.~S.~Chung, C.~Yu, S.~Kim, J.~Lee, Phys.\ Rev.\ D {\bf 81} (20010) 014020.

\bibitem{Lansberg:2010vq} J.~P.~Lansberg, Phys.\ Lett.\  B {\bf 695} (2011) 149.


\bibitem{Dahms} T. Dahms [CMS], talk at Quarkonium 2010.

\bibitem{Price:QQ10} D. Price [ATLAS], these proceedings. 

\bibitem{Arnaldi:HP10} R. Arnaldi [ALICE], talk at Hard Probes 2010.

\bibitem{Robbe:QQ10} P. Robbe [LHCb], these proceedings.

\bibitem{Scomparin:QQ10} E. Scomparin [ALICE], these proceedings.

\bibitem{boyer} B. Boyer [ALICE], Talk at RQW 2010, Oct. 25-28 2010, Nantes, France
[\href{http://indico.cern.ch/getFile.py/access?contribId=3&sessionId=0&resId=0&materialId=slides&confId=93174}{slides}]

\bibitem{ATLAS-JPsi} ATLAS~Coll., {\bf ATLAS-CONF-2010-062} (2010).

\bibitem{CMS:2010yr}  CMS~Coll.,  arXiv:1011.4193 [hep-ex].

\bibitem{LHCb-JPsi} LHCb~Coll.,  {\bf LHCb-CONF-2010-010} (2010).





\bibitem{Kopeliovich:1991pu} B.~Z.~Kopeliovich, B.~G.~Zakharov, Phys.\ Rev.\  D {\bf 44}  (1991) 3466.

\bibitem{Vogt:2001ky} R.~Vogt, Nucl. Phys. A {\bf 700}  (2002) 539 .

\bibitem{deFlorian:2003qf} D.~de~Florian, R.~Sassot, Phys. Rev. D {\bf 69}  (2004) 074028.

\bibitem{Eskola:1998iy} K.~J.~Eskola, V.~J.~Kolhinen, P.~V.~Ruuskanen, Nucl.\ Phys.\  B {\bf 535}  (1998) 351.

\bibitem{Eskola:1998df} K.~J.~Eskola, V.~J.~Kolhinen, C.~A.~Salgado, Eur.\ Phys.\ J.\  C {\bf 9}  (1999) 61.

\bibitem{Eskola:2008ca} K.~J.~Eskola, H.~Paukkunen,  C.~A.~Salgado, JHEP {\bf 0807}  (2008) 102.

\bibitem{Eskola:2009uj} K.~J.~Eskola, H.~Paukkunen,  C.~A.~Salgado, JHEP {\bf 0904} (2009) 065.

\bibitem{Ferreiro:2008wc} E.~G.~Ferreiro, F.~Fleuret, J.~P.~Lansberg,  A.~Rakotozafindrabe, Phys.\ Lett. B {\bf 680} (2009) 50.


\bibitem{Gavin:1991qk} S.~Gavin,  J.~Milana, Phys.\ Rev.\ Lett.\  {\bf 68}  (1992) 1834.

\bibitem{Matsui:1986dk} T.~Matsui,  H.~Satz, Phys. Lett. B {\bf 178}  (1986) 416.

\bibitem{Alessandro:2004ap} B.~Alessandro et al.[NA50], Eur. Phys. J. C {\bf 39}  (2005) 335.

\bibitem{Arnaldi:2007zz} R.~Arnaldi et al. [NA60], Phys. Rev. Lett. {\bf 99}  (2007) 132302.

\bibitem{Kopeliovich:2010jf}
  B.~Z.~Kopeliovich,
  Nucl.\ Phys.\  A {\bf 854} (2011) 187.

\bibitem{Kopeliovich:2001ee} B.~Kopeliovich, A.~Tarasov, J.~H\"ufner, Nucl.\ Phys.\  A {\bf 696} (2001) 669.


\bibitem{Kopeliovich:2010nw}
  B.~Z.~Kopeliovich, I.~K.~Potashnikova, H.~J.~Pirner and I.~Schmidt,
  Phys.\ Rev.\  C {\bf 83} (2011) 014912.

\bibitem{Adare:2007gn}  A.~Adare {\it et al.}  [PHENIX], Phys.\ Rev.\  C {\bf 77}  (2008) 024912, 
Erratum-ibid. C {\bf 79} (2009) 059901.


\bibitem{Kopeliovich:1999am} B.Z.~Kopeliovich, A.~Sch\"afer,  A.V.~Tarasov, Phys. Rev. D {\bf 62} (2000) 054022.


\bibitem{Strikman:2010ew}
  M.~Strikman,
  Nucl.\ Phys.\  A {\bf 854} (2011) 144.

\bibitem{Gribov:69zzz} V. N. Gribov, Sov. Phys. JETP {\bf 29} (1969) 483 [Zh. Eksp. Teor. Fiz. {\bf 56} (1969) 892].

\bibitem{Karmanov:1973va} V.~A.~Karmanov,  L.~A.~Kondratyuk, Pisma Zh.\ Eksp.\ Teor.\ Fiz.\  {\bf 18}, 451 (1973).

\bibitem{Lourenco:1996wn} C.~Lourenco, Nucl.  Phys.  A {\bf 610} (1996) 552c.

\bibitem{Gonin:1996wn} M.~Gonin, Nucl. Phys.  A {\bf 610} (1996) 404c.

\bibitem{Abreu:1997jh} M.C.~Abreu et al., Phys.  Lett. B {\bf 410} (1997) 337.

\bibitem{Scomparin:2009tg} E.~Scomparin  [NA60], Nucl.\ Phys.\  A {\bf 830}, (2009) 239c.

\bibitem{Arnaldi:2009ph}   R.~Arnaldi  [NA60], Nucl.\ Phys.\  A {\bf 830},  (2009) 345c.

\bibitem{Cortese:HP08} P. Cortese et al. (Na60), talk presented at the 3rd International Conference on Hard and Electromagnetic Probes of High-energy Nuclear Collisions,  Illa de A Toxa, Galicia, Spain, June 8-14, 2008.

\bibitem{Hufner:1998hf} J.~H\"ufner,  B.Z.~Kopeliovich, Phys.  Lett.  B {\bf 445} (1998) 223.

\bibitem{Arleo:2010rb} F.~Arleo et al., arXiv:1006.0818.

\bibitem{Dolejsi:1993iw} J.~Dolejsi, J.~Hufner,  B.~Z.~Kopeliovich, Phys.\ Lett.\  B {\bf 312}  (1993) 235.

\bibitem{Johnson:2000dm} M.~B.~Johnson, B.~Z.~Kopeliovich,  A.~V.~Tarasov, Phys.\ Rev.\  C {\bf 63} (2001) 035203.

\bibitem{GolecBiernat:1999qd} K.~J.~Golec-Biernat,  M.~Wusthoff, Phys.\ Rev.\  D {\bf 60}  (1999) 114023.

\bibitem{Alde:1991sw} D.~M.~Alde {\it et al.}, Phys.\ Rev.\ Lett.\  {\bf 66}  (1991) 2285.

\bibitem{Leitch:2000zzz} M.~J.~Leitch {\it et al.}  [FNAL E866/NuSea], Phys.\ Rev.\ Lett.\  {\bf 84}  (2000) 3256.

\bibitem{Badier:1983zzz} J.~Badier {\it et al.}  [NA3], Z.\ Phys.\  C {\bf 20} (1983) 101.

\bibitem{Kopeliovich:1984bf} B.~Z.~Kopeliovich,  F.~Niedermayer, Dubna preprint JINR-E2-84-834 [\href{http://ccdb4fs.kek.jp/cgi-bin/img\_index?8504113}{KEK Library}].

\bibitem{Brodsky:1992nq} S.~J.~Brodsky,  P.~Hoyer, Phys.\ Lett.\  B {\bf 298}  (1993) 165.

\bibitem{Kopeliovich:2005ym} B.~Z.~Kopeliovich, J.~Nemchik, I.~K.~Potashnikova, M.~B.~Johnson,  I.~Schmidt, Phys.\ Rev.\  C {\bf 72}  (2005) 054606.

\bibitem{Abt:2008ya} I.~Abt et al. [HERA-B], Eur. Phys. J. C {\bf 60}  (2009) 525.

\bibitem{Leitch:1999ea} M.J.~Leitch et al. [E866], Phys. Rev. Lett. {\bf 84}  (2000) 3256.

\bibitem{Alessandro:2006jt} B.~Alessandro et al. [NA50], Eur. Phys. J. C {\bf 48}  (2006) 329.

\bibitem{Arnaldi:2008er} R.~Arnaldi et al. [NA60], Eur. Phys. J. C {\bf 59}  (2009) 607. 

\bibitem{Glauber:1959zzz} R.J.~Glauber et al., Lectures on theoretical physics, Inter-Science, New York, (1959) Vol.~I.

\bibitem{DeVries:1987zzz} H.~DeVries, C.W.~DeJager,  C.~DeVries, Atomic Data and Nucl. Data Tables {\bf 36}  (1987) 495.

\bibitem{Vogt:1999dw} R.~Vogt, Phys.\ Rev.\  C {\bf 61}  (2000) 035203.

\bibitem{Woehri:QQ10} H.K.~Woehri, these proceedings.

\bibitem{Badier:1984pm} A.~Badier et al. [NA3], Zeit. Phys. C {\bf 26}  (1984) 489.

\bibitem{Boreskov:2003ck} K.G.~Boreskov,  A.B.~Kaidalov, JETP Lett. D{\bf 77} (2003) 599.

\bibitem{Arnaldi:2010ky} R.~Arnaldi et al., arXiv:1004.5523, submitted to Phys. Rev. Lett.

\bibitem{Fle10} S.J.~Brodsky, F.~Fleuret, J.P Lansberg, to appear.

\bibitem{tonyect} A. D. Frawley, talk at ECT* workshop on Quarkonium Production in Heavy-Ion Collisions, Trento (Italy), May 25-29, 2009 and at Joint CATHIE-INT mini-program ``Quarkonia in Hot QCD'', June 16-26, 2009 
[\href{http://www.int.washington.edu/talks/WorkShops/int\_09\_42W}{slides}].

\bibitem{Vogt:2004dh} R.~Vogt, Phys. Rev. {\bf C71}, 054902 (2005).

\bibitem{Lourenco:2008sk} C.~Lourenco, R.~Vogt,  H.~K.~Woehri, JHEP {\bf 0902} (2009) 014.


\bibitem{Ferreiro:2009ur}   E.~G.~Ferreiro, F.~Fleuret, J.~P.~Lansberg,  A.~Rakotozafindrabe, Phys.\ Rev.\  C {\bf 81} (2010) 064911.

\bibitem{Haberzettl:2007kj}   H.~Haberzettl,  J.~P.~Lansberg,
  Phys.\ Rev.\ Lett.\  {\bf 100} (2008) 032006;  J.~P.~Lansberg, J.~R.~Cudell,  Yu.~L.~Kalinovsky,
  Phys.\ Lett.\ B {\bf 633} (2006) 301.


\bibitem{Adare:2010fn} A.~Adare et al. [PHENIX], arXiv:1010.1246.

\bibitem{Hadjidakis:QQ10} C.~Hadjidakis [ALICE], these proceedings.

\bibitem{Fujii:2006ab} H.~Fujii, F.~Gelis,  R.~Venugopalan, Nucl. Phys. A{\bf 780} (2006) 146.


\bibitem{Brambilla:2004wf} N.~Brambilla, et al, CERN Yellow Report, CERN-2005-005, arXiv:hep-ph/0412158. 


\bibitem{Kramer:2001hh} M.~Kramer,  Prog.\ Part.\ Nucl.\ Phys.\  {\bf 47}  (2001) 141.

\bibitem{Lansberg:2006dh}  J.~P.~Lansberg,  Int.\ J.\ Mod.\ Phys.\  A {\bf 21}  (2006) 3857.


\bibitem{Lansberg:2008zm}
  J.~P.~Lansberg {\it et al.},  AIP Conf.\ Proc.\  {\bf 1038} (2008) 15.

\bibitem{Beneke:1995yb} M. Beneke,  I.Z. Rothstein, Phys.\ Lett.\ {\bf B372}  (1996) 157 , [Erratum-ibid. B{\bf 389} (1996) 769].

\bibitem{Beneke:1996yw}M. Beneke,  M. Kr\"amer, Phys.\ Rev.\ D{\bf 55}  (1997) 5269.

\bibitem{Braaten:1999qk} E. Braaten, B.A. Kniehl,,  J. Lee, Phys.\ Rev.\ D{\bf 62}  (2000) 094005.

\bibitem{Kniehl:2000nn}  B. A. Kniehl,  J. Lee, Phys.\ Rev.\ D{\bf 62}  (2000) 114027.

\bibitem{Leibovich:1996pa}  A.~K. Leibovich, Phys.\ Rev.\  D {\bf 56} (1997) 4412.

\bibitem{Faccioli:2008dx} P.~Faccioli, C.~Lourenco, J.~Seixas,  H.~K.~Wohri, Phys.\ Rev.\ Lett.\  {\bf 102} (2009) 151802.

\bibitem{Faccioli:2010ej}  P.~Faccioli, C.~Lourenco,  J.~Seixas, Phys.\ Rev.\ Lett.\  {\bf 105}  (2010) 061601.

\bibitem{Faccioli:2010ji}  P.~Faccioli, C.~Lourenco,  J.~Seixas, Phys.\ Rev.\  D {\bf 81}  (2010) 111502.

\bibitem{Faccioli:2010kd}  P.~Faccioli, C.~Lourenco, J.~Seixas,  H.~K.~Wohri, Eur.\ Phys.\ J.\  C {\bf 69} (2010) 657.



\bibitem{Butenschoen:2010px} M.~Butenschoen,  B.~A.~Kniehl, arXiv:1011.5619 [hep-ph].


\bibitem{Baranov:2007ay}  S.~P.~Baranov,  N.~P.~Zotov,  JETP Lett.\  {\bf 86} (2007) 435.

\bibitem{Baranov:2008yk}  S.~P.~Baranov,  N.~P.~Zotov,  JETP Lett.\  {\bf 88} (2008) 711.


\bibitem{cdf-note-9966}  CDF~Coll., CDF Note 9966.

\bibitem{Artoisenet:2009xh} P.~Artoisenet, J.~M.~Campbell, F.~Maltoni,  F.~Tramontano, Phys.\ Rev.\ Lett.\  {\bf 102}
  (2009) 142001.

\bibitem{Chang:2009uj} C.~H.~Chang, R.~Li,,  J.~X.~Wang, Phys.\ Rev.\ D {\bf 80} (2009) 034020.

\bibitem{Jungst:2008ip} M.~Jungst  [H1 and ZEUS], arXiv:0809.4150 [hep-ex].

\bibitem{Li:2008ym} R.~Li,  J.~X.~Wang, Phys.\ Lett.\  B {\bf 672}  (2009) 51.

\bibitem{Lansberg:2009db} J.~P.~Lansberg, Phys.\ Lett.\  B {\bf 679} (2009) 340.

\bibitem{baranov-these-proc} S.~P.~Baranov, these proceedings.


\bibitem{Jung:2009eq} Z.~J.~Ajaltouni {\it et al.}, ``Proceedings of the workshop: HERA and the LHC workshop series on the implications of HERA for LHC physics,'' arXiv:0903.3861 [hep-ph].

\bibitem{Lipka:2010xr} G. Corcella,  K. Lipka, `Heavy Flavours in DIS,  Hadron Colliders: Working Group Summary', arXiv:1008.2281 [hep-ph], Proceedings of Science, DIS 2010.

\bibitem{Nason:1987xz} P. Nason, S. Dawson,  R.K. Ellis, Nucl. Phys. B {\bf 303} (1988) 607.

\bibitem{Bonciani:1998vc} R. Bonciani, S. Catani,  M.L. Mangano, Nucl. Phys. B  {\bf 529} (1998) 424. 

\bibitem{Czakon:2009zw} M. Czakon, A. Mitov,  G. Sterman, Phys. Rev. D {\bf 80} (2009) 074017.

\bibitem{Cacciari:2005rk} M. Cacciari, P. Nason,  R. Vogt, Phys. Rev. Lett. {\bf 95} (2005) 122001.

\bibitem{Dokshitzer:1995ev} Yu.L. Dokshitzer, V.A. Khoze, S.I. Troian, Phys.Rev. D {\bf 53} (1996) 89.

\bibitem{Dokshitzer:1995qm} Yu.L. Dokshitzer, G. Marchesini,  B.R. Webber, Nucl. Phys. B {\bf 469} (1996) 93. 

\bibitem{Aglietti:2006yf} U. Aglietti, G. Corcella,  G. Ferrera, Nucl. Phys. B {\bf 775} (2007) 162.

\bibitem{Webber:1984zzz} B.R. Webber, Nucl. Phys. B {\bf 238} (1984) 492.

\bibitem{Andersson:1983ia} B. Andersson, G. Gustafson, G. Ingelman, T. Sj\"ostrand, Phys. Rept. {\bf 97} (1983) 31.

\bibitem{Corcella:2000bw} G. Corcella et al., JHEP {\bf 0101} (2001) 010.

\bibitem{Sjostrand:2006za} T. Sj\"ostrand, S. Mrenna,  P. Skands, JHEP {\bf 0605} (2006) 036.

\bibitem{Frixione:2002ik} S. Frixione,  B.R. Webber, JHEP {\bf 0206} (2002) 029.

\bibitem{Corcella:2005dk} G. Corcella,  V. Drollinger, Nucl. Phys. B {\bf 730} (2005) 82.

\bibitem{Armesto:2009fj} N. Armesto, L. Cunqueiro,  C.A. Salgado, Eur. Phys. J.  C  {\bf 63 }(2009) 679.

\bibitem{Armesto:2009ab} N. Armesto, G. Corcella, L. Cunqueiro,  C.A. Salgado, JHEP {\bf 0911} (2009) 122.

\bibitem{Lokhtin:2005px} I.P. Lokhtin,  A.M. Snigirev, Eur. Phys. J.\ C {\bf 45} (2006) 211.

\bibitem{Zapp:2008gi}  K. Zapp, G. Ingelman, J. Rathsman, J. Stachel,  U.A. Wiedemann, Eur. Phys. J. C {\bf 60} (2009) 617.

\bibitem{Renk:2008pp} T. Renk,  Phys. Rev. C {\bf 78} (2008) 034908.

\bibitem{Zapp:2008af} K. Zapp, J. Stachel,  U.A. Wiedemann, Phys. Rev. Lett. {\bf 103} (2009) 152302.

\bibitem{Schenke:2009gb} B. Schenke, C. Gale,  S. Jeon,  Phys. Rev. C {\bf 80} (2009) 054913.

\bibitem{kn} Y. Dokshitzer, talk at Quarkonium 2010.

\bibitem{Armesto:2003jh} N. Armesto, C.A. Salgado,  U.A. Wiedemann, Phys. Rev. D {\bf 69} (2004) 114003.

\bibitem{StarWhitePaper} J. Adams et al, Nucl. Phys. A {\bf 757} (2005) 102.

\bibitem{PhobosWhitePaper}  B.B. Back et al., Nucl. Phys. A {\bf 757} (2005) 28.

\bibitem{BrahmsWhitePaper}  I. Arsene et al.,Nucl. Phys. A {\bf 757} (2005) 1.

\bibitem{PhenixWhitePaper}  K. Adcox et al., Nucl. Phys. A {\bf 757} (2005) 184.

\bibitem{Adare:2010de} A.~Adare {\it et al.}  [PHENIX], arXiv:1005.1627 [nucl-ex].

\bibitem{Garishvili:2009ei} I.~Garishvili [PHENIX], Nucl. Phys. A {\bf 830} (2009) 625c-626c, arXiv:0907.5479 [nucl-ex]. 

\bibitem{Abelev:2006db} B.I. Abelev et al. [STAR], Phys. Rev. Lett. {\bf 98} (2007) 192301.

\bibitem{Abelev:2008hja} B.~I.~Abelev {\it et al.}  [STAR], arXiv:0805.0364 [nucl-ex].

\bibitem{Frawley:2008kk} A.~D.~Frawley, T.~Ullrich,  R.~Vogt, \PRep {\bf 462} (2008) 125.

\bibitem{Czakon:2007ej} M. Czakon, A. Mitov,  S. Moch, Phys. Lett. B {\bf 651} (2007) 147.

\bibitem{Czakon:2007wk} M. Czakon, A. Mitov,  S. Moch, Nucl. Phys. B {\bf 798} (2008) 210. 

\bibitem{Cacciari:2001cw} M. Cacciari,  S. Catani, Nucl. Phys. B {\bf 617} (2001) 253. 

\bibitem{Cacciari:2002pa} M. Cacciari,  P. Nason, Phys. Rev. Lett. {\bf 89} (2002) 122003.

\bibitem{Adare:2009ic} A.~Adare {\it et al.}  [PHENIX], Phys.\ Rev.\ Lett.\  {\bf 103} (2009) 082002.

\bibitem{Aggarwal:2010xp} M.~M.~Aggarwal {\it et al.}  [STAR], arXiv:1007.1200 [nucl-ex].

\bibitem{JPhysG17.1602} Yu.L. Dokshitzer, V.A. Khoze,  S.I. Troian, J. Phys. G {\bf 17} (1991) 1602. 

\bibitem{Dokshitzer:2005ri} Yu.L. Dokshitzer, F. Fabbri, V.A. Khoze, W. Ochs, Eur. Phys. J. C {\bf 45} (2006) 387.

\bibitem{PhysRev110.974} F.E. Low, Phys. Rev. {\bf 110} (1958) 974.

\bibitem{PhysRevLett20.86} T.H. Burnett,  N.M. Kroll, Phys. Rev. Lett. {\bf 20} (1968) 86.

\bibitem{Dokshitzer:QQ10} Yu.L. Dokshitzer, these proceedings.

\bibitem{Gossiaux:2008jv} P.B. Gossiaux,  J. Aichelin, Phys. Rev. C {\bf 78} (2008) 014904.

\bibitem{Baier:1996sk} R. Baier, Yu.L. Dokshitzer, A.H. Mueller, S. Peigne,  D. Schiff, Nucl. Phys. B {\bf 484} (1997) 265.

\bibitem{PhysRevD25.746} J.F. Gunion,  G. Bertsch, Phys. Rev. D {\bf 25} (1982) 746. 

\bibitem{Stavreva:2009vi} T.P. Stavreva,  J.F. Owens, Phys. Rev. D {\bf 79} (2009) 054017. 

\bibitem{PhysLettB93.451} S.J. Brodsky, P. Hoyer, C. Peterson, N. Sakai, Phys. Lett. B {\bf 93} (1980) 451. 

\bibitem{Abazov:2008er} V. Abazov, et al, [D0], Phys. Lett. B {\bf 666} (2008) 435.

\bibitem{Hirai:2007sx} M. Hirai, S. Kumano,  T.-H. Nagai, Phys. Rev. C {\bf 76} (2007) 065207.

\bibitem{Schienbein:2009kk} I. Schienbein, \etal, Phys. Rev. D 
{\bf 80} (2009) 094004.

\bibitem{Digal:2001iu} S.\ Digal, P.\ Petreczky,  H.\ Satz, \PL B {\bf 514} (2001) 57.

\bibitem{Karsch:1990wi}  F.\ Karsch,  H.\ Satz, \ZP C {\bf 51} (1991) 209.

\bibitem{Gupta:1992cd} S.~Gupta, H.~Satz, Phys.\ Lett.\  B {\bf 283} (1992) 439.

\bibitem{Digal:2001ue} S.~Digal, P.~Petreczky, H.~Satz, Phys.\ Rev.\  {\bf D64} (2001) 094015.

\bibitem{Karsch:2005nk} F.\ Karsch, D.\ Kharzeev,  H.\ Satz, \PL B {\bf 637}(2006) 75.

\bibitem{Karsch:1988ri} F.\ Karsch,  R.\ Petronzio, \PL B {\bf 212} (1988) 255.

\bibitem{Gavin:1988tw}  S.\ Gavin,  M.\ Gyulassy, \PL B {\bf 214} (1988) 241.

\bibitem{Hufner:1988wz} J.\ H\"ufner, Y.\ Kurihara,  H.\ J.\ Pirner, \PL B {\bf 215} (1988) 218.

\bibitem{Kharzeev:1997ry} D.\ Kharzeev, M.\ Nardi,  H.\ Satz, \PL B {\bf 405} (1997) 14.

\bibitem{Abreu:2002qs} M.\ C.\ Abreu et al.(NA50), \NP A{\bf 698} (2002) 127c.

\bibitem{BraunMunzinger:2000ep} P.\ Braun-Munzinger,  J.\ Stachel, \NP A {\bf 690} (2001) 119. 

\bibitem{Thews:2000rj} R.\ L.\ Thews, M.\ Schroedter,  J.\ Rafelski, \PR C {\bf 63} (2001) 054905.

\bibitem{Grandchamp:2002wp}  L.\ Grandchamp,  R.\ Rapp, \NP A {\bf 709} (2002) 415.

\bibitem{Thews:2005vj} M.\ Mangano,  R.\ L.\ Thews et al., \PR C {\bf 73} (2006) 014904. 


\bibitem{Kraan:2008hb} A.~C.~Kraan, AIP Conf.\ Proc.\ {\bf 1038} (2008) 45, arXiv:0807.3123 [hep-ex].

\bibitem{Albajar:1987ke} C.~Albajar {\it et al.}  [UA1], Phys.\ Lett.\  B {\bf 200}  (1988) 380.

\bibitem{Albajar:1990hf} C.~Albajar {\it et al.}  [UA1], Phys.\ Lett.\  B {\bf 256}  (1991) 112.

\bibitem{Abe:2002rb} K.~Abe {\it et al.} [Belle], Phys.\ Rev.\ Lett.\ {\bf 89} (2002) 142001.

\bibitem{Artoisenet:2008tc} P.~Artoisenet, ``Proceedings of  9th Workshop on Non-Perturbative Quantum Chromodynamics, Paris, France, 4-8 Jun 2007'', pp 21. arXiv:0804.2975 [hep-ph].

\bibitem{Klasen:2008mh} M.~Klasen,  J.~P.~Lansberg, Nucl.\ Phys.\ Proc.\ Suppl.\ {\bf 179-180} (2008) 226.


\bibitem{Artoisenet:2007qm}  P.~Artoisenet, F.~Maltoni, T.~Stelzer,  JHEP {\bf 0802} (2008) 102.


\bibitem{Alwall:2007st}  J.~Alwall, P.~Demin, S.~de Visscher {\it et al.},  JHEP {\bf 0709 } (2007)  028.

\end{thebibliography}
\end{document}